\documentclass[a4paper,11pt,hyper]{article}
\pdfoutput=1
\usepackage{jcappub}
\usepackage{amsfonts,latexsym,graphicx,amssymb,amsmath,mathrsfs,diagbox}
\usepackage[normalem]{ulem}
\usepackage[utf8]{inputenc}
\usepackage{comment}
\usepackage[T1]{fontenc}
\usepackage{lmodern}
\usepackage{dcolumn}

\newcommand{\np}{n_{\rm L}}

\newcommand{\pol}{\iota}

\newcommand{\bm}[1]{\hbox{\boldmath{$#1$}}}
\newcommand{\sbm}[1]{\hbox{\boldmath{\scriptsize$#1$}}}

\newcommand{\ITL}{\tilde{I}}

\newcommand{\IWL}{\tilde{I}^{\rm WL}_{i_1i_2\cdots i_n} (\bar{\bm{\theta}})}
\newcommand{\tlth}{{\rm TL}_3}
\newcommand{\tltw}{{\rm TL}_2}

\def\ba#1\ea{\begin{align}#1\end{align}}
\def\Pi#1{\Pi^{[#1]}}

\newcommand{\refeq}[1]{Eq.~(\ref{eq:#1})}

\newcommand{\update}[1]{\textcolor{black}{#1}}

\newif\iffigure
\figuretrue

\newif\ifcomment
\commenttrue

\newif\iffigwl
\figwlfalse

\begin{document}
\title{
  Galaxy imaging surveys as spin-sensitive detector for cosmological colliders
}

\author[a]{Kazuhiro Kogai,}
\emailAdd{kogai@nagoya-u.jp}
\affiliation[a]{Department of Physics and Astrophysics, Nagoya University,
Chikusa, Nagoya 464-8602, Japan}

\author[b]{Kazuyuki Akitsu,}
\affiliation[b]{Kavli Institute for the Physics and Mathematics of the Universe (WPI), UTIAS
The University of Tokyo, Kashiwa, Chiba 277-8583, Japan}
\emailAdd{kazuyuki.akitsu@ipmu.jp}

\author[c]{Fabian~Schmidt,}
\emailAdd{fabians@mpa-garching.mpg.de}

\affiliation[c]{Max--Planck--Institut f\"ur Astrophysik, Karl--Schwarzschild--Stra\ss e 1, 85748 Garching, Germany}

\author[a, d]{Yuko~Urakawa}
\emailAdd{yuko@physik.uni-bielefeld.de}
\affiliation[d]{Fakult\"at f\"ur Physik, Universit\"at Bielefeld, 33501 Bielefeld, Germany}

\abstract{Galaxy imaging surveys provide us with information on both the galaxy distribution and their shapes. In this paper, we systematically investigate the
  sensitivity of galaxy shapes to new physics in the initial conditions. For this purpose, we decompose the galaxy shape function into spin components, and compute the contributions to each spin component from both intrinsic alignment and weak lensing.
  We then consider the angular-dependent primordial non-Gaussianity, which is generated by a non-zero integer spin particle when active during inflation, and show that a galaxy imaging survey essentially functions as a spin-sensitive detector of such particles in the early universe. We also perform a forecast of the PNG generated from a higher spin particle, considering a Rubin Observatory LSST-like galaxy survey. 
}


\arxivnumber{}

\maketitle

\section{\label{sec:introduction}Introduction}
Over the past few decades, our understanding of the Universe has improved dramatically by virtue of various cosmological observations. In particular, the cosmic microwave background (CMB) observations have played an important role in establishing the concordance cosmology, i.e. $\Lambda$CDM model. The observed existence of dark matter and dark energy urges us to seek for a new idea beyond standard model of particle physics. 
In the next decade, their properties will be investigated more profoundly through
many galaxy surveys such as the \textit{Rubin Observatory Legacy Survey of Space and Time} (LSST)~\citep{Abell:2009aa}, \textit{Euclid}~\citep{Laureijs:2011gra}, the \textit{Nancy Grace Roman Space Telescope}, which is formerly known as \textit{WFIRST}~\citep{Spergel:2015sza}, and \textit{Spectro-Photometer for the History of the Universe, Epoch of Reionization, and Ices Explorer} (SPHEREx)~\citep{Dore:2014cca}. While the CMB only observes the two-dimensional surface, galaxy surveys chart the three-dimensional distribution of galaxies. 
Since enlarging the survey volume increases the constraining power, the upcoming galaxy surveys with a huge volume may enable a more precise measurement than the CMB.
More importantly, by probing smaller-scale perturbations than the CMB, galaxy surveys allow us to explore the structure of the Universe on a broader range of scales.

Through the galaxy imaging surveys, we can observe not only the galaxy distribution but also galaxy shapes. Similarly to the CMB, for which measuring the polarization provides complementary information to measuring the temperature fluctuation~\cite{Hanson:2013hsb, Ade:2014afa, Ade:2015tva, Adam:2016hgk}, measuring galaxy shapes supplies us with unique information about the structure of the Universe. 

In particular, the cosmic shear caused by the weak lensing effect (see \cite{Bartelmann:1999yn} for a review) is a powerful tracer of the dark matter distribution, which was reported e.g. in Dark Energy Survey~\citep{Troxel:2017xyo}, Kilo-Degree Survey~\citep{Asgari:2020wuj}, and Subaru HSC observation \citep{Hikage:2018qbn}.
Broadly speaking, the primary target of galaxy imaging surveys has been measuring the weak lensing effect or equivalently the dark matter distribution. Nevertheless, the observed galaxy shear can be generated also by the external tidal fields, which is known as `Intrinsic Alignment (IA)'~\cite{Coutts:1996, Lee:1999ii, Catelan:2000vm, Pen:2012ft} (for reviews, see Refs.~\cite{Troxel:2014dba,Joachimi:2015mma,Kiessling:2015sma,Kirk:2015nma}). The measurements of the cosmic shear have been used to constrain the cosmological parameters such as the equation of state for dark energy~\cite{Bridle:2007ft,Kirk:2011aw,Joachimi:2009ez,Johnston:2018nfi}, the fluctuation amplitude $\sigma_8$ and the matter density $\Omega_{\rm m0}$~\cite{Mandelbaum:2009ck,Hirata:2007np,Johnston:2018nfi,Kirk:2010zk,Samuroff:2018xuo}, and parameters for a deviation from general relativity~\cite{Dossett:2015nda}. Although the IA has been usually treated as a contamination to the weak lensing contribution, the IA itself includes precious information on the model of the Universe~\cite{Chisari:2013dda}.

One may wonder if measuring the IA can also be a powerful tool to seek a fossil of the primordial Universe. Inflation is the most successful scenario of the early universe, which consistently explains the seed of the fluctuations measured by the CMB and the large scale structure observations. A detection of the primordial gravitational waves (PGWs) generated during inflation can pin down the energy scale of the inflation, giving the benchmark value for building models of inflation. The measurement of the CMB $B$-mode~\cite{Ade:2014afa, Ade:2015tva, Matsumura:2013aja} is a promising way to detect the PGWs. Meanwhile, a consequence of the PGWs is also encoded in the cosmic shear through the gravitational lensing~\cite{Dodelson:2010qu} and also through the IA~\cite{Schmidt:2012nw, Schmidt:2013gwa} (see also Ref.~\cite{Chisari:2014xia}).

Without a detection of the PGWs, the energy scale of the inflation is just bounded by above. According to the latest Planck data~\cite{Akrami:2018odb}, the energy scale of inflation can be as high as $H_{\rm inf} < 2.7 \times 10^{-5} M_{\rm pl} \sim 6.6 \times 10^{13}$GeV in 95\%CL, which by far exceeds the accessible energy scale by any ground-based accelerator experiments. In such a high energy environment, an extremely massive particle, which cannot be explored by the conventional accelerator experiments, can be excited. If the particles had interacted with the inflaton, that drove the inflationary expansion, their imprints can be encoded in the primordial perturbations. The cosmological collider program~\cite{Chen:2009zp, Noumi:2012vr, Arkani-Hamed:2015bza} aims at probing massive particles which had existed in inflationary Universe through a precise measurement of the primordial perturbations. The information of the mass and spin for such heavy particles are encoded in the squeezed primordial non-Gaussianity (PNG) as the distinct oscillatory feature~\cite{Chen:2009zp} and angular dependence~\cite{Arkani-Hamed:2015bza}, respectively. In particular, it was shown that a particle with a different spin leads to the PNG with a different angular dependence~\cite{Arkani-Hamed:2015bza, Lee:2016vti, Arkani-Hamed:2018kmz} (see Ref.~\cite{Kim:2019wjo} for a different approach).

Detecting an imprint of these massive particles can open a unique window to explore new physics beyond standard model such as string theory. One may explore an imprint of grand unification encoded in the angular dependent PNG generated by the predicted spin-2 Kaluza-Klein gravitons~\cite{Kumar:2018jxz}. Going one step further, a typical prediction of string theory includes an infinite tower of higher spin particles, whose spins $s$ are larger than 2 (see e.g., the review article \cite{Rahman:2015pzl} and references therein). The seminal works by Vasiliev~\cite{Vasiliev:1990en, Vasiliev:2003ev} show that the no-go theorem for higher spin gauge theories known in the flat spacetime~\cite{Weinberg:1964ew, Weinberg:1980kq, Aragone:1979hx} can be evaded by introducing a negative or positive cosmological constant, including the de Sitter spacetime, which serves a good approximation of an inflationary spacetime. Probing a higher spin particle which might have existed during inflation is the most challenging goal of the cosmological collider program, which nevertheless can lead to a direct evidence of string theory.

The angular dependent PNG has been searched through CMB~\citep{Akrami:2019izv, Bartolo:2017sbu,Franciolini:2018eno, Bordin:2019tyb} and LSS~\cite{Schmidt:2015xka, Chisari:2016xki, Bartolo:2017sbu, Kogai:2018nse,Ballardini:2019wxj,MoradinezhadDizgah:2018ssw,MoradinezhadDizgah:2018pfo, MoradinezhadDizgah:2017szk,Biagetti:2020lpx}.
Since the squeezed PNG generated by a scalar field, the inflaton\footnote{The squeezed PNG from the inflaton in the large scale limit is known to be plagued with gauge issues~\cite{Tanaka:2011aj, Pajer:2013ana}.} or other spectator scalar fields, is typically bigger than the one by a non-zero spin particle, we need to establish an observational method which can selectively pick up the latter without being hidden by the former.  In this regard, observing galaxy shapes from imaging surveys has a compelling advantage~\cite{Schmidt:2015xka}. It is widely known that the angular independent squeezed PNG, whose amplitude is often expressed as $f_{\rm NL}$, causes the scale dependent bias of the galaxy number density~\cite{Dalal:2007cu,Slosar:2008hx,Jeong:2011as,Afshordi:2008ru,Baldauf:2011bh,Cabass:2018roz}. 
Similarly, the angular dependent PNG which can be generated by a spin-2 particle induces a scale-dependent bias for the IA correlation of the 2nd shape moment, which is the usual cosmic shear~\cite{Schmidt:2015xka, Chisari:2016xki, Kogai:2018nse}. Recently, this was confirmed in full $N$-body simulation by Ref.~\cite{Akitsu:2020jvx}. Remarkably, no matter how large $f_{\rm NL}$ is, it does not contaminate the second shape moment at large scales where the non-linear evolution is negligible. Furthermore, even if there exists an infinite tower of higher spin particles that generate different types of angular dependent (squeezed) PNG, they do not contribute to the 2nd shape moment either. This leads us to an ambitious speculation that an ideal galaxy imaging survey that can separately observe different galaxy shape moments may enable us to measure the PNG generated by massive particles with different spins separately, playing the role of a spin-resolved spectroscopy. In this paper, we confirm this speculation, generalizing the analysis in Ref.~\cite{Schmidt:2015xka} to an arbitrary $n$th galaxy shape moment.

The goal of this paper has two folds; First, we compute various effects that contribute to the $n$th galaxy shape moment, assuming the $\Lambda$CDM cosmology with the adiabatic Gaussian initial condition,
based on the formalism of Refs.~\citep{Schmidt:2012ne,Schmidt:2012nw}
(see \cite{Bernardeau:2009bm,Andrianomena:2014sya,Clarkson:2016zzi,Yoo:2018qba} for related approaches).
This generalizes known results for the 2nd moment to all higher moments,
similarly to the recent Ref.~\cite{Fleury:2018odh}.
In order to find new physics by observing galaxy shapes, the imprint of new physics needs to dominate these contributions at least over a certain range of scales. Second, as an example of new physics, we consider the angular-dependent PNG generated by higher spin particles, showing that galaxy imaging surveys indeed work as spin-sensitive detectors of the cosmological collider, which can separately measure the PNG generated by particles with different spins. 

The outline of this paper is as follows.
In Sec.~\ref{sec:projected}, we introduce the definition of the galaxy shape moment and its spin decomposition. In Sec.~\ref{sec:bias}, we estimate various contributions to the galaxy shape moments in the $\Lambda$CDM model with the adiabatic Gaussian initial condition, providing a benchmark value for a signal of new physics to exceed for its discovery. In Sec.~\ref{sec:IA}, we compute the scale-dependent bias generated by the angular dependent PNG from non-zero integer spin particles. In Sec.~\ref{sec:nr}, we forecast the detectability of the PNG from spin-4 particles, considering the future galaxy imaging surveys.  
In this paper, we assume the Planck fiducial \citep{Ade:2015xua} flat $\Lambda$CDM cosmology with $\Omega_{\rm b0}h^2=0.022$, $\Omega_{\rm CDM0}h^2=0.12$, $h=0.67$, $n_{\rm s}=0.9645$, $A_{\rm s}=2.2\times10^{-9}$ and $k_{\rm p}=0.05$~${\rm Mpc}^{-1}$.
With $\Omega_{\rm b0}$ and $\Omega_{\rm CDM0}$, $\Omega_{\rm m0}$ is defined as $\Omega_{\rm m0}\equiv\Omega_{\rm b0}+\Omega_{\rm CDM0}$.
We have used the public Boltzmann code ${\tt CAMB}$ \citep{Lewis:1999bs} to calculate the transfer function and the matter power spectrum at present.

\section{Spin decomposition of galaxy shapes} \label{sec:projected}

In this section, we introduce the galaxy shape function which we will then
use as a window to explore new physics.

\subsection{Definition and spin decomposition}  \label{SSec:Def}
First, let us define the shape function for a galaxy located at $(\tau, \bm{x})$ with $\tau$ being the conformal time as
\begin{align}
  g_{i_1i_2\cdots i_n}({\bm x},\tau) \equiv
 \frac{1}{\bar{B}({\bm x},\tau) R_*^n}\int_{y\leq R_*} d^3 {\bm y}\, y_{i_1}y_{i_2}\cdots y_{i_n} B({\bm x}+ {\bm y},\tau)\,,
\end{align}
where $\bm{x}$ and $\bm{y}$ denote the 3D spatial coordinates for the centroid of the galaxy and those measured from the 3D centroid of light emission from the galaxy, respectively. The indices $i_1,\, \cdots,\, i_n$ run from 1 to 3 and are raised and lowered by the 3-dimensional Kronecker delta. We have introduced the weight by using the brightness of the emission, $B({\bm x},\tau)$. Here, $R_*$ is the size of the galaxy and $\bar{B}({\bm x},\tau)$ is the total brightness of the galaxy, defined as $\bar{B}({\bm x}, \tau)=\int d^3 {\bm y} B({\bm x}+\bm{y},\tau)$. As is clear from the definition, $g_{i_1i_2\cdots i_n}$ is the rank-$n$ symmetric tensor in 3D. The centroid is defined so that $g_i({\bm x},\tau)$ vanishes. For a localized galaxy, the integral in $g_{i_1 i_2\cdots i_n}$ converges.

There are two different origins for the deformation of the galaxy shape,
\begin{align}
    g_{i_1 i_2\cdots i_n} ({\bm x},\, \tau)  = g^{\rm int}_{i_1 i_2\cdots i_n} ({\bm x},\, \tau) + g^{\rm WL}_{i_1 i_2\cdots i_n} ({\bm x},\, \tau)\,.  \label{Exp:I2orignes}
\end{align}
The first term is the intrinsic galaxy shape deformation and the second term is the deformation due to the weak lensing, which is caused by the gravitational field between galaxies and us. We will further discuss each component in the next section.

In galaxy imaging surveys, we can only observe the galaxy shapes projected onto the 2D sky which is orthogonal to the line of sight $\hat{\bm n}$. Operating with the projection tensor, defined as 
\begin{align}
    {\cal P}_{ij} \equiv \delta_{ij}-\hat{n}_i\hat{n}_j\,,
\end{align}
with $\hat{\bm{n}}$ being the unit vector along the line of sight, we obtain the galaxy shape on the projected plane as
\begin{align}
    I_{i_1i_2\cdots i_n}({\bm x},\, \tau)
    &= {\cal P}_{i_1}^{\;j_1}{\cal P}_{i_2}^{\;j_2}\cdots{\cal P}_{i_n}^{\;j_n} g_{j_1j_2\cdots j_n}({\bm x},\, \tau)\,.  \label{eq:relgI}
\end{align}
Over the area of an individual galaxy, which subtends a very small angle on the sky, we can approximate the line of sight $\hat{\bm{n}}$ as constant, so that the sky can be approximated as a two-dimensional plane with coordinates $\bm{\theta}$, where we choose the origin to correspond to the centroid of the galaxy image.

The surface brightness can be obtained by integrating the emission along the line of sight, 
\begin{align}
    I( \bar{\bm{\theta}} + \bm{\theta} ,\, \tau)= \int d y_{\parallel}\, B (\bm{x}+ \bm{y},\,\tau), \label{Def:I}
\end{align}
where we ignore any absorption or other radiative transfer effects. However, our results below are based solely on symmetry considerations, so they continue to hold even in the presence of a more complicated local mapping from emissivity to surface brightness. In Eq.~(\ref{Def:I}), using the source comoving distance $\chi$, defined as
\begin{align}
    \chi(z) \equiv \int_0^z \frac{dz'}{H(z')}\,, 
\end{align}
we have introduced $y_{\parallel}$, $\bar{\theta}_i$ , and $\theta_i$ as $y_{\parallel} \equiv \bm{y} \cdot \hat{\bm{n}}$,
\begin{align}
     \bar{\theta}_i \equiv \frac{1}{\chi} {\cal P}_{i}^{\;j} x_j\,, \qquad  \theta_i \equiv  \frac{1}{\chi} {\cal P}_{i}^{\;j} y_j  \,,
\end{align}
respectively. Here, the dependence on $\bm{x} \cdot \hat{\bm{n}}$ was identified with $\tau$ dependence. We then obtain the projected galaxy moment as
\begin{align}
I_{i_1i_2\cdots i_n}(\bar{\bm{\theta}},\, \tau)=\frac{\chi^n}{\bar{I}(\bar{\bm{\theta}},\, \tau) R_*^n}\int d^2\bm{\theta}\, \theta_{i_1} \cdots  \theta_{i_n}
I(\bar{\bm{\theta}}+ \bm{\theta},\,  \tau)\,,\label{eq:moment}
\end{align}
which is normalized by the total intensity $\bar{I}$,
\begin{align}
\bar{I}(\bar{\bm{\theta}},\, \tau)\equiv\int d^2\bm{\theta}\, I(\bar{\bm{\theta}} + \bm{\theta} ,\,  \tau) = \frac{1}{\chi^2} \int d^3 \bm{y} B (\bm{x} + \bm{y},\,\tau) \,.
\end{align}
Since $g_i$ vanishes, so do $I_i$, satisfying $I_i(\bar{\bm{\theta}},\, \tau) =0 $.

Next, let us introduce the spin$\pm n$ components of $I_{i_1 \cdots i_n}$.
Here, we generalize the approach of \cite{Schmidt:2012ne}.
An alternative, closely related approach to higher-order moments has recently
been presented in Ref.~\cite{Fleury:2018odh}. 
It will turn out that the contribution of the PNG generated by a spin-$n$ particle selectively appears in the spin$\pm n$ components at large scales. Using the orthonormal basis, $(\hat{\bm{n}},\, {\bm e}_\pol,\,{\bm e}_\psi)$, let us introduce the spin$\pm 1$ unit basis vectors as
\begin{align}
& \bm{m}_{\pm} \equiv \frac{1}{\sqrt{2}} \left( \bm{e}_\pol \mp i
\bm{e}_\psi \right) = \frac{1}{\sqrt{2}} \left(
\begin{array}{c}
\cos \pol \cos \psi \pm i \sin \psi \\
\cos \pol \sin \psi \mp i \cos \psi \\
- \sin \pol
\end{array}
\right)\,,
\end{align}
which satisfy
\begin{align}
& m^i_{\pm} m_{\pm\, i} = 0\,, \quad m^i_{\pm} m_{\mp\, i} = 1\,, \quad 
m^i_{\pm} \hat{n}_i=0. \label{Cond:mpm}
\end{align}
The projected $n$th moment tensor $I_{i_1i_2\cdots i_n}$ can be expanded by a product of $m_{\pm i}$ as
\begin{align}
    m_{\pm i_1} m_{\pm i_2} \cdots m_{\pm i_n},\quad (m_{\pm i_1} m_{\mp i_2} \cdots m_{\mp i_n} + ({\rm perms})),\quad  \cdots \,. \label{Exp:2Dbases}
\end{align}
Apart from the first terms, $m_{\pm i_1} m_{\pm i_2} \cdots m_{\pm i_n}$, all other terms, which contain both $\bm{m}_+$ and $\bm{m}_-$, have trace part contributions, as one can confirm by using Eq.~(\ref{Cond:mpm}). Therefore, the traceless components of $I_{i_1i_2\cdots i_n}$, $\ITL_{i_1i_2\cdots i_n}$, which satisfy, for any $p, q \in [1, n]$,
\begin{align}\label{eq:3d-traceless}
    \ITL_{i_1 \cdots i_p \cdots i_q \cdots i_n} = \ITL_{i_1 \cdots i_q \cdots i_p \cdots i_n}, \qquad {\cal P}^{i_p i_q}  \ITL_{i_1 \cdots i_p \cdots i_q \cdots i_n}=\delta^{i_p i_q}  \ITL_{i_1 \cdots i_p \cdots i_q \cdots i_n}= 0\,,
\end{align}
can be expanded by the first terms of Eq.~(\ref{Exp:2Dbases}) as
\begin{align}
\ITL_{i_1i_2\cdots i_n} & = {_n}\gamma\, m_{+i_1} m_{+i_2} \cdots m_{+i_n} + {_{-n}}\gamma\, m_{-i_1} m_{-i_2} \cdots m_{-i_n}\,,
\end{align}
where the coefficients ${_{\pm n}}\gamma$ are given by
\begin{align}
    {_{\pm n}}\gamma & \equiv m^{i_1}_{\mp} m^{i_2}_{\mp} \cdots m^{i_n}_{\mp} \ITL_{i_1i_2\cdots i_n} = m^{i_1}_{\mp} m^{i_2}_{\mp} \cdots m^{i_n}_{\mp} I_{i_1i_2\cdots i_n}\,. \label{Def:gamman} 
\end{align}
In the second equality, we inserted the trace part of $I_{i_1i_2\cdots i_n}$, which does not contribute to ${_{\pm n}}\gamma$, satisfying $m^i_{\pm} m^j_{\pm} \delta_{ij} = m^i_{\pm} m^j_{\pm} {\cal P}_{ij} = 0$. Since $\bm{m}_{\pm}$ transform as spin$\pm 1$ vectors under a rotation by an angle $\beta$, i.e. $\bm{m}_{\pm} \to e^{\pm i \beta} \bm{m}_{\pm}$, the $n$ products of $\bm{m}_{\pm}$ transform as spin$\pm n$ basis tensors, $\bm{m}_{\pm} \cdots \bm{m}_{\pm} \to e^{ \pm i n \beta} \bm{m}_{\pm} \cdots \bm{m}_{\pm}$,
being invariant under the rotation by an angle $\beta= 2 \pi N/n$ where $N$ is an integer.

\subsection{Visual image of spin-$n$ distortion}  \label{SSec:Image}
In the previous subsection, we argued that the traceless components, $\ITL_{i_1 \cdots i_n}$, can be expressed by the spin$\pm n$ components, $ {_{\pm n}}\gamma$. Before we calculate ${_{\pm n}}\gamma$, let us further discuss the spin decomposition to develop an intuitive understanding. Relevant discussions can be found e.g., in Refs.~\cite{Fleury:2017owg, Fleury:2018odh}. For this purpose, here we introduce the 2D coordinates on the projected plane along the directions of $\bm{e}_\pol$ and $\bm{e}_\psi$. In this paper, we express the 2D coordinates on the projected plane as $a_1,\, a_2,\, \cdots\in \{ \pol,\, \psi\}$, distinguishing them from the 3D spatial coordinates, expressed as $i_1,\, i_2,\, \cdots = 1,\, 2,\, 3$. Introducing 
\begin{align} 
    I_{a_1 \cdots a_n} (\bm{x},\, \tau) \equiv {e_{a_1}}^{i_1} \cdots {e_{a_n}}^{i_n} I_{i_1 \cdots i_n}(\bm{x},\, \tau)\,,  \label{Exp:2DI}
\end{align}
where ${e_{a}}^{i}$ is the $i$th component of $\bm{e}_a$, we can rewrite the spin$\pm n$ components, $ {_{\pm n}}\gamma$ as
\begin{align}\label{eq:gamman_exp}
      {_{\pm n}}\gamma 
      & = \frac{1}{2^{\frac{n}{2}}} \left[\, \sum_{l=0}^{[\frac{n}{2}]}  (-1)^l \dbinom{n}{2l}\, I_{\scriptsize \underbrace{\pol \pol \cdots \psi \psi \psi \psi}_{(n-2l, 2l)}} \pm i \sum_{l=0}^{[\frac{n-1}{2}]}  (-1)^l \dbinom{n}{2l+1}\, I_{\!\!\!\!\scriptsize\underbrace{\pol \pol \pol \cdots \psi \psi \psi}_{(n- 2l-1, 2l+1)}} \right]\,,
\end{align}
where $[\cdot]$ denotes the floor function, and
$(m, \, n)$ below the brace denotes the numbers of the indices $\pol$ and $\psi$, respectively. The coefficients are given by the binomial factors
\begin{align}
    \dbinom{n}{k} \equiv \frac{n!}{k! (n-k)!} .
\end{align}
Using the traceless component of $I_{a_1a_2\cdots a_n}$, $\ITL_{a_1a_2\cdots a_n} \equiv [I_{a_1a_2\cdots a_n}]^{\tltw}$, which satisfies 
\begin{align} \label{eq:2d-traceless}
    \ITL_{a_1 \cdots a_p \cdots a_q \cdots a_n} = \ITL_{a_1 \cdots a_q \cdots a_p \cdots a_n}, \qquad \delta^{a_p a_q}  \ITL_{a_1 \cdots a_p \cdots a_q \cdots a_n}= 0\,,
\end{align}
where $p$ and $q$ run over $p,\, q = 1,\, \cdots,\, n$, $ {_{\pm n}}\gamma$ can be given by a more compact expression as
\begin{align}\label{eq:gamman_simp}
    {_{\pm n}}\gamma = 2^{\frac{n}{2} -1}\left( 
    \ITL_{\scriptsize \underbrace{\pol \pol \cdots \pol \pol}_{(n, 0)}}
      \pm i 
    \ITL_{\scriptsize \underbrace{\pol \pol \cdots \pol \psi}_{(n-1, 1)}} \right)\,.
\end{align}
The derivation of Eq.~(\ref{eq:gamman_simp}) is explained in detail in App.~\ref{sec:comment_tl}.
Here and hereafter, we express traceless components of a tensor in 2D, which satisfy the second equation in Eq.~(\ref{eq:2d-traceless}), as $[\cdot]^{\tltw}$ and traceless components of arbitrary tensor in 3D, which satisfy (the second equality of) the second equation in Eq.~(\ref{eq:3d-traceless}), as $[\cdot]^{\tlth}$. Notice that an index for the former runs $a_1,\, a_2,\, \cdots = \pol,\, \psi$ and the one for the latter runs $i_1,\, i_2,\, \cdots = 1,\, 2,\, 3$. As discussed above, the spin $\pm n$ components, $ {_{\pm n}}\gamma$, can be expressed by the two traceless components of $\ITL_{a_1a_2\cdots a_n}$, which are the only independent components of $\ITL_{a_1a_2\cdots a_n}$. In fact, with a use of the symmetric and traceless condition (\ref{eq:2d-traceless}), all other components of $\ITL_{a_1a_2\cdots a_n}$ can be expressed by these two components in the right hand side of Eq.~(\ref{eq:gamman_simp}) as 
\begin{align}
   \ITL_{\scriptsize \underbrace{\pol \pol \pol \cdots \pol \psi \psi\cdots \psi}_{(n-2k, 2k)}} =  (-1)^k \ITL_{\scriptsize \underbrace{\pol \pol \pol \cdots \pol \pol \pol}_{(n, 0)}}  
    \label{eq:ITLeven}
\end{align}
for $k= 0,\, 1,\, \cdots , [\frac{n}{2}]$ and
\begin{align}
  \ITL_{\scriptsize \underbrace{\pol \pol \pol \cdots \pol \psi \psi\cdots \psi}_{(n-(2k+1), 2k+1)}}= (-1)^k  \ITL_{\scriptsize \underbrace{\pol \pol \pol \cdots \pol \pol \psi}_{(n-1, 1)}} \label{eq:ITLodd}
\end{align}
for $k= 0,\, 1,\, \cdots, [\frac{n-1}{2}]$. It will turn out that the imprint of the PNG generated by the spin-$n$ particles is encoded in the spin-$n$ component of the galaxy shape function without being mixed with the contributions of the PNG from particles with other spins.

Next, let us discuss a visual image of the spin-$n$ distortion, described by $\ITL_{a_1a_2\cdots a_n}$. For this purpose, using the 2D polar coordinates $(r,\, \phi)$ with which $\theta_\pol \equiv {e_{\pol}}^i  \theta_i$ and $\theta_\psi \equiv {e_{\psi}}^i \theta_i$ are given by
\begin{align}
     \theta_\pol  = r \cos \phi \,, \qquad   \theta_\psi = r \sin \phi \,,  \label{Exp:2Dcoordinates}
\end{align}
we express the surface brightness $I(\theta_a)$ in the Fourier series expansion as  
\begin{align}
    I(\theta_a) = \frac{c_0(r)}{2}+\sum_{n=1}^\infty \left[  c_n(r)\cos(n\phi)+s_n(r)\sin(n\phi) \right] \label{Exp:IFex}
\end{align}
with
\begin{align}
  c_n (r) \equiv \frac{1}{\pi}\int_{-\pi}^{\pi} d\phi I( \theta_a)\cos (n\phi)\,, \qquad s_n(r) \equiv  \frac{1}{\pi}\int_{-\pi}^{\pi} d\phi I(\theta_a)\sin (n\phi)\,.
\end{align}
Apart from $\theta_a = (\theta_{\pol},\, \theta_{\psi})$, the surface brightness also depends on $\tau$ and the centroid, while we don't write them explicitly. Using Eqs.~(\ref{Exp:2DI}) and (\ref{Exp:2Dcoordinates}), we can express the projected 2D galaxy moment $I_{a_1 \cdots a_n}$ as an $n$th moment described by the integral of $ \theta_{a_1} \cdots  \theta_{a_n}$, which has been used, e.g., in Ref.~\cite{Bartelmann:1999yn}. The condition $I_a = 0$ leads to $c_1= s_1 = 0$. Further, we have $s_0(r) =0$. Notice that the presence of a non-zero $c_n$ or $s_n$ with $n \geq 2$ indicates a deviation from the circular symmetric distribution $I(\theta_a) = I(|\theta_a|)$.

\begin{figure}
	\centering
	\includegraphics[width=1\linewidth]{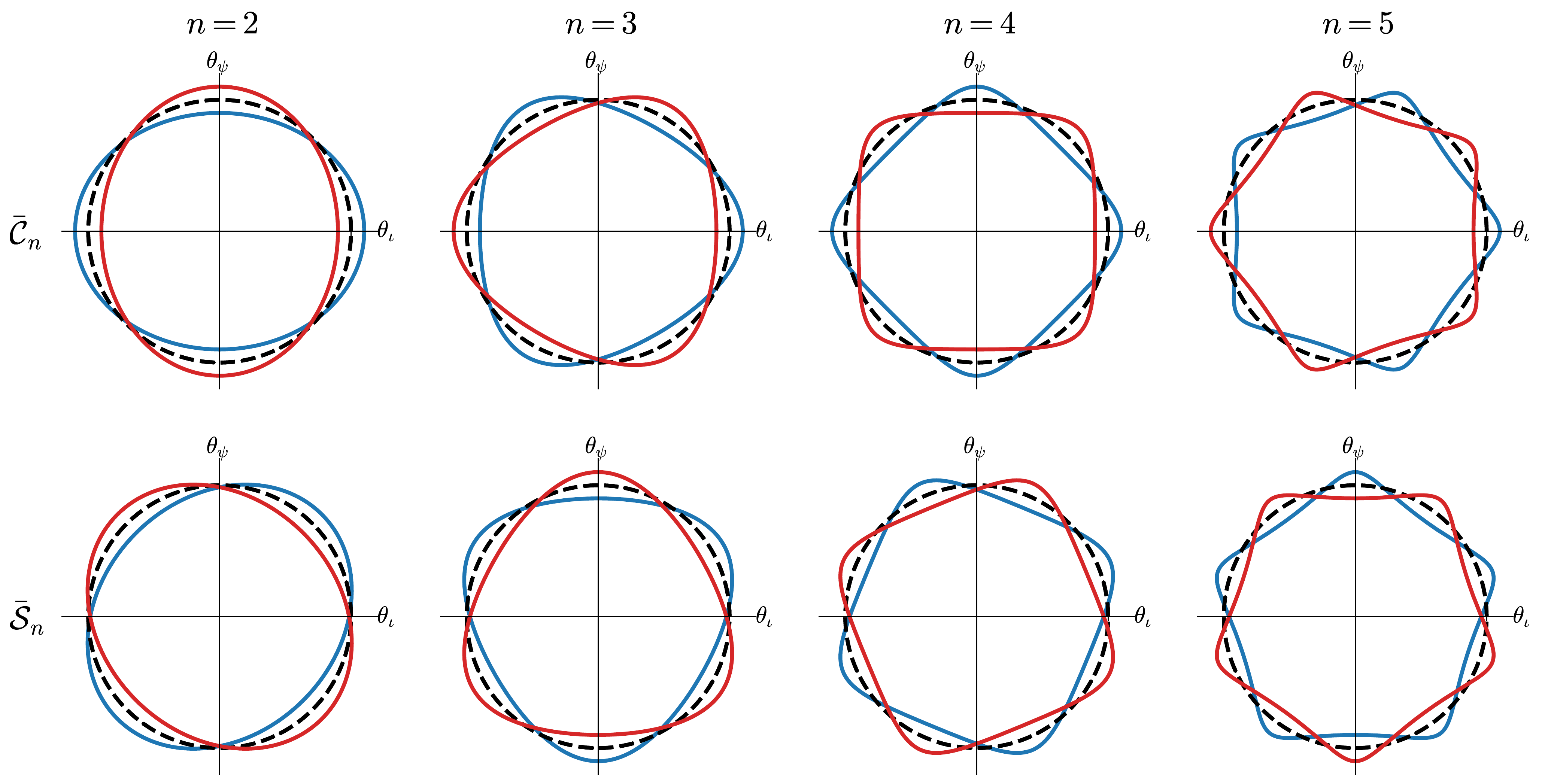}
	\caption{Contours of constant intensity $I(\bm{\theta})$, illustrating the distortions of galaxy images from a circularly symmetric distribution $I(\bm{\theta})= I(|\bm{\theta}|)$ corresponding to the different moments; specifically, we show nonzero $\bar{\cal C}_n\equiv{\cal C}_n/(2^{n-1}{\cal C}_0)$ (the upper row) and $\bar{\cal S}_n\equiv{\cal S}_n/(2^{n-1}{\cal C}_0)$ (the lower row) with $n=2,\,3,\, 4,\, 5$, related to the moments via Eq. (\ref{eq:momentsCS}). The blue lines correspond to positive values  and the red lines correspond to negative values. The dotted lines denote the unit circle. In Refs.~\citep{Goldberg:2004hh, Okura:2006fi}, the $n=3$ component generated by the weak lensing is called ``flexion.'' Ref.~\cite{Fleury:2018odh} refers to a very similar decomposition as ``Fourier decomposition'' of the image.}
	\label{fig:momentshape}
\end{figure}

We now show that the two independent spin-$n$ components of $\ITL_{a_1a_2\cdots a_n}$ with $n \geq 2$, $\ITL_{\pol \pol  \cdots \pol \pol}$ and $\ITL_{\pol \pol \cdots \pol \psi}$, correspond to $c_n(r)$ and $s_n(r)$, respectively. As a simple exercise, let us start with $n=2$, which corresponds to the usual cosmic shear. Inserting Eq.~(\ref{Exp:IFex}) into $\ITL_{a_1 a_2} = I_{a_1 a_2} - \delta_{a_1 a_2}  I_{bb}/2$, we obtain  
\begin{align}
   \ITL_{a_1 a_2} = \frac{1}{2{\cal C}_0}\left(
   \begin{array}{cc}
        {\cal C}_2 & {\cal S}_2 \\
        {\cal S}_2 & -{\cal C}_2
   \end{array}
   \right)\,,  
\end{align}
where we have introduced
\begin{align}
    {\cal C}_n \equiv \pi \int dr r^{n+1} c_n(r) \,,\qquad {\cal S}_n \equiv  \pi \int dr r^{n+1} s_n(r)\,.
\end{align}
Extending this discussion to $n \geq 3$, we obtain
\begin{align}
  \ITL_{\pol \pol \cdots \pol \pol} = \frac{{\cal C}_n}{2^{n-1} {\cal C}_0} , \qquad \ITL_{\pol \pol \cdots \pol \psi} = \frac{{\cal S}_n}{2^{n-1} {\cal C}_0} \,,
  \label{eq:momentsCS}
\end{align}
where we can avoid a messy computation by virtue of the formulae (the derivation can be found in App.~\ref{Sec:Apptensor})
\begin{align}
 r^n\cos(n\phi) &= 2^{n-1}\left[\theta_{a_1} \theta_{a_2} \cdots  \theta_{a_n} \right]^{\tltw} \big|_{a_1=\cdots=a_n=\pol}\label{eq:cosTL}\,,\\
 r^n\sin(n\phi) &= 2^{n-1}\left[ \theta_{a_1}  \theta_{a_2} \cdots \theta_{a_n} \right]^{\tltw} \big|_{a_1=\cdots=a_{n-1}=\pol ,a_n= \psi}\label{eq:sinTL}\,.
\end{align}
Thus, we find that the spin-$n$ components of the galaxy shape function are directly related to a deviation from the circular symmetric distribution which is proportional to $\cos (n\phi)$ or $\sin (n \phi)$. As it should be, these terms remain invariant under the rotation by an angle $\Delta \phi= 2 \pi N/n$ where $N$ is an integer. Figure.~\ref{fig:momentshape} shows the deviation of $I(\bm{\theta})$ from the circular symmetric distribution, described by $c_n(r)$ and $s_n(r)$ for $n \geq 2$.

The spin decomposition of the distortion due to the gravitational lensing was discussed e.g., in Refs.~\cite{Goldberg:2004hh, Bacon:2005qr, Fleury:2017owg, Fleury:2018odh, Schneider:2007ks, Clarkson:2016zzi, Clarkson:2016ccm}. In Refs.~\cite{Goldberg:2004hh, Bacon:2005qr}, the spin-3 components were dubbed the second flexion, ${\cal G}$. Ref.~\cite{Goldberg:2004hh} discusses how the distortion induced by the flexion can be measured based on the Shapelet formalism~\cite{Refregier:2001fd,Refregier:2001fe,Kelly:2003ws}.

\subsection{Angular power spectrum}  \label{SSec:Cl}
After having described how an individual galaxy image is decomposed into spin-$n$ components, we now generalize to spin-$n$ fields on the sky $\ITL_{i_1\cdots i_n}(\hat{\bm{n}})$, where $\hat{\bm{n}}$ is a unit vector denoting a location on the sky. This field can be estimated by dividing the survey area of a galaxy imaging survey into pixels, and measuring the average shape component of the many galaxies in each pixel.

Assuming only the adiabatic perturbation to the FLRW metric, and working to linear order in perturbations, we can relate $\ITL_{i_1\cdots i_n}$ to the matter density perturbation $\delta_{\rm m}(\bm{k})$ without loss of generality via
\begin{align}
  \ITL_{i_1 \cdots i_n}(\hat{\bm{n}}) = \left[{\cal P}_{i_1}\,^{j_1} \cdots {\cal P}_{i_n}\,^{j_n} \right]^{\tlth} i^n \int\frac{d^3{\bm k}}{(2\pi)^3} \int dz\frac{dN}{dz} \frac{D(z)}{D(0)}F^{(n)}(k,\mu,z) \hat{k}_{j_1} \cdots \hat{k}_{j_n} \delta_{\rm m}(\bm{k})e^{ix\mu}, \label{Exp:In}
\end{align}
where $z$ is the redshift, $dN/dz$ is the redshift distribution function of galaxies, $\delta_{\rm m}(\bm{k})$ is the total matter density fluctuation at present, $D(z)$ is the growth factor normalized as $(1+z)^{-1}$ during matter domination, $\mu=\hat{\bm{k}} \cdot \hat{\bm{n}}$, and $x=k\chi(z)$.
All physical effects are encoded in the kernel $F^{(n)}(k,\mu,z)$ which will be derived in the next two sections. As discussed above, there are two different contributions to $F^{(n)}(k,\mu,z)$, which stem from the intrinsic alignment and the weak lensing, respectively (at linear order in perturbations, there is no coupling between these two contributions). As a consequence of the global rotation symmetry, for scalar perturbations the tensor indices of $\ITL_{i_1 \cdots i_n}$ can be expressed only in terms of the (projected) wavenumbers $k_{j_1} \cdots k_{j_n}$. In this paper, we further assume that the matter power spectrum preserves the global rotation symmetry as
\begin{align}
    \langle \delta_{\rm m}({\bm k})\delta_{\rm m}({\bm k'}) \rangle = (2\pi)^3 P_{\rm m}(k)\delta_{\rm D}({\bm k}+{\bm k}').  \label{Exp:Pm}
\end{align}

Inserting Eq.~(\ref{Exp:In}) into Eq.~(\ref{Def:gamman}), we obtain $ {_{\pm n}}\gamma(\hat{\bm n})$. Multiplying the spin-lowering operator $\bar{\dh}$, defined as
\begin{align}
    \bar{\dh}\,{_{+n} f}(\hat{\bm n}) \equiv (1-\mu^2)^{\frac{1-n}{2}}\left[\frac{\partial}{\partial \mu}+\frac{i}{1-\mu^2}\frac{\partial}{\partial\psi}\right]\left[(1-\mu^2)^\frac{n}{2} \,{_{+n} f}(\hat{\bm n})\right],
\end{align}
on $_{+n}\gamma$, we obtain a scalar quantity $\gamma^{(n)}$ as
\begin{align}
  \gamma^{(n)}(\hat{\bm n})=\bar{\dh}^{n} {_{+n}\gamma}(\hat{\bm n}) = \int\frac{d^3{\bm k}}{(2\pi)^3} \int dz\frac{dN}{dz} \frac{D(z)}{D(0)}\delta_{\rm m}(\bm{k}) i^n\, \bar{\dh}^{n}\!\left[ F^{(n)}(k,\mu, z) \left(\hat{k}_+\right)^n e^{ix\mu} \right]\,, \label{Exp:gamman}
\end{align}
where we have introduced $\hat{k}_\pm \equiv m^i_{\mp} \hat{k}_i$. Using this expression, we compute the angular power spectrum defined as
\begin{align}
     \langle a^{(n)}_{lm} a^{(n')}_{l'm'}\rangle \equiv   C^{(n,\, n')}_l \delta_{ll'}\delta_{mm'}, \label{Def:Clnn}
\end{align}
with
\begin{align}
    a_{lm}^{(n)} \equiv \sqrt{(l-|n|)! \over (l+|n|)!} (-1)^n \int d \Omega_{\hat{\sbm{n}}}\,  Y_{lm}^* (\hat{\bm{n}}) \gamma^{(n)}(\hat{\bm n})\,, \label{Def:alm}
\end{align}
where $d \Omega_{\sbm{n}}$ denotes solid angle element.

Inserting Eq.~(\ref{Exp:gamman}) into Eq.~(\ref{Def:alm}) and exchanging the order of the integrals, we obtain
\begin{align*}
    a_{lm}^{(n)} &= \sqrt{(l-|n|)! \over (l+|n|)!} (-i)^n \int\frac{d^3{\bm k}}{(2\pi)^3} \int dz\frac{dN}{dz} \frac{D(z)}{D(0)}\delta_{\rm m}(\bm{k})  \int d \Omega_{\hat{\sbm{n}}}  Y_{lm}^* (\hat{\bm{n}}) \, \bar{\dh}^{n}\!\left[ F^{(n)}(k,\mu, z)  \left(\hat{k}_+\right)^n e^{ix\mu} \right] \,. \label{Exp:alm} 
\end{align*}
Following Ref.~\cite{weinberg2008cosmology}, let us define the $E$-mode and $B$-mode of the spin-$n$ components as
\begin{align}
    a_{lm}^{E(n)} \equiv \frac{a_{lm}^{(n)} + a_{l\, -m}^{*(n)} (-1)^{m}}{2}\,, \qquad a_{lm}^{B(n)} \equiv \frac{a_{lm}^{(n)} - a_{l\, -m}^{*(n)} (-1)^{m}}{2i} \,,
\end{align}
where we use the spherical harmonics which satisfies $Y_{lm}^* =(-1)^m Y_{l -m}$. The $E$-mode and $B$-mode distortions transform under the parity transformation as
\begin{align}
    a_{lm}^{E(n)}\rightarrow (-1)^l a_{l\,m}^{E(n)}\,, \qquad a_{lm}^{B(n)}\rightarrow -(-1)^l a_{l\,m}^{B(n)}\,. \label{Exp:parity}
\end{align}
Because of the factor $(-1)^l$, the $EE$ and $BB$ auto correlation $\langle a_{lm}^{X(n)} a_{l'm'}^{X(n)} \rangle$ with $X=E,\, B$ transforms as parity-even for $l-l'=2N$ and parity-odd for $l-l'=2N+1$, where $N$ is an integer. Meanwhile, their cross correlation $\langle a_{lm}^{E(n)} a_{l'm'}^{B(n)} \rangle$ transforms as parity-odd for $l-l'=2N$ and parity-even for $l-l'=2N+1$.

The matter density perturbation $\delta_{\rm m}(\bm{k})$ will be replaced with the matter power spectrum at present $P_{\rm m}(k)$, which preserves the global rotation symmetry, after taking the correlation in the angular power spectrum (\ref{Def:Clnn}). Then, since the contribution from each $\bm{k}$ to $a_{lm}^{(n)}$ does not depend on the angular direction of $\bm{k}$, we can simply compute the angular integral in $a_{lm}^{(n)}$, choosing $k^i=(0,\, 0,\, k)$ as
\begin{align*}
    & \int d \Omega_{\hat{\sbm{n}}}  Y_{lm}^* (\hat{\bm{n}}) \, \bar{\dh}^{n}\!\left[ \left(\hat{k}_+\right)^n F^{(n)}(k,\mu, z) e^{ix\mu} \right] \cr
    &\qquad \qquad =  \int d \Omega_{\hat{\sbm{n}}}  Y_{lm}^* (\hat{\bm{n}}) \frac{\partial^n}{\partial \mu^n} \left[ \left( \frac{1-\mu^2}{\sqrt{2}}\right)^nF^{(n)}(k,\mu, z) e^{i x \mu} \right] \,.
\end{align*}
Here, we used that a spin-$s$ function ${_s}f$ with $s>0$ which only depends on the polar angle $\theta$ satisfies~\citep{Newman:1966ub,Goldberg:1966uu,Thorne:1980ru,Schmidt:2012ne}
\begin{align}
\bar{\dh}^s {_s}f = \frac{\partial^s}{\partial \mu^s} \left[(1-\mu^2)^{s/2} {_s}f\right]\,,
\end{align}
where $\mu = \hat{\bm{k}} \cdot \hat{\bm{n}} = \cos \theta$. We can further rewrite the angular integral, using
\begin{align}
\frac{\partial^n}{\partial \mu^n}[(1-\mu^2)^n F^{(n)}(k,\mu, z) e^{ix\mu}]&=\frac{\partial^n}{\partial \mu^n}[(1+\partial_x^2)^n F^{(n)}(k,-i\partial_x, z) e^{ix\mu}] \cr
& =i^n(1+\partial_x^2)^n F^{(n)}(k,-i\partial_x, z) [x^ne^{ix\mu}]\,,
\end{align}
where the argument $\partial_x$ in $F^{(n)}$ means that the derivative operates on all $x$-dependent terms to the right in $F^{(n)}$.
Finally, using the partial wave expansion and the following formula
\begin{align}
(1+\partial_x^2)^s[x^sj_l(x)]=\frac{(l+s)!}{(l-s)!}\frac{j_l(x)}{x^s} \quad(s\geq0\,,s\in\mathbb{Z})\label{eq:bessle_spins},
\end{align}
whose derivation is summarized in App.~\ref{Sec:Appjl}, we obtain 
\begin{align}
    a_{lm}^{(n)} \sim \sqrt{\frac{2}{\pi}}\sqrt{(l+|n|)!\over (l-|n|)!} \int dk k^2 \int dz\frac{dN}{dz}\frac{D(z)}{D(0)} \delta_{\rm m}({\bm k})  F^{(n)}(k,- i \partial_x, z)\frac{j_l(x)}{x^s}  \,,  \label{Exp:alm_sim}
\end{align}
where we use $\sim$ instead of the equality, since we have already employed the global rotation symmetry of the matter power spectrum, which becomes manifest only after taking the correlation. Using this expression, we find that when $\tilde{I}_{i_1 \cdots i_n}$ is given by Eq.~(\ref{Exp:In}) and the matter power spectrum preserves the global rotation symmetry as given in Eq.~(\ref{Exp:Pm}), the $B$-mode distortion identically vanishes. The $B$-mode distortion can be generated from the primordial gravitational waves (PGWs)~\cite{Dodelson:2010qu, Schmidt:2013gwa} and also from a violation of the global rotation symmetry in the PNG~\cite{Kogai:2018nse}. In Ref.~\cite{Kogai:2018nse}, it was shown that the off-diagonal $EB$ correlation (parity-even) can be generated from the violation of the global rotation symmetry and in Ref.~\cite{Biagetti:2020lpx}, it was shown that the diagonal $EB$ correlation (parity-odd) can be generated from the helical PGWs. (Recall the argument below Eq.~(\ref{Exp:parity}).)

Using Eq.~(\ref{Exp:alm_sim}), we obtain the angular power spectrum for the $E$-mode of the spin-$n$ galaxy shape component as
\begin{align}
C_l^{(n,n')} &= \frac{2}{\pi}\sqrt{\frac{(l-|n|)!}{(l+|n|)!}\frac{(l-|n'|)!}{(l+|n'|)!}} \int dk k^2 P_{\rm m}(k)F^{(n)}_l(k)F^{(n')}_l(k) , \label{Exp:Clnn}
\end{align}
with
\begin{align}
    F^{(n)}_l(k) &= \left(\frac{1}{2}\right)^{\frac{n}{2}}\frac{(l+|n|)!}{(l-|n|)!}\int dz\frac{dN}{dz}\frac{D(z)}{D(0)} F^{(n)}(k,- i \partial_x, z) \left[\frac{j_l(x)}{x^n}\right]\,.\label{eq:fln}
\end{align}
Here and hereafter, for our notational brevity, we drop the index $E$, since we only consider the $E$-mode. Unless stated, we use the linear spectrum $P_{\rm L}(k)$ for the matter power spectrum $P_{\rm m}(k)$. For a later use, we introduce
\ba
P^{(n, n)}(k,\, z)\equiv \left(\frac{D(z)}{D(0)}\right)^2[F^{(n)}(k,z)]^2P_{\rm m}(k)\,,
\ea
which captures the rough structure of the integrand of $C_l^{(n,n')}$ for a given $k$ and $z$.

\section{Galaxy shapes as observable} \label{sec:bias}
In order to probe a consequence of new physics through shape observation, we first derive the power spectrum of the galaxy shape function $\tilde{I}_{i_1 \cdots i_n}$ predicted in the $\Lambda$CDM cosmology with adiabatic Gaussian initial condition. The galaxy distribution is known to represent a biased tracer of the matter density distribution. Similarly, the galaxy shape represents a biased tracer of the tidal field. All the contributions to $\tilde{I}_{i_1 \cdots i_n}$ can be summarized as
\begin{align*}
	\text{Shape observable $\tilde{I}_{i_1 \cdots i_n}$}
	\left\{\begin{array}{l}
	\text{Intrinsic shape}	\left\{\begin{array}{l}\text{(A) Linear alignment}\\
	\text{(B) Non-linear alignment}\\
	\text{(C) Noise}
        \end{array}\right.\\
	\text{(D) Weak lensing shear}\\
	\end{array}
	\right.
        \quad .
\end{align*}

The observed galaxy shape can be decomposed into the intrinsic shape (as would be measured by an observer close to the galaxy) and the weak lensing shear due to the propagation of light through the inhomogeneous universe (projection effect). 
We further divide the former into the linear alignment (A), which behaves as a linearly biased tracer of the tidal field, and the non-linear alignment (B), which corresponds to the higher order correction to the linear alignment.
In addition, since galaxy shapes are intrinsically non-spherical, there is a random (noise) contribution to the shape. For the 2nd moment, this is known as shape noise. For higher order shape contributions, the magnitude of the noise is uncertain. In the rest of this paper, we will ignore the noise in the galaxy shapes, considering a noise-free ideal observation.
Weak lensing also adds contributions to the shape both at linear and non-linear orders.

In this section, we compute the scale dependence of each contribution one by one. 
In Sec.~\ref{SSec:shapefn}, we will discuss the bias expansion of an arbitrary moment galaxy shape function. In Sec.~\ref{SSec:nla}, using this expansion, we will compute the linear and non-linear alignment contribution. In Sec.~\ref{SSec:WLsummery}, we briefly summarize the weak lensing contributions, deferring the detailed computation to App.~\ref{sec:wl}.

\subsection{Bias expansion}\label{SSec:shapefn}

The goal of the bias expansion of galaxy density and shapes is to capture
the effect of long-wavelength modes on these observables. By completely parametrizing
this dependence with free parameters, it is possible to describe the large-scale statistics of galaxies and their shapes rigorously \cite{Desjacques:2016bnm,Vlah:2019byq}, even though the formation of galaxies is extremely complex and nonlinear. This approach works via an expansion in perturbations and derivatives, and thus applies only on scales where the matter density $\delta$ is much less than one, which can be phrased as $k < k_{\rm NL}$ where $k_{\rm NL}$ is the nonlinear wavenumber, and which are larger than the length scale $R_*$ on which galaxies form ($k < k_* = 1/R_*$). 

The position and shape of a galaxy at a given time $\tau$ depend on the trajectory of the non-linear matter density around the galaxy under consideration. The matter density is affected by the small scale physics such as the gas cooling, radiation and star formation, and the past trajectory of matter around the focused galaxy. However, when we consider a much larger scale than $R_*$, which is the typical size of galaxies, the position and shape of galaxies are predominantly determined by the coarse-grained gravitational potential $\Psi$. In this paper, choosing the conformal Newtonian gauge, we express the line element for the perturbed FLRW spacetime as 
\begin{align}
    ds^2=a^2(\tau)[-(1+2\Psi)d\tau^2+(1+2\Phi)\delta_{ij}dx^idx^j]\,,
\end{align}
where $a(\tau)$ is the scale factor and $\Psi, \Phi$ correspond to the gravitational potential and the curvature perturbation in this gauge. In the absence of the anisotropic pressure, the Einstein equation relates $\Psi$ and $\Phi$ as $\Psi+\Phi=0$. Here, we set the background spatial curvature to 0 and ignore the tensor perturbation and the non-linear metric perturbations.

Focusing on much larger scales than $R_*$, we express the number density and shape of galaxies evaluated at the Eulerian spacetime coordinates $( \bm{x},\, \tau)$ as
\begin{align}
    \delta_{\rm n}({\bm x},\tau) &= {\cal N}[\delta({\bm x},\tau), K_{i_1 i_2}({\bm x},\tau), \{s_{i_1 \cdots i_n}({\bm x},\tau) \}_{n=3,\,4,\, \cdots}]({\bm x},\tau)\,,  \label{Exp:deltan}\\
    g_{i_1 \cdots i_n}({\bm x},\tau) &= {\cal G}_{i_1 \cdots i_n}[\delta({\bm x},\tau),  K_{i_1 i_2}({\bm x},\tau), \{s_{i_1\cdots i_n}({\bm x},\tau)\}_{n=3,\,4,\, \cdots}]({\bm x},\tau) \,,  \label{Exp:g} 
\end{align}
where $\delta$, $K_{ij}$, and $s_{i_1 \cdots i_n}$ are defined as
\begin{align}
    & \delta ({\bm x},\tau)\equiv \frac{\rho_{\rm m}({\bm x},\tau)}{\bar{\rho}_{\rm m}(\tau)}-1 =  \frac{1}{4\pi G\bar{\rho}_{\rm m}(\tau) a^2} \partial^2 \Psi(\bm{x}, \tau) \,, \label{Def:delta}\\
    &K_{i_1i_2}({\bm x},\tau) \equiv  \frac{1}{4\pi G\bar{\rho}_{\rm m}a^2}\left[\partial_{i_1}\partial_{i_2} \right]^{{\rm TL}_3}\Psi(\bm{x}, \tau) = {\cal D}_{i_1i_2}\delta(\bm{x},\tau)\,, \label{Def:Kij} \, \\
    &s_{i_1i_2\cdots i_n}({\bm x},\tau) \equiv \frac{(R_*)^{n-2}}{4\pi G\bar{\rho}_{\rm m}a^2} \partial_{i_1}\partial_{i_2}\cdots\partial_{i_n} \Psi(\bm{x}, \tau)\quad (n\geq3)\,, \label{Def:sij}
\end{align}
where $\bar{\rho}_{\rm m}$ denotes the background matter energy density and ${{\cal D}_{i_1i_2}}$ is defined as
\begin{align}
    {\cal D}_{i_1i_2} \equiv \frac{\partial_{i_1}\partial_{i_2}}{\partial^2}-\frac{1}{3}\delta_{i_1i_2}\,.
\end{align}
On the second equality of Eq.~(\ref{Def:delta}), we used the Poisson equation. The symmetric and traceless tensor $K_{i_1 i_2}$ is called the (second-rank) tidal tensor. The equivalence principle states that gravity can be locally eliminated, making the system equivalent to a local inertial system. This ensures that we can eliminate the gravitational potential $\Psi$ and its derivative $\partial_i \Psi$ by performing a local coordinate transformation to choose a free-falling observer frame. Therefore, in Eqs.~(\ref{Exp:deltan}) and (\ref{Exp:g}), we did not include the gravitational potential and its first derivative. As a result, the energy density $\delta$ and the tidal tensor $K_{i_1 i_2}$ give the leading order local contributions. 

Note that the galaxy density and shape in general depend on the matter density and tidal field on their entire past history. \update{A key feature of the perturbative approach is that this dependence can be incorporated order by order in perturbation theory. Here, one uses the fact that time and spatial dependence of the matter density and tidal field at each order in perturbation theory factorize. Hence, one can capture the non-locality in time by allowing for the contributions at each order to appear separately in the bias expansion \cite{Mirbabayi++:2014}; in case of the galaxy density, contributions from the non-locality in time} appear at third order. The same holds for the $n=4$ moment of galaxy shapes, while in case of the 2nd moment, there is a contribution at second order \cite{Vlah:2019byq}. Since we are interested in the leading contribution to $n=4$ moments, which is given by the second-order contribution, it is sufficient to work with the local, Eulerian density and tidal field in the bias expansion.

To determine the functional form of ${\cal N}$ and ${\cal G}_{i_1 \cdots i_n}$, we need to trace the history of the galaxy formation. Instead, here, along the line with Refs.~\cite{Assassi:2015jqa,Assassi:2015fma,Schmidt:2015xka,Vlah:2019byq}, we expand $\delta_{\rm n}$ and the traceless part of $g_{i_1 i_2}$, using the dominant local contributions as
\begin{align}
    \delta_{\rm n}({\bm x},\tau) &= b_{\delta}^{(0)}({\rm \tau})\delta({\bm x},\tau)+\frac{1}{2}b_{\delta^2}^{(0)}({\rm \tau})\delta^2({\bm x},\tau) + \frac{1}{2}b_{K^2}^{(0)}(\tau)[K_{ij}^2]({\bm x},\tau)+\cdots\,, \label{Exp:deltan2}\\
    \tilde{g}_{i_1 i_2}({\bm x},\tau) &= b_{K}^{\rm (2)}({\rm \tau})K_{i_1 i_2}({\bm x},\tau)+\frac{1}{2}b_{\delta K}^{\rm (2)}({\rm \tau})[\delta K_{i_1 i_2}]({\bm x},\tau) + \frac{1}{2}b_{K^2}^{\rm (2)}(\tau)[K_{i_1 j}K^{j}_{\;i_2}]^{{\rm TL}_3}({\bm x},\tau)+\cdots\,, \label{Exp:gij}
\end{align}
where $b^{(n)}_{X}$ with $X=\delta, \delta^2, \cdots$ and $n= 0, 2$ denote the bias parameters which are defined as the response of the mean $n$th moment galaxy shape to the change of $X$. 
For example, the linear bias parameter for the number density, $b_\delta^{(0)}$, is given by the response of the mean galaxy density with respect to the background matter density as $\partial \ln \bar{n}_g/\partial \ln {\bar \rho}_{\rm m}$ and the one for the 2nd moment galaxy shape, $b^{(2)}_{K}$, is given by the response of the mean 2nd galaxy shape to the external tidal field $K_{i_1 i_2}$ and so on. Here and hereafter we put a tilde to denote the traceless part of $g_{i_1 \cdots i_n}$, i.e. $\tilde{g}_{i_1 \cdots i_n} \equiv \left[ g_{i_1 \cdots i_n} \right]^{\tlth}$. 
In Eqs.~(\ref{Exp:deltan2}) and (\ref{Exp:gij}), the higher order terms in perturbation and the terms suppressed by $|R_* \partial| \sim R_*k$ with $k$ being the Fourier mode are abbreviated.

Similarly, we can expand $\tilde{g}_{i_1 \cdots i_n}$ with $n \geq 3$, using $\delta$, $K_{i_1 i_2}$, and $s_{i_1 \cdots i_n}$. For example, the 4th moment $\tilde{g}_{ijkl}$ up to the quadratic order in perturbation is given by
\begin{align}
	\tilde{g}_{i_1 i_2 i_3 i_4}({\bm x},\tau) &= b^{\rm (4)}_{K^2} \left[K_{i_1 i_2}({\bm x},\tau)K_{i_3 i_4}({\bm x},\tau)\right]^{{\rm TL}_3, {\rm sym}} + {\cal O}\left( (k R_*)^2 \right) \label{eq:4thbiasex}\,,
\end{align}
where the coefficient $b^{(4)}_{K^2}$ describes the response to the quadratic external tidal field\footnote{The explicit form of $\left[K_{ij}({\bm x},\tau)K_{kl}({\bm x},\tau)\right]^{{\rm TL}_3, {\rm sym}}$ is given by
\begin{align}
    &3\left[K_{ij}({\bm x},\tau)K_{kl}({\bm x},\tau)\right]^{{\rm TL}_3, {\rm sym}}\notag\\
    =&K_{ij} K_{kl}+K_{ik} K_{jl}+K_{il} K_{jk}\notag\\
    &-\left[\frac{2}{7}\sum_p\left(\delta_{ij}K_{kp}K_{pl}+\delta_{ik}K_{jp}K_{pl}+\delta_{il}K_{jp}K_{pk}+\delta_{jk}K_{ip}K_{pl}+\delta_{kl}K_{ip}K_{pj}+\delta_{jl}K_{ip}K_{pk}\right)\right.\notag\\
    &\left.+\frac{2}{35}\sum_{p,q}\left(\delta_{ij}\delta_{kl}+\delta_{ik}\delta_{jl}+\delta_{il}\delta_{jk} \right)K_{pq}K_{qp}\right]\,.
\end{align}}. The terms with $s_{i_1 i_2 i_3 i_4}$, $\delta\, s_{i_1 i_2 i_3 i_4}$, and $s_{i_1 i_2 j} {s^j}_{i_3 i_4}$ are all suppressed by $(k R_*)^2$.  

\subsection{Linear and non-linear intrinsic alignment}\label{SSec:nla}
Using the bias expansion, discussed in the previous subsection, we can compute the linear and non-linear intrinsic alignment. The linear alignment term which contributes to $F^{(n)}$, introduced in Eq.~(\ref{Exp:In}), is given by
\begin{align}
     F^{(n)}_{\rm LA}(k,z) &= 
    \begin{cases}
    b_\delta^{(0)} & (n=0)\\
    b_K^{(2)} & (n=2)\\
    b_s^{(n)} \left(\dfrac{k}{k_*}\right)^{n-2}& (n\geq3)
    \end{cases}
    \,,\label{eq:int_linear}
\end{align}
with $k_* \equiv 1/R_*$. We set $k_*$ as $k_*=1\,{\rm Mpc}^{-1}$ for our numerical results. Notice that for $n \geq 3$, the leading linear alignment term is a higher-derivative term and suppressed for $k \ll k_*$. Therefore, for $n \geq 3$, the most significant contribution of the intrinsic alignment on large scales comes from the non-linear alignment.

Non-linearity in the galaxy alignment appears in two different ways. First, the relations (\ref{Exp:deltan}) and (\ref{Exp:g}) are already non-linear, as shown in Eqs.~(\ref{Exp:deltan2}) and (\ref{Exp:gij}). Second, each argument of Eqs.~(\ref{Exp:deltan}) and (\ref{Exp:g}), i.e. $\delta,\, K_{i_1 i_2},\, \cdots $ evolves non-linearly, which is the usual non-linear clustering effect. While we consider $\tilde{I}_{i_1 \cdots i_n}$, which is given by projecting $\tilde{g}_{i_1 \cdots i_n}$ to the 2D sky, the loop integrals should be performed in the 3D space. The 3D loops are qualitatively different from the non-linear weak lensing effects, whose loop integrals are performed in the projected 2D space.

The non-linear alignment (NLA) for the 2nd moment was discussed, using simulation in Refs.~\cite{Tenneti:2014bca,Tenneti:2015aqa,Chisari:2016bqi}, galaxy catalogs in Ref.~\cite{Singh:2014kla} and bias expansion in Refs.~\cite{Blazek:2017wbz, Vlah:2019byq}. Here, let us provide a crude estimation of (3D) loop integrals from non-linear alignment for galaxy shape moments. 
In Ref.~\cite{Angulo:2015eqa}, the 1-loop contribution to the power spectrum of the number density of halos was computed based on EFTofLSS. 
\update{A representative term which is dominant in the limit of $k \to 0$ is given by}  
\update{
\ba
P_{\rm hh}(k)\ni 2 c_2^2 \int\frac{d^3{\bm p}}{(2\pi)^3}
\left[ \frac57 + \frac27 \frac{[{\bm p}\cdot (\bm{k}-\bm{p})]^2}{p^2 |\bm{k}-\bm{p}|^2}\right]^2 P_{\rm L}(p)P_{\rm L}(|{\bm k}-{\bm p}|)\,, \label{Eq:Phh}
\ea}where $P_{\rm hh}$ is the halo auto-power spectrum,
\update{$c_2$ is an effective second-order bias coefficient, and the kernel
  corresponds to second order bias contributions from density and tidal field squared. Note that this loop integral scales as $k^0$ in the limit $k\to 0$.}
\update{The scaling of loop contributions in the limit $k \to 0$ is known to be different between matter and biased tracers. While the 1-loop contributions of the former scale as $k^2 P_{\rm L}(k)$ and $k^4$, the one of the latter contain terms that scale as $k^0$.} As discussed in Refs.~\cite{peebles, Abolhasani:2015mra}, the scaling of the stochastic term for matter, $k^4$, can be derived from the mass and momentum conservation. Since the conservation is not imposed in our bias expansion, the stochastic terms with $\propto k^0$ appear without being canceled for biased tracers. 

\update{
  We can use this result to estimate the corresponding contribution to $n=2,4$ shape moments from the second-order bias expansion. While correlation functions of shape moments $\tilde{g}_{i_1 \cdots i_n}$ additionally have dimensionless tensor legs, these simply modify results by an ${\cal O}(1)$ factor, which we can ignore for this order-of-magnitude estimate. Thus, we can use a similar expression to Eq.~(\ref{Eq:Phh}),
\ba\label{eq:NLA}
P^{(n,n)}_{\rm NLA}(k) = 2 (b^{(n)}_{\rm NLA})^2 \int\frac{d^3{\bm p}}{(2\pi)^3}
\left[ \frac57 + \frac27 \frac{[{\bm p}\cdot (\bm{k}-\bm{p})]^2}{p^2 |\bm{k}-\bm{p}|^2}\right]^2
P_{\rm L}(p)P_{\rm L}(|{\bm k}-{\bm p}|)\,,  
\ea
to provide a crude estimation of the NLA contribution for the $n$th shape moments. Up to a ${\cal O}(1)$ coefficient, this matches loop computations for the 2nd moment previously derived in Refs.~\cite{Blazek:2017wbz, Vlah:2019byq}.}
This contribution is shown in Fig.~\ref{fig:scaling} by brown dot-dashed lines. The bias parameter is set to $|b_{\rm NLA}^{(n)}| = 0.1$ both for the 2nd and 4th shape moments. Since the 1-loop contribution which appears from the linear alignment term of $g_{ijkl}$ is suppressed by $(k/k_*)^2$ for the cross-correlation and $(k/k_*)^4$ for the auto-correlation, the above contributions only appear from $K^2$ type term like the one in Eq.~(\ref{eq:4thbiasex}).

As is known, when the linear power spectrum is given by a simple power law spectrum with
\begin{align}
    \np \equiv d \ln P_{\rm L}(k)/d \ln k|_{k=k_{\rm NL}},
\end{align}
being $k$-independent, the scaling of the $l$-loop contribution can be estimated  as\footnote{Here, the time evolution is computed, considering the EdS Universe. However, according to Ref.~\citep{Desjacques:2016bnm}, the result does not change significantly even if we consider the $\Lambda$CDM Universe. In this paper, we took into account the redshift dependence, using the linear growth factor. In Ref.~\citep{Foreman:2015lca}, analyzing up to the 2loop EFTofLSS predictions for matter power spectrum, it was shown that the counter term additionally introduces tiny time dependence (see Fig.~13 in Ref.~\citep{Foreman:2015lca}).} (see e.g., \cite{Pajer:2013jj, Assassi:2015jqa, Desjacques:2016bnm})
\begin{align}
 &  \left( \frac{k}{k_{\rm NL}} \right)^{(l+1)( \np+3)} \frac{1}{k^3} \sim \left(\frac{k}{k_{{\rm NL},z=0}}\right)^{(l+1)(\np+3)}\left(\frac{D(z)}{D(0)}\right)^{2(l+1)}  \frac{1}{k^3} \,.\label{eq:scalingloops}
\end{align}
Here, $k_{\rm NL}$ denotes the non-linear scale, at which the dimensionless matter power spectrum becomes unity. The 1-loop contribution shown in Fig.~\ref{fig:scaling} \update{or given by Eq.~(\ref{eq:NLA})} follows this scaling around $k_{\rm NL}$, which explicitly corresponds to $n_{\rm L}=\update{-1.7}$ at $k_{{\rm NL},z=0} = 0.25h\,{\rm Mpc}^{-1}$.
For the actual Universe, since the power spectrum does not scale with a single power, the estimation becomes more complicated (see e.g., Ref.~\cite{Foreman:2015uva}). Since $k_* > k_{\rm NL}$, these loop contributions dominate the linear alignment terms, suppressed by $(k/k_*)^2$ for the 4th moment as given in Eq.~(\ref{eq:int_linear}).

For the $n$th moment function with $n=2m\,(m=1,\, 2,\, \cdots)$, the leading contributions, which are not suppressed by powers of $k/k_*$, only appear from the non-linear bias expansion terms schematically in the form
$\tilde{g}_{i_1 \cdots i_{2m}} \sim  (K)^m$, where the tensor indices of $K_{ij}$ are abbreviated. For a larger $n$, the non-linear alignment contributions without $k/k_*$ suppression start with higher loops. For instance, the auto-correlation of $\tilde{g}_{i_1 \cdots i_6}$ starts from 2loop contributions.

\subsection{Weak lensing}\label{SSec:WLsummery}
In this subsection, we briefly summarize the weak lensing contribution to the galaxy shape function, whose detailed computation can be found in App.~\ref{sec:wl}. At leading order in the lensing deflection (Born approximation), the weak lensing contribution can be expressed by using the deformation matrix $A_{ij}$~\cite{Mellier:1998pk} as 
\begin{align}
    A_{ij} (\tilde{\bm{\theta}}) \equiv  \frac{\partial \tilde{\theta}_{{\rm s} i}}{\partial \tilde{\theta}^j}=\delta_{ij} + \int^{\chi}_0 d\chi' \frac{\chi - \chi'}{\chi} \chi' \partial_{\perp i}\partial_{\perp j} (\Phi (\chi',\, \tilde{\bm{\theta}})-\Psi (\chi',\, \tilde{\bm{\theta}})), \label{Exp:Aij}
\end{align}
with $\partial_{\perp i} \equiv {{\cal P}_i}^{j} \partial_j$. Here, $\tilde{\theta}$ and $\tilde{\theta}_{\rm s}$ denote the 2D coordinates on the image plane and the source plane, respectively. We put a tilde to distinguish them from the offset from the centroid, introduced in the previous section. Because of the spatial inhomogeneity of $\Phi$ and $\Psi$, the deformation matrix $A_{ij}$ depends on $\tilde{\bm{\theta}}$.

The weak lensing contribution to the $n$th moment function, $\tilde{I}^{\rm WL}_{i_1 \cdots i_n}$, should have $n$ tensor indices which are projected to the 2D plane. For the 2nd moment, the lensing contribution, $\tilde{I}^{\rm WL}_{ij}$ is nothing but the deformation matrix $A_{ij}$. Meanwhile, for $n \geq 3$, the additional tensor indices are supplied either by acting with angular derivatives or by multiplying several $A_{ij}$s. The former yields a suppression by powers of $k/k_*$, while the latter contributes as two-dimensional loop integrals.

The linear weak lensing (LWL) contribution, whose indices can thus only be supplied by operating with the spatial derivative operator, are suppressed as
\begin{align}
 F^{(n)}_{\rm LWL}(k,z) & \propto 
    \left(\dfrac{k}{k_*}\right)^{n-2}  & (n \geq 2)\label{eq:treewl_power}\,,
\end{align}
where we have dropped the redshift dependence. This result is in agreement with Eq.~(84) of \cite{Fleury:2018odh}. 
Therefore, similarly to the intrinsic alignment, for $n > 2$, the leading contribution for $k \ll k_*$ stems from the loop contributions. Notice that since $A_{ij}$, which is given by integrating along the line of sight, is a 2D object, loop integrals of the weak lensing, expressed by non-linear terms of $A_{ij}$, should be performed in the projected 2D space. Therefore, weak lensing loops are qualitatively different from the loop contributions in the intrinsic alignment, which are given by first computing loop integrals in 3D space and subsequently projecting it to the 2D space. As discussed in App.~\ref{sec:wl}, the lensing loops obtained after projection are much smaller than the latter for the auto-correlation of the 4th moment.

\subsection{Summary}
In order to probe an imprint of new physics encoded in galaxy shapes, we need to understand the contributions predicted in the concordance cosmology, which is $\Lambda$CDM cosmology, so in general relativity, with the Gaussian adiabatic initial condition. The sweet spot of hunting new physics is located at scales or angular multipoles where a signal of new physics can be exposed without being hidden by the contributions predicted in the standard cosmology.

In this section, estimating each contribution to the $n$th moment shape function, we have found that the situation is very different between $n=2$, which has been widely investigated, and $n\geq 3$. This is essentially because for $n \geq3$, both 
the intrinsic alignments and the weak lensing are suppressed by a factor of $(k/k_*)^2$ at linear order in perturbations. Then, the leading contributions appear from the loop contributions, especially those in intrinsic alignment. Meanwhile, as is widely known, the 2nd moment or the spin$\pm2$ component includes the linear alignment and the linear weak lensing effect which are not suppressed at large scales.

\begin{figure}
    \centering
    \includegraphics[width=155mm]{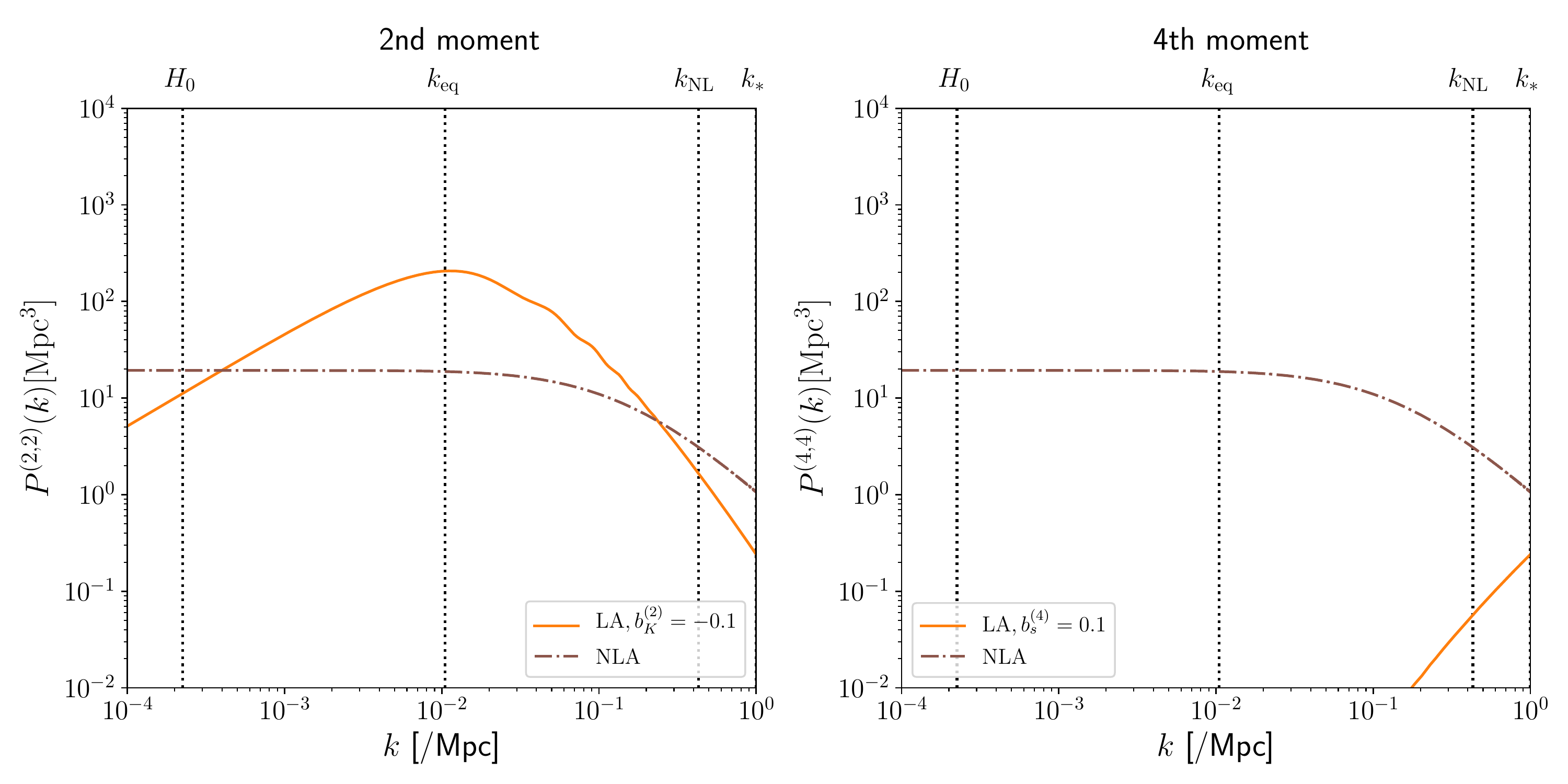}
    \caption{The 3D power spectra of each contribution at $z=1.5$ for the 2nd moment (left) and the 4th moment galaxy shape function (right).  The orange solid lines are the linear alignment and the brown dash-dotted lines are the non-linear alignment contributions. The bias parameters are set as $(b_K^{(2)},b_K^{(4)})=(-0.1, 0.1)$. We used $k_{{\rm NL},z=0} = 0.25[h{\rm /Mpc}]$.}
    \label{fig:scaling}
\end{figure}

Figure~\ref{fig:scaling} shows each contribution to the auto-correlation of the 2nd moment, $P^{(2,2)}$, and 4th moment galaxy shape, $P^{(4,4)}$, respectively, evaluated at $z=1.5$. 
As the redshift becomes closer to $z=0$, LA and $l$ loop NLA in Fig.~\ref{fig:scaling} change as $D^2(z)$ and $D^{2(l+1)}(z)$, respectively.
Since Fig.~\ref{fig:scaling} shows results for the 3D power spectrum, it is mostly relevant for considering shape correlations in spectroscopic surveys, where the redshift space distortion is yet to be included.  We will later perform
the projection to angular correlations.

\section{Intrinsic alignment from angular dependent PNG} \label{sec:IA}
As a candidate of new physics, in the rest of the paper, we consider the PNG generated by the higher spin particle excited in inflationary Universe. In Refs.~\cite{Arkani-Hamed:2015bza, Lee:2016vti, Kumar:2017ecc}, it was shown that the primordial bispectrum generated from the massive particles with non-zero spins exhibits the characteristic dependence on the angle between the wave vectors of the bispectrum. In this section, we show that such an angular dependent PNG generates an intrinsic distortion of the galaxy shape, extending the analysis in Ref.~\cite{Schmidt:2015xka} to an arbitrary spin.

\subsection{PNG from higher spin particles}
Chen and Wang showed that the imprints of additional massive scalar particles can be encoded as an oscillatory contribution in the squeezed bispectrum~\cite{Chen:2009zp}. This was extended to a general even spin particle by Arkani-Hamed and Maldacena in Ref.~\cite{Arkani-Hamed:2015bza} and to a general integer spin particle, including odd spin particles, by Lee et al. in Ref.~\cite{Lee:2016vti} (see also Refs.~\cite{Ghosh:2014kba, Arkani-Hamed:2018kmz}). Such a higher spin particle is naturally predicted in string theory as a part of the infinite tower of the higher spin states~\cite{Rahman:2015pzl} (see also Ref.~\cite{Camanho:2014apa}). Since the oscillatory feature in the PNG is characterized by the non-analytic scaling determined by the particle's spin $s$ and mass $M_s$, a precise measurement of the squeezed limit PNG may allow us to measure the mass and spin of the particles excited almost around the GUT scale. This program was dubbed the cosmological collider~\cite{Arkani-Hamed:2015bza}.

When the mass of the exchanged particle, $M_s$, is much heavier than $H_{\rm inf}$, integrating out the massive particle yields a local contribution to the Lagrangian of the inflaton. Unless we know the original Lagrangian of the inflaton, we cannot selectively capture the information on the exchanged massive particle by analyzing the local contribution. On the other hand, when $M_s$ is of ${\cal O}(H_{\rm inf})$, the contribution of the particle exchange becomes non-local.

The squeezed bispectrum with the soft exchange of the even spin particles in the slow-roll limit is given by~\cite{Arkani-Hamed:2015bza}
\begin{align}
B_{\Phi}({\bm k}_{\rm S}, {\bm k}_{\rm L})=\sum_{s = 0, 2, 4, \cdots}
{\cal A}_{s} \mathcal{P}_{s}(\hat{\bm k}_{\rm L} \cdot\hat{\bm k}_{\rm S})
f_s\left(\frac{k_{\rm L}}{k_{\rm S}} \right)P_{\Phi}(k_{\rm L})P_{\Phi}(k_{\rm S})
\left[1+\mathcal{O}\left(\frac{k^2_{\rm L}}{k^2_{\rm S}} \right) \right]\,,
\label{eq:NGIC}
\end{align}
with $k_{\rm L}/k_{\rm S}\ll 1$, $\hat{\bm k}_{\alpha} \equiv {\bm k}_\alpha/k_\alpha\,(\alpha = {\rm L,S})$ and ${\cal P}_s$ denotes the Legendre polynomials. For example, $f_s$ for a particle in the principal mass series is given by
\begin{align}
	f_s\left(\frac{k_{\rm L}}{k_{\rm S}}\right) = \left(\frac{k_{\rm L}}{k_{\rm S}}\right)^\frac{3}{2}\cos\left[\nu_s\ln\left(\frac{k_{\rm L}}{k_{\rm S}}\right)+\psi_s\right]\,,  \label{Exp:fs}
\end{align}
with
\begin{align}
\nu_s=\left\{ \begin{array}{ll}
\sqrt{\left(\dfrac{M_0}{H}\right)^2-\dfrac{9}{4}} & (s=0) \\
\sqrt{\left(\dfrac{M_s}{H}\right)^2-\left(s-\dfrac{1}{2}\right)^2} & (s=2,4,\cdots)
\end{array} \right.\,.
\end{align}
Reflecting the fact that the contribution of the massive particle is non-local, the bispectrum acquires the oscillatory contribution as a consequence of the non-analytic scaling. The contribution from the odd spin particles is cancelled in the leading order of $k_{\rm L}/k_{\rm S}$~\cite{Arkani-Hamed:2015bza}, but appears in the sub-leading order with $(k_{\rm L}/k_{\rm S})^{5/2}$~\cite{Lee:2016vti} (see also Ref.~\cite{Kumar:2017ecc}).

In what follows, we show that the contribution of the spin-$s$ particles to the PNG, described by Eq.~(\ref{eq:NGIC}), results in the generation of the intrinsic spin-$s$ shape moment, $\tilde{I}^{\rm int}_{i_1 i_2 \cdots i_s}$, keeping the spin-$n$ shape moment with $n \neq s$ intact as far as the non-linear evolution is negligible. To keep the generality, in the following, we use a generalized ansatz of the squeezed bispectrum (\ref{eq:NGIC}) with $s=0,\, 1,\, 2,\, \cdots$, including the odd $s$ and leaving $f_s$ as a general function of $k_{\rm L}/k_{\rm S}$.  As was shown in Ref.~\cite{Lee:2016vti}, the sub-leading contribution from a spin-$s$ particle includes ${\cal P}_l$ with $l \neq s$. Therefore, at the sub-leading orders, the $s$th galaxy shape moment does not selectively single out the contribution of the spin-$s$ particle.

When the function $f_s$ can be rewritten into the separable form as
\begin{align}
    f_s \left( \frac{k_{\rm L}}{k_{\rm S}} \right) = g_s \left( \frac{k_{\rm L}}{k_{\rm p}} \right) \, h_s \left( \frac{k_{\rm p}}{k_{\rm S}}\right)\,, \label{Exp:fsfact}
\end{align}
with a pivot scale $k_{\rm p}$,
the squeezed bispectrum (\ref{eq:NGIC}) modifies the local power spectrum at ${\bm x}$ as~\cite{Assassi:2015jqa,Assassi:2015fma} 
\begin{align}
    P_{\Phi}({\bm k}_{\rm S};{\bm x}|\Phi_{\rm L})
    =\left[
    1+\sum_{s = 0, 1, 2, 3, \cdots}
    h_s\!\left(\frac{k_{\rm p}}{k_{\rm S}} \right) [\hat{k}_{\rm S}^{i_1}\cdots \hat{k}_{\rm S}^{i_s}]^{\rm TL_3}\alpha_{{\rm L}\,i_1\cdots i_s}
    \right]P_{\Phi}(k_{\rm S}),\label{eq:alps}
\end{align}
where $\alpha_{{\rm L}\,i_1\cdots i_s}$ is given by 
\begin{align}
    \alpha_{{\rm L}\,i_1\cdots i_s} \equiv \frac{(2s-1)!!}{s!} \int \frac{d^3 {\bm k}_{\rm L}}{(2\pi)^3}\, {\cal A}_s g_s\!\left(\frac{k_{\rm L}}{k_{\rm p}}\right)[\hat{k}_{{\rm L},i_1}\cdots \hat{k}_{{\rm L},i_s}]^{\rm TL_3}\Phi({\bm k}_{\rm L})e^{i{\sbm k}_{\rm L} \cdot {\sbm x}}.
    \label{eq:along}
\end{align}
Here, we used Eq.~(\ref{eq:formula_inn_prod}) in App.~\ref{ssec:appB}.
These results are straightforwardly generalized to the case where the squeezed
bispectrum is given by a sum of separable contributions.
When $f_s$ is given by a linear combination of a power of $k_{\rm L}/k_{\rm S}$, e.g., Eq.~(\ref{Exp:fs}), which is given by
\begin{align}
  f_s \left( \frac{k_{\rm L}}{k_{\rm S}} \right) =  \frac{1}{2} \left[  \left( \frac{k_{\rm L}}{k_{\rm S}} \right)^{\Delta_s} e^{i\psi_s} + {\rm c.c.} \right],
  \label{Exp:fsfactex}
\end{align}
with $\Delta_s = 3/2 + i \nu_s$, each term can be rewritten as in Eq.~(\ref{Exp:fsfact}).

In a simple setup, the squeezed PNG from massive particles is suppressed by the smallness of the coupling, the Boltzmann factor, $e^{- M_s/T_{\rm H}}$ with $T_{\rm H}$ being the Hawking temperature $T_{\rm H} = H_{\rm inf}/(2\pi)$, and the dilution factor $(k_{\rm L}/k_{\rm S})^{3/2}$ (see Eq.~(\ref{Exp:fs})). A larger coupling between the inflaton and the massive particles enhances the radiative corrections, which can lead to the break down of the EFT description. In Ref.~\cite{Kumar:2019ebj}, Kumar and Sundrum showed that the PNG mediated by heavy particles in several scenarios are too small to be detected, when we require the validity of the EFT description. They also showed that this difficulty can be circumvented in a curvaton scenario, where two different cutoff scales can be introduced. The Boltzmann suppression is not significant for a model where $M_s$ is around $H_{\rm inf}$ as discussed in Refs.~\cite{Kumar:2017ecc, Kumar:2018jxz} by considering explicit models. In Ref.~\cite{Wang:2019gbi}, Wang and Xianyu pointed out that a coupling between the inflaton and the massive sector is strictly restricted to evade the Boltzmann suppression without fine-tuning, since a large coupling leads to a large radiative correction to $M_s$. This restriction has guided their attention to parity odd dimension 5 operators, which can relax the Boltzmann suppression~\cite{Wang:2019gbi, Wang:2020ioa} (see also Refs.~\cite{Chen:2018xck, Hook:2019zxa}). Finally, the suppression by the dilution factor $(k_{\rm L}/k_{\rm S})^{3/2}$ for $M_s \geq {\cal O}(H_{\rm inf})$ can be evaded by lowering $M_s$, while the straightforward attempt conflicts with the Higuchi bound for $s \geq 2$~\cite{Higuchi:1986py}. In Refs.~\cite{Kehagias:2017cym, Bordin:2018pca}, it was argued that the Higuchi bound and subsequently the suppression by the dilution can be avoided, when the de Sitter symmetry is explicitly broken. The model proposed by Kehagias and Riotto in Ref.~\cite{Kehagias:2017cym} violates the global rotation symmetry.

In this paper, deferring an attempt to build a model that predicts a large angular dependent PNG to elsewhere, we simply take ${\cal A}_s$ as a parameter. For our forecast, we consider both cases with and without the dilution, i.e. ${\rm Re}[\Delta_s] = 3/2$ and $\Delta_s=0$, respectively. 
The tightest constraint on the angular dependent PNG parameters has been obtained from the CMB observation by Planck satellite. The Planck result put the limit on ${\cal A}_2$ for $\Delta_2=0$ as $\sigma({\cal A}_2) \sim 77$~\citep{Akrami:2019izv}. In Ref.~\citep{Franciolini:2018eno}, this analysis was extended to a more general angular dependence with even numbers of $s$, including the PNG generated by higher spin particles.  
However, the CMB constraint almost reaches the cosmic variance limit~\citep{Shiraishi:2013vja}. In Ref.~\citep{MoradinezhadDizgah:2018pfo}, it was argued that from the bispectrum of the galaxy number density, ${\cal A}_s$ can be constrained as $\sigma({\cal A}_2) \sim 15$ for $\Delta_2=0$ and $\sigma({\cal A}_4) \sim 21$ for $\Delta_4=0$.
Ref.~\citep{MoradinezhadDizgah:2017szk} discussed the constraint on ${\cal A}_s$ for ${\rm Re}[\Delta_s]=3/2$ by combining the galaxy power spectrum and bispectrum.
These analyses require the bispectrum information, which is more complex than the power spectrum analysis, especially in galaxy surveys (e.g. the decomposition of anisotropic signals in the galaxy bispectrum~\citep{Assassi:2015fma,Slepian:2017lpm,Sugiyama:2018yzo}),
and the angular dependent PNG signals are mixed up in the bispectrum. Here, one can find an advantage to use galaxy shape \citep{Schmidt:2015xka,Kogai:2018nse,Akitsu:2020jvx}, whose observation enables us to pick up the imprints of particles with different spins separately as shown in this paper.

\subsection{PNG contribution to galaxy shape function}\label{sec:PNG}
Now, we are ready to calculate the imprint of the angular dependent PNGs encoded in the galaxy shape function.

\subsubsection{Imprint of spin-4 particles}\label{sec:imprint4}
In Ref.~\cite{Schmidt:2015xka}, it was shown that the contribution of the PNG (\ref{eq:NGIC}) in the second shape moment $\tilde{g}_{i_1 i_2}$, is only that generated by the spin-2 particles, as far as we consider the large scales where the non-linear secondary evolution is negligible. Here, we will show that similarly only the PNG generated by the spin-4 particle contributes to the 4th shape moment, $\tilde{g}_{i_1 i_2 i_3 i_4}$. Along the line with Ref.~\cite{Schmidt:2015xka}, we compute the contribution of the PNG to $\tilde{g}_{i_1 i_2 i_3 i_4}$ by evaluating the correlation between $\delta$ and $\tilde{g}_{i_1 i_2 i_3 i_4}$.

In Sec.~\ref{sec:bias}, various contributions to the 4th shape moment for the Gaussian initial condition were computed. The angular dependent PNG leads to an additional contribution to the correlation $\langle \delta({\bm x}) \tilde{g}_{i_1 i_2 i_3 i_4}({\bm y})\rangle$. For example, when the PNG is given by Eq.~(\ref{eq:NGIC}) with $\Delta_4=0$, we obtain 
\begin{align}
	\langle \delta({\bm x}) \tilde{g}_{i_1 i_2 i_3 i_4}({\bm y}) \rangle 
	= \frac{2}{9}b_{K^2}^{(4)}{\cal A}_4 \mathcal{D}_{i_1 i_2 i_3 i_4}\xi_{\delta\Phi}(|{\bm x}-{\bm y}|)\langle \delta^2 \rangle\,,  \label{eq:ccdeltag}
\end{align}
with $\mathcal{D}_{i_1 i_2 i_3 i_4} \equiv [\mathcal{D}_{i_1 i_2}\mathcal{D}_{i_3 i_4}]^{\rm TL_3,sym}$ and 
$\xi_{\delta\Phi}$ being the cross-correlation between the linear $\delta$ and $\Phi$. A more detailed computation of Eq.~(\ref{eq:ccdeltag}) can be found in App.~\ref{sec:appB}. Performing the Fourier transformation, we obtain 
\begin{align}
	\langle \delta({\bm k}) \tilde{g}_{i_1 i_2 i_3 i_4}({\bm k'}) \rangle 
	= \frac{2}{9}b_{K^2}^{(4)}{\cal A}_4 [\hat{k}_{i_1}\hat{k}_{i_2}\hat{k}_{i_3}\hat{k}_{i_4}]^{\tlth}(2\pi)^3{\cal M}^{-1}(k,z)P_{\rm m}(k,z)\langle \delta^2 \rangle\delta_{\rm D}({\bm k}+{\bm k}') \,,  \label{Exp:CCwNG}
\end{align}
where $P_{\rm m}(k,z)$ is the linear matter power spectrum at $z$ and ${\cal M}(k,z)$, which relates $\delta(\bm{k},z)$ to the primordial curvature perturbation $\Phi_{\sbm k}$ as 
\begin{align}
    \delta(\bm{k},z) &={\cal M}(k,z)\Phi_{\sbm k}\,,
\end{align}
is given by
\begin{align}
    {\cal M}(k,z) = \frac{2}{3}\frac{k^2T(k)D(z)}{H_0^2\Omega_{\rm m0}},
\end{align}
with $T(k)$ being the transfer function.

Since Eq.~(\ref{Exp:CCwNG}) depends on $\langle \delta^2 \rangle$, which diverges when we send the UV cutoff to the infinity, we need to perform the renormalization to compute the observable effect. The renormalization proceeds analogously to the one for the PNG generated by the spin-0 and spin-2 particles discussed in Refs.~\cite{Schmidt:2015xka,Assassi:2015jqa,Assassi:2015fma}. As one can see in Eq.~(\ref{eq:alps}), the counter term for the contribution of the PNG generated by the spin-4 particle should be in the form, $\mathcal{D}_{i_1 i_2 i_3 i_4}\Phi$. In the presence of the PNG generated by the spin-4 particle, the local matter density for the short mode $\bm{k}$ is modified as
\begin{align}
	\delta^{\rm loc}({\bm k};{\bm x}) = \left(1+\frac{1}{2}\alpha_{{\rm L}\,i_1 i_2 i_3 i_4}({\bm x}) [\hat{k}^{i_1}\hat{k}^{i_2}\hat{k}^{i_3}\hat{k}^{i_4}]^{\tlth}\right)\delta^{\rm iso}(\bm{k})\,, \label{Eq:deltaloc}
\end{align}
with
\begin{align}
	\alpha_{{\rm L}\,i_1 i_2 i_3 i_4}({\bm x}) \equiv \int\frac{d^3{\bm k}_{\rm L}}{(2\pi)^3}\frac{35}{8} {\cal A}_4[\hat{k}_{{\rm L},{i_1}} \hat{k}_{{\rm L},{i_2}} \hat{k}_{{\rm L},{i_3}} \hat{k}_{{\rm L},{i_4}}]^{\tlth} \Phi({\bm k}_{\rm L})e^{i{\sbm k}_{\rm L}\cdot{\sbm x}}\,.
\end{align}
The renormalized bias is then defined through the response of 4th moment to $\alpha$
\begin{align}
	b^{(4)}_{\rm NG} \equiv \left.\frac{\partial\langle \tilde{g}_{i_1 i_2 i_3 i_4}\rangle_{\alpha_{\rm L}}}{\partial \alpha_{{\rm L}\,i_1 i_2 i_3 i_4}}\right|_{\alpha_{\rm L}=0}\, ,
\end{align}
which leads to 
\begin{align}
\langle \delta({\bm k}) \tilde{g}_{i_1 i_2 i_3 i_4}({\bm k'}) \rangle 
= \frac{35}{4}b^{(4)}_{\rm NG}{\cal A}_4 [\hat{k}_{i_1}\hat{k}_{i_2}\hat{k}_{i_3}\hat{k}_{i_4}]^{\tlth}(2\pi)^3{\cal M}^{-1}(k,z)P_{\rm m}(k,z) \delta_{\rm D}({\bm k}+{\bm k}') \,.
\end{align}
The detailed computation is summarized in App.~\ref{sec:appc}.

This result can be straightforwardly extended to the PNG with $\Delta_4 = 3/2\pm i\nu_4$ as
\begin{align}
	\langle \delta({\bm k}) \tilde{g}_{i_1 i_2 i_3 i_4}({\bm k'}) \rangle 
	&= \frac{35}{4}b_{\rm NG}^{(4)}{\cal A}_4\left(\frac{k}{k_*}\right)^{3/2} [\hat{k}_{i_1}\hat{k}_{i_2}\hat{k}_{i_3}\hat{k}_{i_4}]^{\tlth}\cos\left[\nu_4\ln\left(\frac{k}{k_*}\right)+\Theta_4\right]\notag\\
	&\quad\times(2\pi)^3{\cal M}^{-1}(k,z)P_{\rm m}(k,z) \delta_{\rm D}({\bm k}+{\bm k}')\,.
\end{align}
Here and hereafter, we set the pivot scale $k_{\rm p}$ to $k_*$. For a different choice of $k_*$, the corresponding $b_{\rm NG}^{(4)}$ and $\Theta_4$ differ.

\subsubsection{Imprint of particle with a general integer spin} \label{SSSec:otherspins}
We can also evaluate the imprint of the PNG generated by the spin-$n$ particle with a general integer $n$, encoded in $\langle \delta(\bm{x}) \tilde{g}_{i_1 \cdots i_n} (\bm{y}) \rangle$. The PNG from the spin-$n$ particle selectively appears in the traceless part of the $n$th shape moment, $\tilde{g}_{i_1 \cdots i_n}$. Therefore, even if ${\cal A}_0$, which corresponds to $f_{\rm NL}$, is much larger than ${\cal A}_s$ with $s \geq 2$, the contribution of ${\cal A}_0$ does not contaminate $\tilde{g}_{i_1 \cdots i_n}$ with $n \geq 2$ in the linear regime.

Repeating a similar computation to App.~\ref{sec:appB}, we find that the spin-$n$ contribution in the squeezed PNG (\ref{eq:NGIC}) yields the additional intrinsic alignment contribution given by 
\begin{align}
    F^{(n)}_{\rm PNG}(k,z) &= C_n b_{\rm NG}^{(n)}{\cal A}_n\left(\dfrac{k}{k_*}\right)^{{\rm Re}[\Delta_n]} {\cal M}^{-1}(k,z) \cos\left[{\rm Im}[{\Delta_n}]\ln\left(\dfrac{k}{k_*}\right)+\Theta_n\right]\,,\label{eq:FnPNG}
\end{align}
where $F^{(n)}_{\rm PNG}$ denotes the PNG contribution in Eq.~(\ref{Exp:In}) and $\Theta_n$ is the phase determined for a given halo model, i.e. it is a function of $k_*$. The leading PNG contribution appears from the quadratic terms in the bias expansion of the $n$th shape moment, which are accompanied with additional spatial gradient for $n> 4$. However, the spatial gradient does not yield an additional suppression by $k/k_*$ in Eq.~(\ref{eq:FnPNG}), since it is replaced with $(k_{\rm S}/k_*)^{n-4}$, where $k_{\rm S}$ is the short mode. Therefore, as written in Eq.~(\ref{eq:FnPNG}), the leading PNG contribution scales as $\propto k^{{\rm Re}[\Delta_n]} {\cal M}^{-1}(k,z)$ for a general integer $n$. The coefficient $C_n$ is determined by conducting the renormalization. The counter term needed for the renormalization differs for a different $n$. For $n=0,\, 2,\, 4$, $C_n$ is given by $C_0 =1/2$, $C_2 = 3$ (\cite{Schmidt:2015xka}) and $C_4 = 35/4$ (Sec.~\ref{sec:imprint4}).

Let us emphasize that the separability of the PNG contributions from different spins no longer holds, once the loop contributions become important, because the kernel functions also induce the angular dependence. Then, the PNG generated by the spin-$s$ particles can contribute to the $n$th moment of the galaxy shape function only for $n=s$. Having considered this, in the next section, we explore whether there is a scale where the contribution from the PNG becomes dominant, keeping the late time non-linear contributions subdominant.

\section{Forecast on PNG from higher spin particles}\label{sec:nr}
In the previous section, we have computed the contribution of the PNG generated by a spin-$n$ particle, to the $n$th moment galaxy shape function, $\tilde{I}_{i_1 \cdots i_n}$. In this section, comparing it to other contributions predicted in $\Lambda$CDM cosmology with the adiabatic Gaussian initial condition, let us discuss whether we can observe the imprint of the higher spin particles from future observations.

\subsection{Dominant contribution at different scales}\label{sec:scaling}
\begin{figure}
    \centering
    \includegraphics[width=150mm]{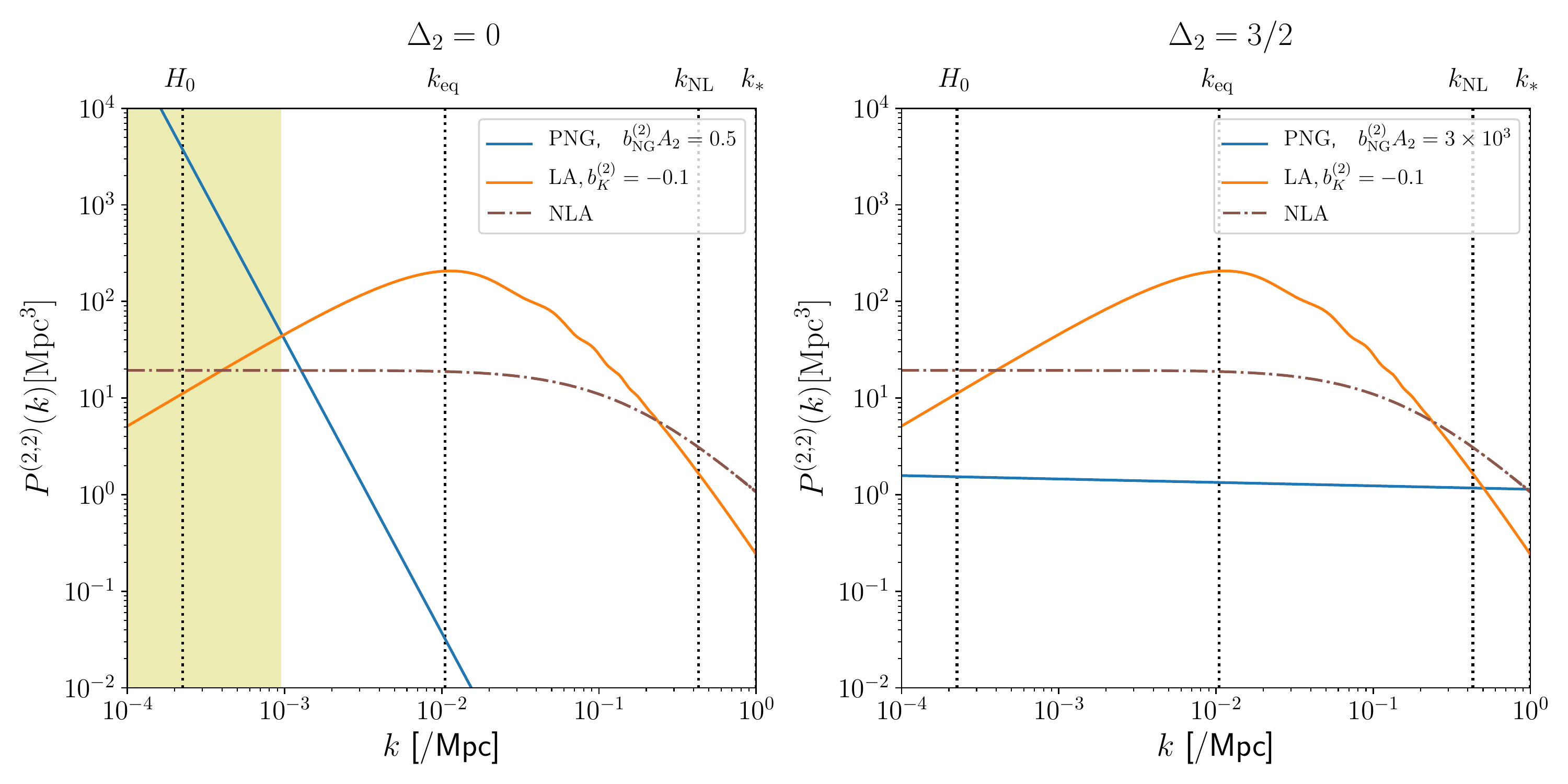}
    \caption{The $k$-dependence of each contribution at $z=1.5$ for the 2nd moment galaxy shape function. The LA and NLA are the same as Fig.~\ref{fig:scaling}. The left and right panels additionally include the PNG contribution, given in Eq.~(\ref{eq:FnPNG}), for $\Delta_2=0$ and $b_{\rm NG}^{(2)}{\cal A}_2=0.5$ (left) and for $\Delta_2=3/2$ and $b_{\rm NG}^{(2)}{\cal A}_2=3\times10^3$ (right), respectively.
    The yellow shade shows the range of $k$ at which the PNG contribution exceeds those of LA and NLA. }
    \label{fig:scale-dependence_2}
\end{figure}

Combining the results obtained in Sec.~\ref{sec:bias} and Sec.~\ref{sec:IA}, we can evaluate which effect is dominant at each scale or at each multipole moment. Figure \ref{fig:scale-dependence_2} compares the contributions of the PNG from the spin-2 particle to LA and NLA, computed in Sec.~\ref{sec:bias} for the 2nd galaxy moment. Here, considering the $\Lambda$CDM Universe, we set $\np\sim-1.7$ and $k_{{\rm NL},z=0}=0.25h\mathrm{/Mpc}$~\cite{Desjacques:2016bnm}. For $\Delta_2=0$, the contribution of the PNG dominates at the large scales as pointed out in Ref.~\cite{Schmidt:2015xka} (see also Ref.~\cite{Akitsu:2020jvx}). The colored region shows the range of $k$ at which the PNG dominates the other contributions. Meanwhile, for ${\rm Re}[\Delta_2] = 3/2$, the contribution of the PNG is dominated by NLA at all scales with $k < k_{\rm eq}$, where the galaxy imaging surveys work as a spin-sensitive detector. When the perturbative expansion holds,
satisfying $b^{(2)}_{\rm NG} {\cal A}_2 \ll 10^4$ (assuming $b^{(2)}_{\rm NG}=0.1$), there is no range of $k$ where the PNG generated from the massive spin-2 particle with ${\rm Re}[\Delta_2] = 3/2$ dominates the 2nd moment of the galaxy shape function. The PNG from the spin-2 particle in the principle mass series, which is suppressed by the dilution, ${\rm Re}[\Delta_2] = 3/2$, exhibits the oscillatory resonance feature as shown in Eqs.~(\ref{eq:NGIC}) and (\ref{Exp:fs}). Since our purpose here is to compare the amplitudes of different contributions, for the illustrative purpose, the oscillatory contribution in \refeq{FnPNG} is ignored.

\begin{figure}
	\centering
	\includegraphics[width=150mm]{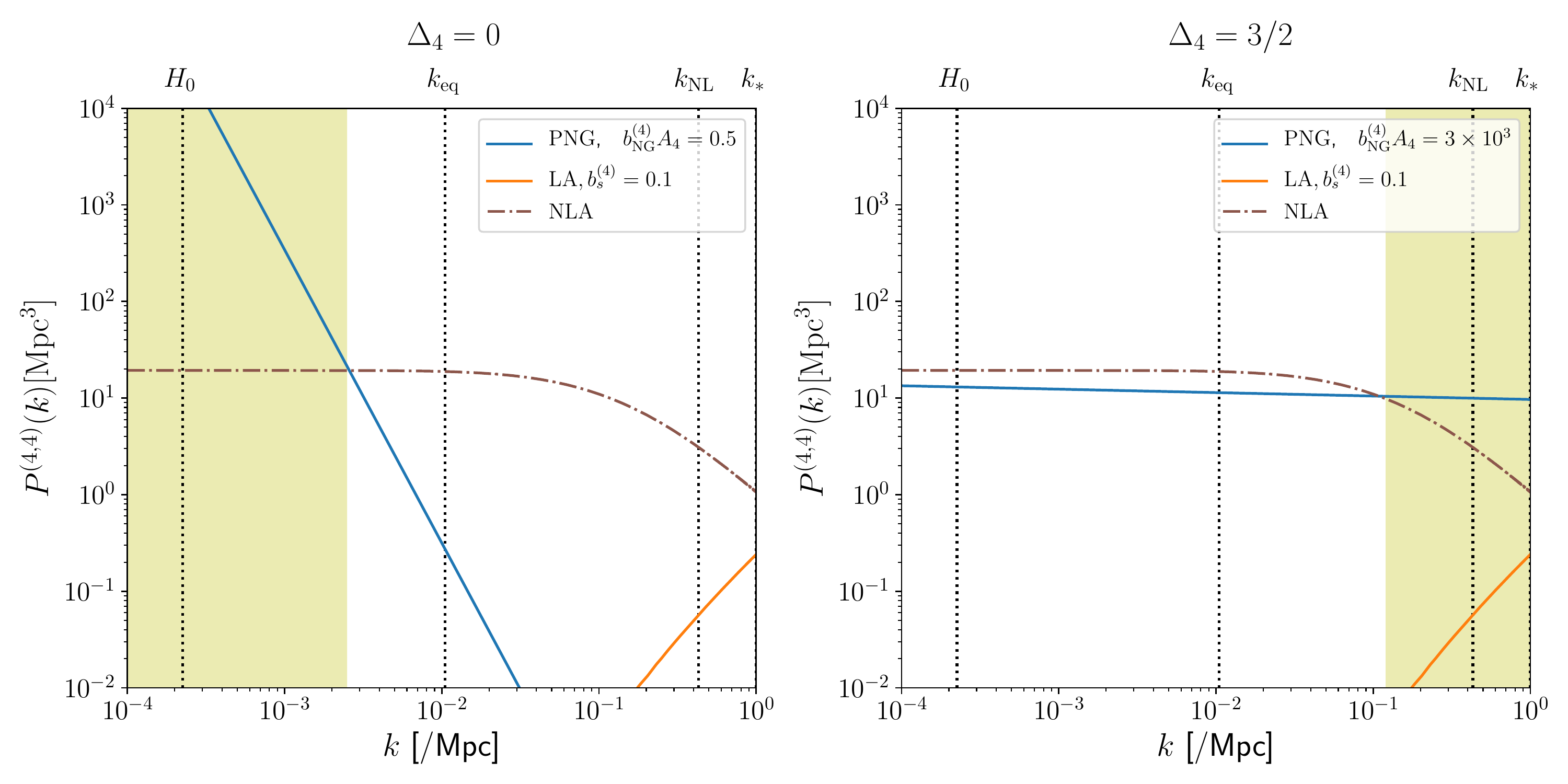}
	\caption{The $k$-dependence of each contribution at $z=1.5$ for the 4th moment galaxy shape function. The LA and NLA are the same as Fig.~\ref{fig:scaling}. The left and right panels additionally include the PNG contribution, given in Eq.~(\ref{eq:FnPNG}), for $\Delta_4=0$ and $b_{\rm NG}^{(4)}{\cal A}_4=0.5$ (left) and for $\Delta_4=3/2$ and $b_{\rm NG}^{(4)}{\cal A}_4=3\times10^3$ (right), respectively. The yellow shades show the range of $k$ at which the PNG contribution exceeds those of LA and NLA. 
    }
	\label{fig:scale-dependence_4}
\end{figure}

The situation is different for the PNG generated from the spin-$n$ particle with $n>2$, because the linear alignment and the linear weak lensing for $\tilde{I}_{i_1 i_2 \cdots i_n}$ are both suppressed by $(k/k_*)^{n-2}$. Meanwhile, as shown in Eq.~(\ref{eq:FnPNG}), the PNG can contribute to $\tilde{I}_{i_1 i_2 \cdots i_n}$ without being suppressed at large scales, especially for $\Delta_s =0$. Figure \ref{fig:scale-dependence_4} compares the different contributions to the auto-correlation of $\tilde{I}_{i_1 i_2 i_3 i_4}$. Since the dominant contamination of the 4th shape moment comes from NLA, i.e. the 3D loops under the Gaussian initial condition (even if we include the lensing contribution), here let us compare it to the PNG contribution. 
Around $k_{\rm NL}$, where the 1-loop NLA acquires the additional factor $(k/k_{\rm NL})^{n_{\rm L}+3}$, the PNG contribution dominates the NLA contribution in the range of $k$ which satisfies
\begin{align}
    (C_4b^{(4)}_{\rm NG}{\cal A}_4)^2\left(\frac{k}{k_*}\right)^{2{\rm Re}[\Delta_4]}{\cal M}^{-2}(k)> \left(\frac{k}{k_{\rm NL}}\right)^{\np+3} \label{eq:range}\,.
\end{align}
Here, focusing on the amplitude, the oscillatory contribution (if any) is again ignored. On the other hand, for $k <  k_{\rm eq}$, the leading contribution of NLA, which corresponds to the stochastic term, is roughly given by
\ba
P_{\rm NLA}(k<k_{\rm eq},z) \sim b_{\rm NLA}^2\left[\frac{D(z)}{D(0)}\right]^4\int_k^\infty dpp^2[P_{\rm L}(p)]^2 \sim b_{\rm NLA}^2\left[\frac{D(z)}{D(0)}\right]^4[P_{\rm L}(k_{\rm eq})]^2k_{\rm eq}^3\,,
\ea
which is almost scale independent like the white noise. In practice, this contribution would be absorbed by the amplitude of the noise in the 4th moment, for which we do not have a reliable estimate. In our idealized setting then, the PNG contribution should exceed this to be detectable.

The PNG contribution is $z$-independent while the linear alignment contribution varies as $\sim(1+z)^{-2}$ and the NLA contribution depends on $\sim(1+z)^{-4}$. Therefore, a smaller amplitude of PNG can be detected for a deeper survey. This point is quantitatively analyzed in Fig.~\ref{fig:SNfromAPS}.

\subsection{Angular power spectrum}
Using Eq.~(\ref{Exp:Clnn}), we can compute the angular power spectrum. Here, we focus on $C_l^{(2,2)}$ and $C_l^{(4,4)}$, by which we can explore the imprint of the spin-2 and spin-4 particles, respectively. \update{In conducting the numerical computation, the Limber approximation is employed for $l>20$. }

\begin{figure}
	\centering
	\includegraphics[width=140mm]{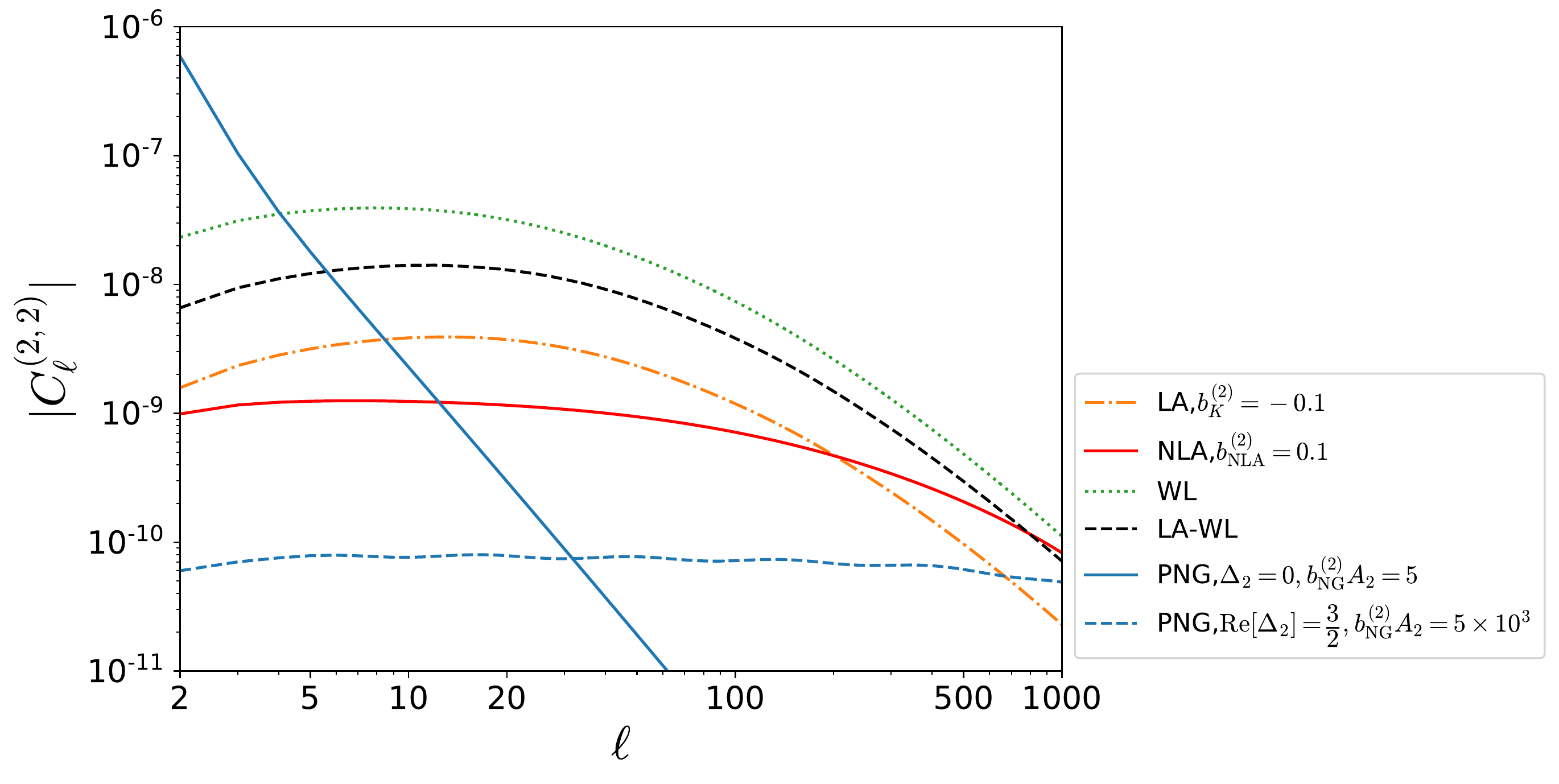}
	\caption{This plot shows the contributions of LA (orange dash-dotted), NLA (red solid), linear WL (green dotted), and PNG (blue), to the angular power spectrum of the second shape moment $C_l^{(2,2)}$ for LSST-like survey, whose $dN/dz$ is given by Eq.~(\ref{eq:dndz}). \update{The black dashed curve shows the cross-correlation between LA and linear WL, which becomes negative, and the other curves show the auto-correlations of each contribution.} The blue solid line corresponds to $\Delta_2=0$
	and the blue dashed line corresponds to ${\rm Re}[\Delta_2]=3/2$, $\nu_2=3$, $\Theta_2=0$.}
	\label{fig:cl2nd}
\end{figure}

\begin{figure}
	\centering
	\includegraphics[width=140mm]{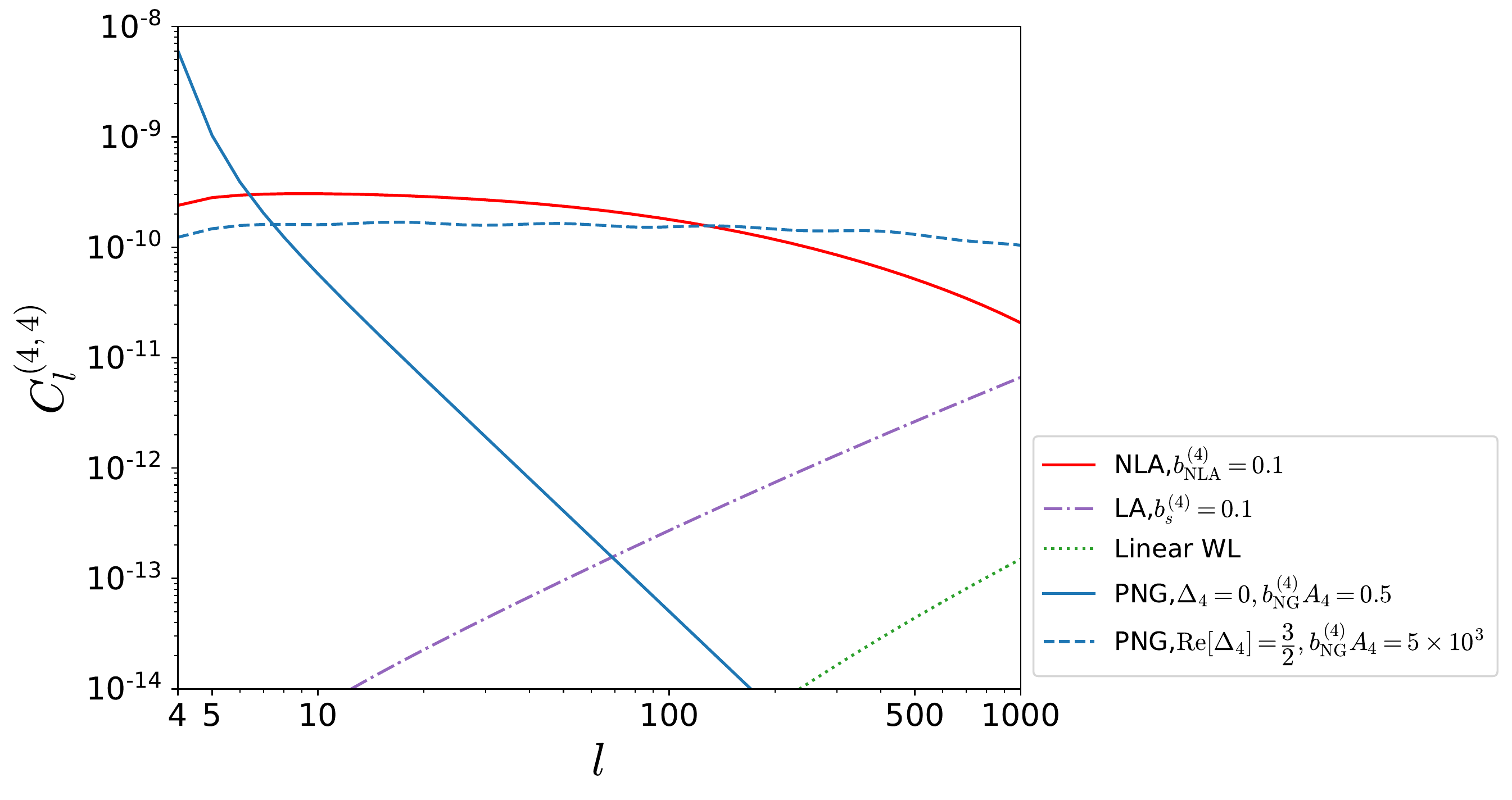}
	\caption{This plot shows each contribution in the auto-correlation of 4th shape moment for LSST-like survey.
		The blue solid line corresponds to $\Delta_4=0$
		and the blue dashed line corresponds to ${\rm Re}[\Delta_4]=3/2$, $\nu_4=3$, $\Theta_4=0$ in Eq.~(\ref{eq:FnPNG}). The purple dash-dotted line is the LA contribution. The red solid line shows the 1-loop NLA contribution evaluated by using Eq.~(\ref{eq:NLA}). The green dotted line corresponds to the linear WL.}
	\label{fig:Cl}
\end{figure}

Inserting Eq.~(\ref{eq:FnPNG}) into Eq.~(\ref{Exp:Clnn}), we can compute the contribution of the PNG from spin-$s$ particle to the angular power spectrum $C_l^{(n, n')}$, where $n$ and (or) $n'$ are (is) equal to $s$. The PNG contribution from the spin-2 particle was addressed in Ref.~\cite{Schmidt:2015xka} ($\Delta_2=0$) and in Ref.~\cite{Kogai:2018nse} (${\rm Re}[\Delta_2] =3/2$). In Fig.~\ref{fig:cl2nd}, we show the contributions of LA, NLA, WL, and PNG to $C_l^{(2, 2)}$, including NLA, which was not taken into account in Refs.~\cite{Schmidt:2015xka, Kogai:2018nse}. Assuming an LSST like lensing survey~\cite{Abell:2009aa,Chang:2013xja}, we have used the function $dN/dz$ given by
\begin{align}\label{eq:dndz}
    \frac{dN}{dz}\propto\left(\frac{z}{0.51}\right)^{1.24}\exp\left[-\left(\frac{z}{0.51}\right)^{1.01}\right]\,,
\end{align}
which is normalized as $\int dz (dN/dz) = 1$.
\update{In Fig.~\ref{fig:cl2nd}, we show the auto-correlations of each contribution and the cross-correlations between the LA and the linear WL.}

Similarly, using Eqs.~(\ref{Exp:Clnn}), (\ref{eq:int_linear}),
and (\ref{eq:FnPNG}), we can compute the auto-correlation of each contribution in the angular power spectra for $n=4$ as
\begin{align}
    C^{(4, 4){X}}_l&=\frac{2}{\pi}\frac{(l-4)!}{(l+4)!}\int k^2dkP_{\rm m}(k)|F^{X}_l(k)|^2\quad\quad(X={\rm LA, PNG, LWL})\,,
\end{align}
with
\begin{align}
  & F_l^{Y}(k) = \frac{1}{4}\frac{(l+4)!}{(l-4)!}\int dz\frac{dN}{dz} F^{(4)}_{Y}(k,z)\frac{D(z)}{D(0)}\left[\frac{j_l(x)}{x^4}\right]_{x=k\chi(z)}
  \quad\quad(Y={\rm LA, PNG})  ,\label{eq:deltab}\\
  & F_l^{\rm LWL}(k) = \frac{1}{4}\frac{(l+4)!}{(l-4)!}\int_0^{\chi_{\rm max}} d\chi' \chi' D_\Phi(z(\chi'))\left(\frac{k}{k_*}\right)^2\left[\frac{j_l(k\chi')}{(k\chi')^4}\right] \cr
  & \qquad \qquad \qquad \qquad \qquad \qquad \times \int_{\chi'}^{\chi_{\rm max}}d\chi H(\chi) \frac{dN}{dz} \left(\frac{\chi'}{\chi}\right)^2\frac{\chi - \chi'}{\chi}\,,\label{eq:wlkernel}  
\end{align}
and
\begin{align}
    D_\Phi(z) \equiv 3H_0^2\Omega_{\rm m0}\frac{(1+z)D(z)}{D(0)}\,.
\end{align}
Equation (\ref{eq:wlkernel}) can be derived by operating the source distribution function on the expression for a source at fixed redshift given in App.~\ref{sec:wl}\footnote{Since we estimated only the scaling of the linear WL for $(kR_*\chi'/\chi)\ll1$, our Eq.~(\ref{eq:wlkernel}) should be additionally multiplied by $1/12$ to match with Eq.~(84) in Ref.~\cite{Fleury:2018odh}.}.

Figure \ref{fig:Cl} compares different contributions to $C^{(4, 4)}_l$. 
The purple dash-dotted line shows the contribution of the LA, which is suppressed by $(k/k_*)^4$.
The green dotted line shows the contribution of the linear weak lensing.
The contribution of the weak lensing 1-loop in the projected 2D space (2D 1LOOP), which is given by Eq.~(\ref{eq:aps_4th_wl}) (App.~\ref{SSec:IWL}), is much smaller and below the shown range in the figure.
The blue solid line shows the contribution of the PNG with $\Delta_4=0$ and the blue dashed line shows the one of the PNG with $\Delta_4= 3/2 + i\nu_4$, which corresponds to the PNG from a spin-4 particle in the principle mass series. For a larger $\nu_4$, the net contribution of the PNG to $C^{(4, 4){\rm PNG}}_l$ becomes smaller, being smoothed out by the integration over $k$. This aspect is common with massive particles in the principal mass series with other spins, as discussed in Ref.~\cite{Kogai:2018nse}.

Since the PNG contribution with ${\rm Re}[\Delta_s] = 3/2$ is almost constant, being independent of $l$, one may think it can be detected at high $l$s even if it is subdominant at lower $l$s. However, at high $l$s, because of the mixing due to the non-linear evolution, we cannot separately pick up the PNG contributions from particles with different spins. 
(Roughly speaking, using the redshift of the peak of $dN/dz$, we can estimate $l_{\rm NL}\sim k_{{\rm NL},z=0.5}\chi(z=0.5)\sim470$ as the non-linear angular scale.)
In the range where the non-linear evolution is negligible, while the PNG contribution does not become dominant for $n=2$, it can be dominant for $n=4$ in $l\gtrsim200$\footnote{The coefficients $C_2, C_4$, which are determined by renormarization in \refeq{FnPNG}, are different between the 2nd moment and the 4th moment. Therefore, even if $b^{(2)}_{\rm NG}{\cal A}_2 = b^{(4)}_{\rm NG}{\cal A}_4$, the corresponding contribution of the PNG for $n=4$ is larger than the one for $n=2$ by the factor $(C_4/C_2)^2=(35/12)^2\sim9$.}.

\subsection{Forecast for future imaging survey}
Finally, we investigate the detectability of the PNG generated by higher spin particles, considering the future galaxy survey, LSST. As discussed in Sec.~\ref{sec:PNG}, on large scales the PNG from each particle with spin-$s$ only contributes to the $s$th galaxy moment of the galaxy shape function, leaving an imprint in the auto-correlation $C_l^{(n,\, n)}$ with $n=s$ and the cross-correlations  $C_l^{(n,\, n')}$ with $n=s$ or $n'=s$. Therefore, no matter how large the angular independent PNG $f_{\rm NL}$ is, this does not disturb probing the signal of the higher spin particle encoded in $\tilde{I}_{i_1 \cdots i_n}$ with the corresponding value of $n$ (see \cite{Chisari:2016xki} for a demonstration including $f_{\rm NL}$ and $\mathcal{A}_2$). Furthermore, even if there exists an infinite tower of higher spin particles as predicted in string theory, we can single out the contribution of the spin-$s$ particle by looking at the $s$th galaxy shape moment.

On the other hand, when the contribution of the loops becomes important, the non-trivial momentum dependence in the kernel functions allows the contribution of the PNG from the spin-$n'$ particle to contaminate $\tilde{I}_{i_1 \cdots i_n}$ with $n' \neq n$. Therefore, in what follows, we focus on $k< k_{\rm NL}$, where the loop contribution remains subdominant.

As discussed in Sec.~\ref{sec:bias}, since the dominant contamination to the $s=4$ signal at $k < k_{\rm NL}$ comes from NLA, in the following, we ignore the contributions of the linear alignment and weak lensing. The signal to noise ratio (S/N) for each redshift, which is not cumulative, is expressed by 
\begin{align}
    [{\rm S}/{\rm N}]^2(z)\sim f_{\rm sky}\sum_{l=4}^{l_{\rm max}}  (2l+1)\left[\frac{C^{(4, 4){\rm PNG}}(l,z)}{C^{(4, 4){\rm PNG}}(l,z)+C^{(4, 4){\rm NLA}}(l,z)}\right]^2\,,
    \label{eq:sn}
\end{align}
where $f_{\rm sky}$ is the sky fraction to the galaxy imaging survey and $C^{(4, 4){\rm NLA}}(l,z)$ denotes the NLA contribution which is crudely estimated by using Eq.~(\ref{eq:NLA}). Here, we have discarded the contribution from the non-linear scales, setting $l_{\rm max}=k_{\rm NL}(z)\chi(z)$. 
For $\Delta_4 =0$ (without dilution), S/N becomes independent of $l_{\rm max}$ when we choose a sufficiently large $l_{\rm max}$, since the shape signal of PNG is significant at the large scales. On the other hand, for ${\rm Re}[\Delta_4] = 3/2$ (with dilution), since the signal dominates the shape noise only at the small scales, S/N depends on the choice of $l_{\rm max}$. However, since we cannot selectively single out the contribution of the spin-4 particle, we expect that increasing $l_{\rm max}$ will not improve the actual S/N very much. \update{Meanwhile, the forecast depends on $k_{\rm NL}$, which determines the scales at which the PNG can be selectively detected. The amplitude of the PNG $b_{\rm NG}^{(4)} {\cal A}_4$ needs to be roughly factor 4 larger for a detection, when we choose $k_{{\rm NL}, z=0}=0.1h$/Mpc.}

\begin{figure}
    \centering
    \includegraphics[width=100mm]{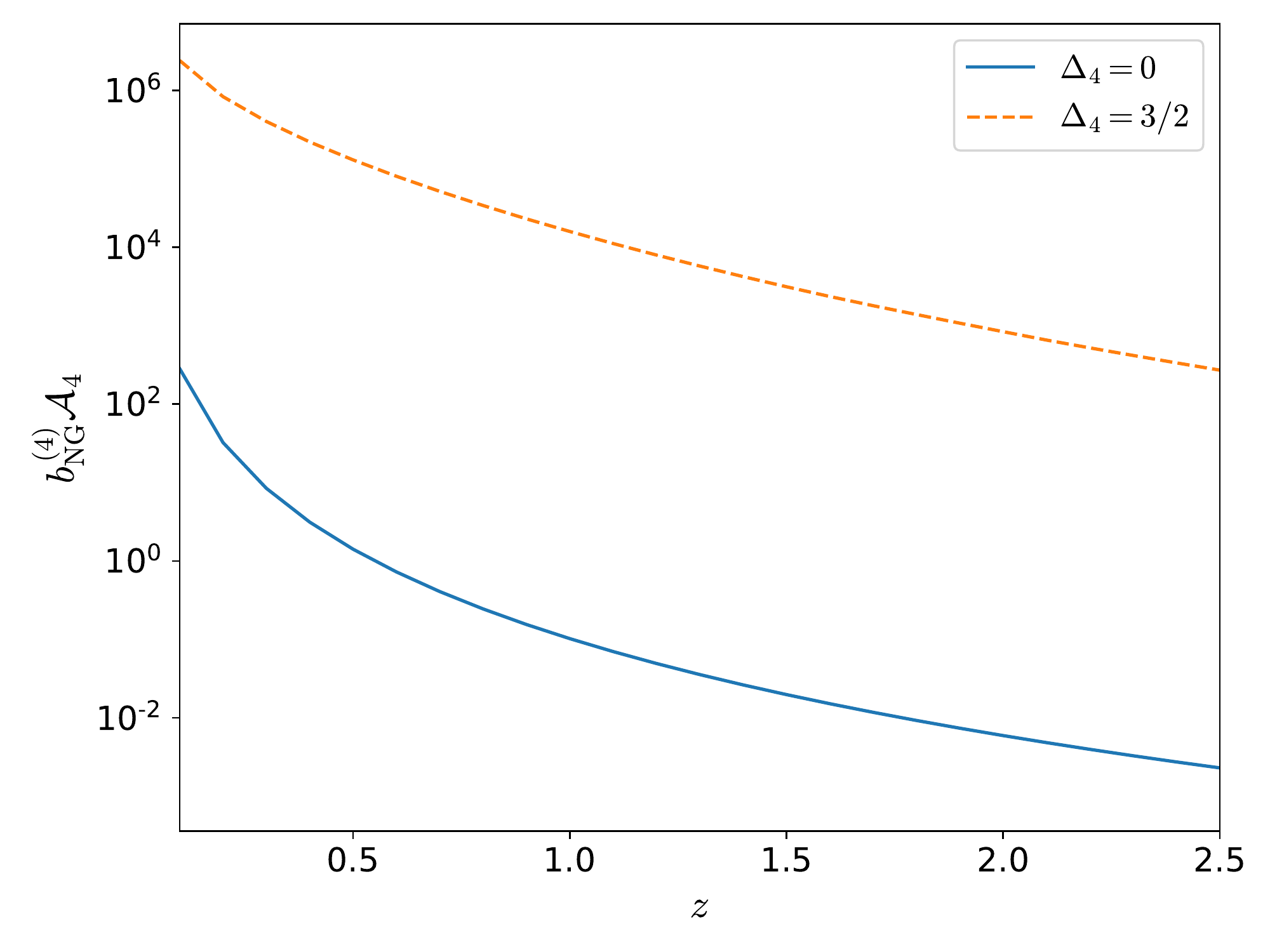}
    \caption{This plot shows the amplitude of the PNG $b^{(4)}_{\rm NG}{\cal A}_4$ which is required for a detection with ${\rm S/N} = 5$ by a survey covering a fraction $f_{\rm sky}=0.5$ of the sky when including scales up to $l_{\rm max} = k_{\rm NL}(z)\chi(z)$. 
    The blue solid line corresponds to $\Delta_4=0$ and the orange dotted line corresponds to $\Delta_4=3/2$.} \label{fig:SNfromAPS}
\end{figure}

Using Eq.~(\ref{eq:sn}) we can compute the corresponding amplitude of the PNG which can be detected for a given ${\rm S/N}$. Figure \ref{fig:SNfromAPS} shows the amplitude of $b^{(4)}_{\rm NG}{\cal A}_4$ which is detectable with ${\rm S/N} = 5$ for a survey covering a fraction $f_{\rm sky}=0.5$ of the sky up to redshift $z$. The increasing volume and number of modes available at higher redshift leads to improved constraints. Further, as discussed in Sec.~\ref{sec:scaling}, since the NLA contribution is smaller at higher redshifts while the PNG contribution remains constant, we find that the high redshift survey works in favor of PNG detection.

\section{Conclusion}\label{sec:discussions}
In this paper, we have explored the possibility that a new physics can be searched through future imaging surveys. For this purpose, we have established a way to compute the angular power spectrum for the $n$th moment galaxy shape function, performing the spin decomposition. This generalizes the conventional analysis for the cosmic shear, the spin-2 component, to an arbitrary spin component of the galaxy shape function.

In order to look for a signal of new physics, we need to understand the prediction in the concordance cosmology. We estimated the contribution to the $n$th shape moment through the linear and non-linear alignment and also the linear and non-linear weak lensing, assuming the $\Lambda$CDM cosmology with the adiabatic Gaussian initial condition. This provides a benchmark value which the signal of the new physics under consideration should exceed. For the 2nd moment, the usual cosmic shear, the dominant contamination to the signal of the new physics comes from the linear weak lensing and the linear alignment. Meanwhile, for the higher shape moment with $n> 2$, since these linear contributions are suppressed by a positive power of $k/k_* \ll 1$, the dominant contamination comes from loop contributions through the non-linear alignment. We have estimated these loop contributions, based on the effective field theory of large scale structure.

As an example of new physics, which can be searched through galaxy imaging surveys, we have studied the angular-dependent primordial non-Gaussianity, which can be generated through a non-zero integer spin particle during inflation. We have shown that an ideal galaxy imaging survey, which can measure the different spin components of the galaxy shape function independently, enables us to detect the imprints of particles with different spins separately, playing the role of a spin-resolved detector of the cosmological collider. This separation only works at smaller wavenumbers than $k_{\rm NL}$, since the angular dependence of the kernel functions mixes up contributions of different spins.

We have considered two different scaling behaviors of PNG, with and without the dilution, corresponding to $\Delta_4 = 0$ and ${\rm Re}[\Delta_4] = 3/2$, respectively. We have found that for $\Delta_4 =0$, an LSST-like survey can detect the PNG of ${\cal O}(1)$ generated by a spin-4 particle. This is comparable to the expected constraints on local primordial non-Gaussianity, which probes the existence of an additional scalar (spin-0) particle. Meanwhile, detecting the PNG for ${\rm Re}[\Delta_4] = 3/2$ is rather challenging, mainly because the signal grows towards smaller scales (larger $k$) and is difficult to distinguish from nonlinear contributions arising from the dominant Gaussian perturbations.

In an actual observation, of course we also need to take into account the shape noise, which is ignored in this paper \update{but is necessary to provide a more realistic forecast.} In case of the 2nd moment, the amplitude of shape noise is well known, while we are not aware of reliable published shape noise estimates for the 4th moment studied here. Instead, we have used the constant, $k^0$, limit of the nonlinear alignment loop contribution as a rough estimate, which is likely to underpredict the actual noise in the 4th moment.
Furthermore, in order to probe the actual amplitude of the PNG, one needs to constrain the shape bias $b_{\rm NG}^{(4)}$ for the 4th moment, as the observed correlation functions only constrain $b_{\rm NG}^{(4)} \mathcal{A}_4$. 
We can approach this issue from both an analytic (e.g. the peak theory~\citep{Bardeen:1985tr}) and numerical approaches. In particular, using $N$-body simulations we can estimate the response of the 4th moment of halo shapes to PNG in analogy to Ref.~\cite{Akitsu:2020jvx}. These issues will be addressed in future work.

\appendix
\section{Formulae}\label{sec:formulae}
Here, we derive several formulae used in the main text. 

\subsection{Derivation of Eq.~(\ref{eq:bessle_spins})} \label{Sec:Appjl}
Using mathematical induction, we prove that for all integers $s\geq0$,
\begin{align}
    P(s): (1+\partial_x^2)^s[x^sj_l(x)]=\frac{(l+s)!}{(l-s)!}\frac{j_l(x)}{x^s}.\notag
\end{align}
We can explicitly confirm $P(s)$ for $s=0,1,2$. Given that $P(s)$ is true for $s=k,k-1\,(k\geq2)$, taking the derivative of $P(k)$ with respect to $x$, we obtain
\begin{align}
    (1+\partial_x^2)^k[x^k j_l'(x)]=\frac{(l+k)!}{(l-k)!}\left[\frac{j_l'(x)}{x^k}-k\frac{j_l(x)}{x^{k+1}}\right]
    -(1+\partial_x^2)\left[k\frac{(l+k-1)!}{(l-k+1)!}\frac{j_l(x)}{x^{k-1}}\right], \label{eq:b_s2}
\end{align}
where we used
\begin{align}
    (1+\partial_x^2)^k[kx^{k-1}j_l(x)+x^kj_x'(x)]=
    (1+\partial_x^2)\left[k\frac{(l+k-1)!}{(l-k+1)!}\frac{j_l(x)}{x^{k-1}}\right]+(1+\partial_x^2)^k[x^k j_l'(x)]. \notag
\end{align}
The left hand side of $P(k+1)$ is given by
\begin{align}
 &(1+\partial_x^2)^{k+1}[x^{k+1}j_l(x)] \notag\\
 =&(1+\partial_x^2)^{k}[\{x^{k+1}+k(k+1)x^{k-1}\} j_l(x)+2(k+1)x^{k}j_l'(x)+x^{k+1} j_l''(x)]. \label{eq:b_s3}
\end{align}
Using $P(k)$, Eq.~(\ref{eq:b_s2}) and the Bessel differential equation:
\begin{align}
    x^2 j_l''(x)+2x j_l'(x) +[x^2-l(l+1)]j_l(x)=0, \notag
\end{align}
we obtain
\begin{align}
 ({\rm \ref{eq:b_s3}}) &= 2k\frac{(l+k)!}{(l-k)!}\left[\frac{j_l'(x)}{x^k}-k\frac{j_l(x)}{x^{k+1}}\right]
                +\frac{(l+k-1)!}{(l-k+1)!}[l(l+1)-k(k-1)](1+\partial_x^2)\frac{j_l(x)}{x^{k-1}}\notag \\
                &=\frac{\left\{l+(k+1)\right\}!}{\left\{l-(k+1)\right\}!}\frac{j_l(x)}{x^{k+1}} \,. \notag
\end{align}
Thus, $P(k+1)$ is true, whenever $P(k)$ and $P(k-1)$ are true.
Hence, by the Principle of Mathematical Induction, $P(s)$ is true for all integers $s\geq0$.

\subsection{Symmetric traceless tensor in 2D}
\label{Sec:Apptensor}
In the following, we show the following identities by mathematical induction;
\begin{align}
    \cos(n\phi) =& 2^{n-1}\left[\theta_{a_1}\cdots\theta_{a_n}\right]^{\tltw}|_{a_\pol=\cdots=a_n=\pol},
    \label{eq:cosnphi_TL}
    \\
    \sin(n\phi) =& 2^{n-1}\left[\theta_{a_1}\cdots\theta_{a_n}\right]^{\tltw}|_{a_\pol=\cdots=a_{n-1}=\pol,a_n=\psi},
    \label{eq:sinnphi_TL}
\end{align}
with $(\theta_\pol, \theta_\psi) = (\cos{\phi}, \sin{\phi}) $.
For $n=2$, it is straightforward to show
\begin{align}
    [\theta_{a_1}\theta_{a_2}]^{\tltw} 
    = \frac{1}{2}
    \begin{pmatrix}
    \cos^2{\phi}-\sin^2{\phi} & 2\cos{\phi}\sin{\phi} \\
    2\sin{\phi}\cos{\phi}     & -(\cos^2{\phi}-\sin^2{\phi})
    \end{pmatrix},
\end{align}
and hence (no summation) $[\theta_\pol\theta_\pol]^{\tltw} = \cos(2\phi)/2$ and $[\theta_\pol\theta_\psi]^{\tltw} = \sin(2\phi)/2$.

Let us assume that for $n=k$ Eq.~(\ref{eq:cosnphi_TL}) and (\ref{eq:sinnphi_TL}) hold. A symmetric traceless rank-$(k+1)$ tensor is related to a symmetric traceless rank-$k$ tensor
\footnote{This relation can be shown as follows.
We can assume a symmetric traceless rank-$(k+1)$ tensor to be of the form:
$\left[\theta_{a_1}\cdots\theta_{a_k}\theta_{a_{k+1}} \right]^{\tltw} 
= \frac{1}{k+1}\left(  \left[\theta_{a_1}\cdots\theta_{a_k}\right]^{\tltw} \theta_{a_{k+1}} 
+ \textrm{perms.} \right) 
- c_k \sum_{a=\pol,\psi}\left(\delta^{a_1 a_2}\left[\theta_{a_3}\cdots\theta_{a_{k+1}}\theta_a\right]^{\tltw} \theta_{a} +\textrm{perms.}\right)$, where $c_k$ is a $k$-dependent constant.
The traceless condition  
$\delta^{a_1a_2} \left[\theta_{a_1}\cdots\theta_{a_k}\theta_{a_{k+1}} \right]^{\tltw} =0$
gives $c_k = 1/k(k+1)$.
}
\begin{align}
    \left[\theta_{a_1}\cdots\theta_{a_k}\theta_{a_{k+1}} \right]^{\tltw} =&
    \frac{1}{k+1}\left(
    \left[\theta_{a_1}\cdots\theta_{a_k}\right]^{\tltw} \theta_{a_{k+1}} + \textrm{perms.}
    \right)
    \nonumber\\
    &-\frac{1}{k(k+1)}\sum_{i=1,2}\left(
    \delta^{a_1 a_2}\left[\theta_{a_3}\cdots\theta_{a_{k+1}}\theta_a\right]^{\tltw} \theta_{a}
    +\textrm{perms.}\right).
\end{align}
Notice that with the assumption we have
\begin{align}
    \sum_{a=\pol,\psi} \left[\theta_{a_1}\cdots\theta_{a_{k-1}}\theta_{a}\right]^{\tltw}
    \theta_{a}|_{a_1=\cdots=a_{k-1}=\pol}
    =&\frac{1}{2^{k-1}}\left(\cos(k\phi)\cos\phi + \sin(k\phi)\sin\phi\right),
    \\
    \sum_{a=\pol,\psi} \left[\theta_{a_1}\cdots\theta_{a_{k-1}}\theta_{a}\right]^{\tltw}
    \theta_{a}|_{a_1=\cdots=a_{k-2}=\pol,a_{k-1}=\psi}
    =&\frac{1}{2^{k-1}}\left(\sin(k\phi)\cos\phi - \cos(k\phi)\sin\phi\right).
\end{align}
Then, we get 
\begin{align}
    &\left[\theta_{a_1}\cdots\theta_{a_k}\theta_{a_{k+1}} \right]^{\tltw}|_{a_1=\cdots=a_{k+1}=\pol} 
    \nonumber\\
    &\hspace{1cm}=\frac{1}{2^{k-1}}\cos(k\phi)\cos\phi 
    - \frac{1}{k(k+1)}\cdot\binom{k+1}{2}\cdot\frac{1}{2^{k-1}}\left(\cos(k\phi)\cos\phi + \sin(k\phi)\sin\phi\right)
    \nonumber\\
    &\hspace{1cm}=\frac{1}{2^k}\left(\cos(k\phi)\cos\phi - \sin(k\phi)\sin\phi\right)
    \nonumber\\
    &\hspace{1cm}=\frac{1}{2^k}\left(\cos(k+1)\phi\right),
    \\
    &\left[\theta_{a_1}\cdots\theta_{a_k}\theta_{a_{k+1}} \right]^{\tltw}|_{a_1=\cdots=a_{k}=\pol,a_{k+1}=\psi} 
    \nonumber\\
    &\hspace{1cm}=\frac{1}{k+1}\cdot\frac{1}{2^{k-1}}
    \left( \cos(k\phi)\sin\phi + k\sin(k\phi)\cos\phi \right)
    \nonumber\\
    &\hspace{2cm}- \frac{1}{k(k+1)}\cdot\binom{k}{2}\cdot\frac{1}{2^{k-1}}
    \left(\sin(k\phi)\cos\phi - \cos(k\phi)\sin\phi\right)
    \nonumber\\
    &\hspace{1cm}=\frac{1}{2^k}\left(\sin(k\phi)\cos\phi + \cos(k\phi)\sin\phi\right)
    \nonumber\\
    &\hspace{1cm}=\frac{1}{2^k}\left(\sin(k+1)\phi\right),
\end{align}
which indicates that Eq.~(\ref{eq:cosnphi_TL}) and Eq.~(\ref{eq:sinnphi_TL}) are true for $n=k+1$. Therefore, Eq.~(\ref{eq:cosnphi_TL}) and Eq.~(\ref{eq:sinnphi_TL}) hold for all $n\geq2$.

\subsection{Symmetric traceless tensor in arbitrary dimension}\label{sec:comment_tl}
Next, we consider an $n$th symmetric traceless tensor in $d$-dimension, expressing it as
\begin{align}  \label{eq:tl}
    \tilde{I}_{i_1i_2\cdots i_n} &= I_{i_1i_2\cdots i_n} + \sum_{k=1}^{[n/2]}\tilde{C}_k\left[I_{\tilde{i}_{1}\tilde{i}_{1}\cdots \tilde{i}_{k}\tilde{i}_{k}i_{2k+1}\cdots i_n}\delta_{i_{1},i_{2}}\cdots\delta_{i_{2k-1},i_{2k}}+({\rm perms.})\right]\,,
\end{align}
where $I_{i_1i_2\cdots i_n}$ is an arbitrary rank-$n$ symmetric tensor and $\tilde{I}_{i_1i_2\cdots i_n}$ is its traceless part. In the square brackets of the right hand side, the indices $I_{i_1i_2\cdots i_n}$ are replaced with the contraction by $\tilde{i}_m~(m=1,\, \cdots,\, k)$ and the replaced indices appear as the indices of the Kronecker delta. Imposing the traceless condition, we obtain
\begin{align}
\tilde{C}_k=\tilde{C}_{k+1}[2k-(d+2n-4)]\,.
\end{align}
For $n \geq 2, d \geq 2$ and $0\leq k\leq [ n/2 ]$, $2k-(d+2n-4)\leq0$. Then, we can obtain
\begin{align}
\tilde{C}_k=\left(-\frac{1}{2}\right)^k\frac{\Gamma(n-k-1+d/2)}{\Gamma(n-1+d/2)!}\,.
\end{align}
For example, the coefficient for $d=2$ is given by
\begin{align}
    \tilde{C}_k=\left(-\frac{1}{2}\right)^k\frac{(n-k-1)!}{(n-1)!}\,,
\end{align}
and the one for $d=3$ is given by
\begin{align}
    \tilde{C}_k=\left(-1\right)^k\frac{(2n-2k-1)!!}{(2n-1)!!}\,.
\end{align}

The number of permutation in Eq.~(\ref{eq:tl}) is determined by counting the possible number of this replacement as
\begin{align}\label{eq:tl_number}
\frac{1}{k!}\dbinom{n}{2}\dbinom{n-2}{2}\cdots\dbinom{n-2(k-1)}{2} = \frac{1}{2^k k!}\frac{n!}{(n-2k)!}\,.
\end{align}
For instance, for $n=2$, Eq.~(\ref{eq:tl}) simply gives the familiar expression as
\begin{align}
    \tilde{I}_{i_1 i_2} = I_{i_1 i_2} + \tilde{C}_1 I_{\tilde{i}_1 \tilde{i}_1}\delta_{i_1, i_2}= I_{i_1 i_2} - \frac{I_{\tilde{i}_1 \tilde{i}_1}}{d} \delta_{i_1, i_2}\,. 
\end{align}

\subsection{Rewriting Eq.~(\ref{eq:gamman_exp}) into Eq.~(\ref{eq:gamman_simp})}
Next, we show the following formulae
\begin{align}
\tilde{I}_{\scriptsize \underbrace{\pol\pol\cdots \pol\pol}_{(n, 0)}} &= \frac{1}{2^{n-1}}\sum^{[n/2]}_{l=0}(-1)^l\dbinom{n}{2l} \, I_{\scriptsize \underbrace{\pol\pol\cdots \psi\psi}_{(n-2l, 2l)}}\,,\label{eq:tl_11}\\
\tilde{I}_{\scriptsize \underbrace{\pol\pol\cdots \pol\psi}_{(n-1, 1)}} &= \frac{1}{2^{n-1}}\sum^{[(n-1)/2]}_{l=0}(-1)^l\dbinom{n}{2l+1}\, I_{\!\!\!\!\!\scriptsize \underbrace{\pol\pol\cdots \psi\psi}_{(n-2l-1, 2l+1)}}\label{eq:tl_12}\,,
\end{align}
with which we can rewrite Eq.~(\ref{eq:gamman_exp}) into Eq.~(\ref{eq:gamman_simp}).

Let us start with expressing $\tilde{I}_{\pol\pol\cdots \pol\pol}$ and $\tilde{I}_{\pol\pol\cdots \pol\psi}$, using Eq.~(\ref{eq:tl}). Counting the numbers of $I_{i_1 \cdots i_n}$ with $2l$ indices being $\psi$ and $(n-2l)$ indices being $\pol$ for $l=1,\, \cdots [n/2]$, we obtain
\begin{align}
\tilde{I}_{\pol\pol\cdots\pol} &= \sum_{k=0}^{[n/2]}(-1)^k2^{-2k}\frac{(n-k-1)!}{(n-1)!}\frac{n!}{k!(n-2k)!}\sum_{q=0}^{k}\dbinom{k}{q} \,I_{\!\!\!\!\!\!\!\!\scriptsize \underbrace{\pol\pol\cdots \psi\psi}_{(n-2k+2q, 2k-2q)}}\notag\\
&=\sum_{k=0}^{[n/2]}\sum_{l=0}^{k}(-1)^k2^{-2k}\frac{n}{l}\dbinom{n-k-1}{k-1}\dbinom{k-1}{k-l}\, I_{\scriptsize \underbrace{\pol\pol\cdots \psi\psi}_{(n-2l, 2l)}}, \label{eq:tl_11_raw}
\end{align}
where the summation over $q$ counts the number of indices with $\tilde{a}_{i_1} = \cdots= \tilde{a}_{i_q} =\pol$ among the $k$ contracted indices of $
I_{\tilde{a}_{1}\tilde{a}_{1}\cdots  \tilde{a}_{k}\tilde{a}_{k}a_{2k+1}\cdots a_n}$ (then the total number of index $\pol$ becomes $(2n-2k) + 2q$ and the one of index $\psi$ becomes $2k-2q$) and the numerical factor, $n!/(2^k k! (n-2k)!)$ comes from the number of permutation, given in Eq.~(\ref{eq:tl_number}). In the second equality, we have changed $k-q\rightarrow l$. Similarly, we obtain
\begin{align}\label{eq:tl_12_raw}
\tilde{I}_{\pol\pol\cdots\psi} 
&=\sum_{k=0}^{[(n-1)/2]}\sum_{l=0}^{k}(-1)^k2^{-2k}\dbinom{n-k-1}{k}\dbinom{k}{k-l}I_{\scriptsize \underbrace{\pol\pol\cdots \pol\psi}_{(n-2l-1, 2l+1)}}\,,
\end{align}
where the number of permutation is now given by
\begin{align}
\frac{1}{k!}\dbinom{n-1}{2}\dbinom{n-3}{2}\cdots\dbinom{n-2(k-1)-1}{2} = \frac{1}{2^k k!}\frac{(n-1)!}{(n-2k-1)!}\,,
\end{align}
since the terms including $\delta_{\pol\psi}$ vanish.

Changing the summation from $0\leq k \leq [n/2]$ and $0\leq l \leq k$ to $0\leq l \leq [n/2]$ and $l\leq k \leq [n/2]$, we can further rewrite \refeq{tl_11_raw} and \refeq{tl_12_raw} as 
\begin{align}
\tilde{I}_{\pol\pol\cdots\pol} &= \sum_{l=0}^{[n/2]}\frac{n}{l}\left(-\frac{1}{4}\right)^l\sum_{k=0}^{[n/2]-l}\left(-\frac{1}{4}\right)^k\dbinom{n-k-l-1}{k+l-1}\dbinom{k+l-1}{k} I_{\scriptsize \underbrace{\pol\pol\cdots \pol\psi}_{(n-2l, 2l)}}\,,\label{eq:tl_11_2}\\
\tilde{I}_{\pol\pol\cdots\psi} &= \sum_{l=0}^{[(n-1)/2]}\left(-\frac{1}{4}\right)^l\sum_{k=0}^{[(n-1)/2]-l}\left(-\frac{1}{4}\right)^k\dbinom{n-k-l-1}{k+l}\dbinom{k+l}{k} I_{\scriptsize \underbrace{\pol\pol\cdots \pol\psi}_{(n-2l, 2l)}}\,,
\end{align}
where we have changed $k-l \rightarrow k$. Rewriting the summation over $k$ by using the following formulae:
\begin{align}
    &\sum_{k=0}^{[n/2]}\left(-\frac{1}{4}\right)^k\dbinom{n-k+l-1}{k+l-1}\dbinom{k+l-1}{k}
    =\frac{1}{2^n}\frac{(n+2l-1)!}{n!(2l-1)!}\,, \label{eq:tl_formula1} \\
    &\sum_{k=0}^{[n/2]}\left(-\frac{1}{4}\right)^k\dbinom{n-k+l}{k+l}\dbinom{k+l}{k}
    =\frac{1}{2^n}\frac{(n+2l+1)!}{n!(2l+1)!} \,,
\end{align}
we arrive at Eqs.~(\ref{eq:tl_11}) and (\ref{eq:tl_12}).

\section{Weak lensing}\label{sec:wl}
As discussed in Sec.~\ref{sec:bias}, the galaxy shape moment also can be generated through the gravitational lensing. In this Appendix, we estimate the spin-$n$ component of the distortion generated through the weak lensing, $\IWL$, providing the detailed computation. As discussed around Eq.~(\ref{Exp:I2orignes}), $\ITL^{\rm WL}_{i_1 \cdots i_n}$ is given by the difference between the observed galaxy shape function, $I_{i_1i_2\cdots i_n}$, and the intrinsic one, $I^{\rm int}_{i_1i_2\cdots i_n}$. The former is defined by using the coordinates on the image plane, $\tilde{\theta}_i = \bar{\theta}_i + \theta_i $, and the latter is defined by using those on the source plane, $\tilde{\theta}_{{\rm s}\, i} = \bar{\theta}_{{\rm s}\, i} + \theta_{{\rm s}\, i}$. In Sec.~\ref{SSec:WL}, solving the geodesic equation, we derive the relation between these two coordinates,  $\tilde{\theta}^i= \tilde{\theta}^i(\tilde{\bm{\theta}}_{\rm s})$. We put a bar on the centroid coordinates and a tilde on the coordinates measured from the common origin on each plane. Using this relation, we can express the deviation from the centroid on the image plane, $\theta^i$, using the coordinates on the source plane ($\to$ Eq.~(\ref{Exp:DeltaTheta})). Using this expression, in Sec.~\ref{SSec:IWL}, we compute $\IWL$.

\subsection{Solving the null geodesics} \label{SSec:WL}
The relation between the intrinsic galaxy shape on the source plane and the apparent galaxy shape on the image plane can be computed by tracing the photon propagation in the perturbed FLRW spacetime. Using the affine parameter $\lambda$, the null geodesic equation, which describes the photon propagation, is given by
\begin{align}
	\frac{d^2 x^\mu}{d\lambda^2}+\Gamma^\mu_{\alpha\beta}\frac{dx^\alpha}{d\lambda}\frac{dx^\beta}{d\lambda} = 0\,,
\end{align}
where $\Gamma^\mu_{\alpha \beta}$ is the Christoffel symbol of the perturbed spacetime. The Greek indices $\mu,\alpha,\beta$ run from 0 to 3. Solving the geodesic equation along the orthogonal direction to $\hat{\bm{n}}$ with a use of the null condition
\begin{align}
	g_{\mu\nu}\frac{dx^\mu}{d\lambda}\frac{dx^\nu}{d\lambda} = 0\,,
\end{align}
we obtain the lens equation as
\begin{align}
\tilde{\theta}_{{\rm s}\,i} (\tilde{\bm{\theta}}) = \tilde{\theta}_i +  \int^{\chi}_0 d\chi' \frac{\chi-\chi'}{\chi}  \partial_{\perp i} (\Phi (\chi',\, \tilde{\bm{\theta}})-\Psi (\chi',\, \tilde{\bm{\theta}})),\,  \label{Exp:WL_single}
\end{align} 
with $\partial_{\perp i} \equiv {{\cal P}_i}^{j} \partial_j$. Here, we have imposed the boundary condition as $\tilde{\theta}^i(\chi =0) = \tilde{\theta}^i$ and $\tilde{\theta}^i(\chi) = \tilde{\theta}^i_{\rm s}$. The projected coordinates $\tilde{\theta}_i$ and $\tilde{\theta}_{{\rm s}\, i}$ correspond to those on the image plane and the source plane, respectively. A detailed computation of Eq.~(\ref{Exp:WL_single}) can be found e.g. in Ref.~\cite{Schmidt:2012ne}. We also use $\tilde{\theta}_{{\rm s}\,i}$ to express the mapping of a coordinate on the image plane $\tilde{\theta}_i$ to the corresponding one on the source plane as $\tilde{\theta}_{{\rm s}\,i} (\tilde{\bm{\theta}})$.

Taking the partial derivative of $\tilde{\theta}_{{\rm s}\, i}$ with respect to $\tilde{\theta}^j$, we obtain the deformation matrix $A_{ij}$~\cite{Mellier:1998pk} as
\begin{align}
    A_{ij} (\tilde{\bm{\theta}}) \equiv  \frac{\partial \tilde{\theta}_{{\rm s} i}}{\partial \tilde{\theta}^j}=\delta_{ij} + \int^{\chi}_0 d\chi' \frac{\chi-\chi'}{\chi} \chi' \partial_{\perp i}\partial_{\perp j} (\Phi (\chi',\, \tilde{\bm{\theta}})-\Psi (\chi',\, \tilde{\bm{\theta}}))\,.  \label{Exp:AijAp}
\end{align}
Because of the spatial inhomogeneity of $\Phi$ and $\Psi$, the deformation matrix $A_{ij}$ depends on $\tilde{\bm{\theta}}$.

\begin{figure}[htbp]
    \centering
    \includegraphics[width=140mm]{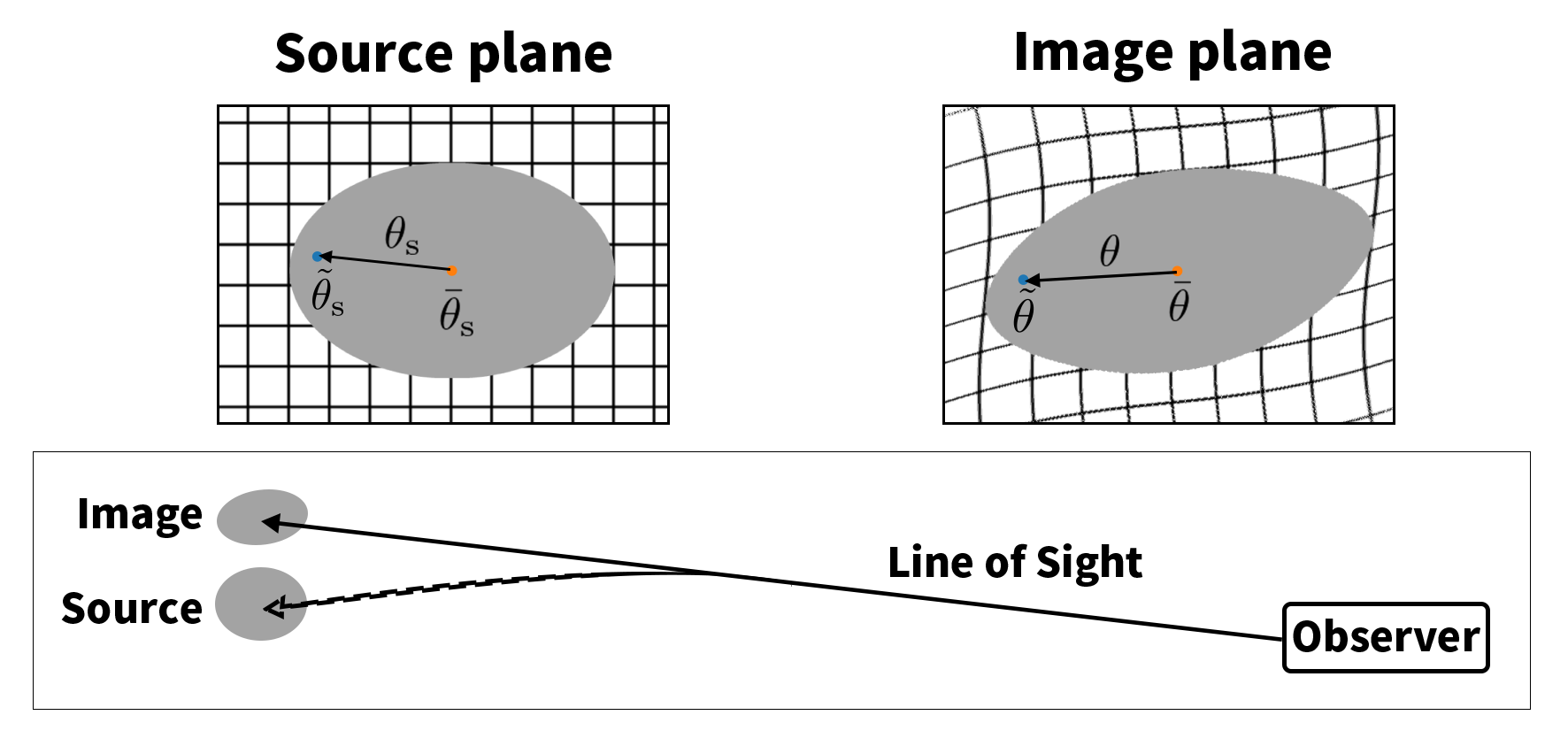}
    \caption{This figure shows the 2D coordinates on the source plane and on the image plane. The orange and blue dots denote the centroids and given points on each plane.}
    \label{fig:my_label}
\end{figure}
Now, let us compute the deviation from the centroid on the source plane, $\theta_{{\rm s}\, i} \equiv \tilde{\theta}_{{\rm s}\, i} - \bar{\theta}_{{\rm s}\, i}$, decomposing it into the two parts as
\begin{align}
     \theta_{{\rm s}\,i} =  \left[\tilde{\theta}_{{\rm s}\,i}(\tilde{\bm{\theta}}) - \tilde{\theta}_{{\rm s}\, i}(\bar{\bm{\theta}}) \right] + \left[ \tilde{\theta}_{{\rm s}\,i}(\bar{\bm{\theta}})   - \bar{\theta}_{{\rm s}\, i} \right] \,. \label{Exp:coordinate}
\end{align}
These coordinates on the source plane and the image plane are visually explained in Fig.~\ref{fig:my_label}. The first square brackets describe the difference between the two coordinates on the source plane which are mapped from $\tilde{\theta}_i$ and $\bar{\theta}_i$ on the image plane, respectively. The second square brackets describe the difference between the coordinates on the source plane mapped from the apparent centroid, $\bar{\theta}_i$, and the actual centroid, $\bar{\theta}_{{\rm s}\, i}$. Since we cannot directly measure the source distribution, the actual centroid $\bar{\bm{\theta}}_{\rm s}$ is left undermined.

Using Eq.~(\ref{Exp:WL_single}), we can compute the first square brackets in Eq.~(\ref{Exp:coordinate}) minus $\theta_i$ as 
\begin{align}
    \delta \theta_i \equiv \tilde{\theta}_{{\rm s}\,i} (\tilde{\bm{\theta}}) -  \tilde{\theta}_{{\rm s}\,i} (\bar{\bm{\theta}}) - \theta_i =  \theta_{{\rm s}\, i} -  \theta_i - \left[ \tilde{\theta}_{{\rm s}\,i}(\bar{\bm{\theta}})   - \bar{\theta}_{{\rm s}\, i} \right]  ,
    \label{Def:deltatheta}
\end{align}
as
\begin{align}
   \delta \theta_i
   &= \int^{\chi}_0 d\chi' \frac{\chi-\chi'}{\chi} \, \partial_{\perp i} \left[ (\Phi (\chi',\, \tilde{\bm{\theta}})-\Psi (\chi',\, \tilde{\bm{\theta}})) - (\Phi (\chi',\, \bar{\bm{\theta}})-\Psi (\chi',\, \bar{\bm{\theta}})) \right]\,.
\end{align}
Here we are evaluating $\Phi$ and $\Psi$ along the unperturbed photon path. This is only correct at linear order in the deflection. If one includes the lower-order deflection in the argument of these metric perturbations 
and expands this effect as well, one obtains additional contributions which involve the integral over products of
$\partial_k\partial_l\Phi$ and $\partial_k\partial_l\Psi$. These contributions, which are known as ``post-Born corrections'' (e.g., \cite{Krause:2009yr}), to the $n$th moment can be roughly approximated as being of order $(\delta A^{(2)})^m$, with $m=n$ if $n$ is even and $m=(2n+1)/2$ if $n$ is odd. This means that they are of the same order as the nonlinear lensing contribution we will discuss below,
and hence we will not derive them in detail here.
  
Expanding $\delta \theta_i$ with respect to $\theta_i = \tilde{\theta}_i - \bar{\theta}_i$, we obtain
\begin{align}
    \delta \theta_i(\bm{\theta}) = \sum_{n=2}^{\infty} \frac{1}{(n-1)!}\, \delta A^{(n)}_{i i_1 \cdots i_{n-1}} \theta^{i_1} \cdots  \theta^{i_{n-1}}\,,  
\end{align}
with
\begin{align}
	\delta A^{(n)}_{i i_1 \cdots i_{n-1}} = 
	\int^{\chi}_0 d\chi' \frac{\chi-\chi'}{\chi} \left[\prod_{k=1}^{n-1} (\chi' \partial)_{\perp i_k} \right] \partial_{\perp i} (\Phi(\chi',\bar{\bm{\theta}})-\Psi(\chi',\bar{\bm{\theta}}))\,.
\end{align}
The spatial variation of $A_{ij}$ yields the higher order terms of $\theta_i$, characterized by $\delta A^{(n)}_{i i_1 \cdots i_{n-1}}$ with $n \geq 3$. For $n=2$, $\delta A^{(2)}_{ij}$ is nothing but the deviation of the deformation matrix from the unit matrix. The traceless part of $\delta A^{(n)}_{i i_1 \cdots i_{n-1}}$ describes the spin-$n$ deformation due to the gravitational lensing, whose visual image is discussed in Sec.~\ref{SSec:Image}. 

\subsection{Estimation of weak lensing effect} ~\label{SSec:IWL}
Using the formula derived in the previous subsection, let us compute the weak lensing contribution in Eq.~(\ref{Exp:I2orignes}), which can be rewritten as
\begin{align}
    I^{\rm WL}_{i_1i_2\cdots i_n} (\bar{\bm{\theta}}) = \left[ I_{i_1i_2\cdots i_n} (\bar{\bm{\theta}}) - I^{\rm int}_{i_1i_2\cdots i_n} (\bar{\bm{\theta}}_{\rm s}) \right] + \left[I^{\rm int}_{i_1i_2\cdots i_n} (\bar{\bm{\theta}}_{\rm s}) - I^{\rm int}_{i_1i_2\cdots i_n} (\bm{\theta}_{\rm s} (\bar{\bm{\theta}})) \right]\,. \label{Exp:IWL}
\end{align}
Similarly to Eq.~(\ref{eq:moment}), using the projected coordinates on the source plane, $\theta_{{\rm s}\,i}$, the $n$-th moment intrinsic galaxy shape function is defined as 
\begin{align}
 I^{\rm int}_{i_1i_2\cdots i_n} (\bar{\bm{\theta}}_{\rm s},\, \tau)=\frac{\chi^n}{\bar{I}^{\rm int}(\bar{\bm{\theta}}_{\rm s}) R_*^n}\int d^2 \bm{\theta}_{\rm s} \prod_{m=1}^{n}  \theta_{{\rm s}\, i_m } I^{\rm int}(\bar{\bm{\theta}}_{\rm s}+\bm{\theta}_{\rm s},\, \tau)\,,
\end{align}
with the normalization
\begin{align}
\bar{I}^{\rm int} (\bar{\bm{\theta}}_{\rm s},\, \tau) \equiv\int d^2 \bm{\theta}_{\rm s}\, I^{\rm int}(\bar{\bm{\theta}}_{\rm s} + \bm{\theta}_{\rm s},\, \tau)\,.
\end{align}
The brightness theorem, which states that the lensing does not change the surface brightness, relates the apparent surface brightness $I(\tilde{\bm{\theta}})$ to the intrinsic surface brightness $I^{\rm int}(\tilde{\bm{\theta}}_{\rm s})$ as $I^{\rm int} (\tilde{{\bm \theta}}_{\rm s}) = I(\tilde{{\bm \theta}})$ or equivalently in our notation,
\begin{align}
    I^{\rm int} (\bar{\bm{\theta}}_{\rm s} + \bm{\theta}_{\rm s},\, \tau) = I(\bar{\bm{\theta}} + \bm{\theta},\, \tau)\,. 
\end{align}

First, let us compute the first square brackets of Eq.~(\ref{Exp:IWL}) by using $\theta_i$ expressed in terms of $\theta_{{\rm s}\,i}$. In the previous subsection, we derived the expression of $\delta \theta_i (\bm{\theta})$, solving the null geodesics. Using this expression, $\theta_{{\rm s}\,i}$ in the left hand side of Eq.~(\ref{Exp:coordinate}) can be expressed in terms of $\theta_i$ and the global shift between the apparent and actual centroids. Solving Eq.~(\ref{Exp:coordinate}) recursively, we can express $\theta_i$ in terms of $\theta_{{\rm s}\,i}$ and the contribution due to the global shift of the centroid as
\begin{align}
    \theta_i = \theta_{{\rm s}\,i} + \sum_{n=2}^{\infty} \frac{1}{(n-1)!}\, \delta B^{(n)}_{i i_1 \cdots i_{n-1}} \theta_{\rm s}^{i_1} \cdots \theta_{\rm s}^{i_{n-1}} + {\cal F}_i (\tilde{\bm{\theta}}_{\rm s} (\bar{\bm{\theta}}) - \bar{\bm{\theta}}_{\rm s})\,,  \label{Exp:DeltaTheta}
\end{align}
where, e.g., the leading contributions of $\delta B^{(2)}_{i_1 i_2}$ and ${\cal F}_i$, which is a function of $\tilde{\bm{\theta}}_{\rm s} (\bar{\bm{\theta}}) - \bar{\bm{\theta}}_{\rm s}$, are given by
\begin{align}
    & \delta B^{(2)}_{i_1 i_2} = - \delta A^{(2)}_{i_1 i_2} + \delta A^{(2)}_{i_1 j} \delta A^{(2)j}_{~~~~i_2} + \cdots \,, \cr
   &   {\cal F}_i (\tilde{\bm{\theta}}_{\rm s} (\bar{\bm{\theta}}) - \bar{\bm{\theta}}_{\rm s}) = \left[ - {\delta_i}^j  + \delta A^{(2)j}_i  + \cdots  \right] \left(  \tilde{\theta}_{{\rm s}\,j} (\bar{\bm{\theta}}) - \bar{\theta}_{{\rm s}\,j} \right) + \cdots \,. 
\end{align}
Since $\delta A^{(n)}_{i_1 \cdots i_n}$ are the functions of $\bar{\bm{\theta}}$, so are $\delta B^{(n)}_{i_1 \cdots i_n}$. Inserting Eq.~(\ref{Exp:DeltaTheta}) into Eq.~(\ref{eq:moment}), we can expand $I_{i_1i_2\cdots i_n} (\bar{\bm{\theta}})$ in terms of $I^{\rm int}_{i_1i_2\cdots i_n} (\bar{\bm{\theta}}_{\rm s})$, where the coefficients are given by products of $\delta B^{(n)}_{i_1 \cdots i_n}$ and ${\cal F}_i$.

The leading lensing effect on the $n$th moment of a galaxy image can be computed by making the lensing operators derived above act on a perfectly symmetric (spherical) intrinsic image, whose moments are given by
\begin{align}
    I^{\rm int}_{i_1i_2\cdots i_n} (\bar{\bm{\theta}}_{\rm s})
    =\left\{\begin{array}{lr}
    C_{2N} [{\cal P}_{i_1i_2} {\cal P}_{i_3i_4}\cdots {\cal P}_{i_{2N-1}i_{2N}}+\mathrm{perms.}] & n=2N \,, \\
    0 & n=2N-1  \,,
    \end{array} \right.\label{odd}
\end{align}
where $N$ is a natural number and $C_{2N}$ is of order the angular size of the image to the $n$th power; Ref.~\cite{Fleury:2018odh} also presents the effects of non-symmetric intrinsic images. 
In the following, we calculate the terms in the first brackets of Eq.~(\ref{Exp:IWL})
for $n= 2N$. For the moment, let us ignore the contributions due to the shift of the centroid, which turns out to be sub-leading. Inserting Eq.~(\ref{Exp:DeltaTheta}) into Eq.~(\ref{eq:moment}) and using Eq.~(\ref{odd}), we obtain
\begin{align}
 1 {\rm st~brackets~of~Eq.(\ref{Exp:IWL})}
 &\ni \Bigl( \frac{R_*}{\chi} \Bigr)^{2(M-1)} 
 \delta A^{(2M)}_{i_1 \cdots
 i_m} , \cr
 &\,\,\,\,\,\delta A^{(2)}_{i_1 j_1} {\delta A^{(2)j_1}}_{i_2} \cdots \delta A^{(2)}_{i_{2N-1} j_N} {\delta A^{(2)j_N}}_{i_{2N}}\,,\, ({\rm perms})\,, \cr
 &\,\,\,\,\, \cdots\,, \label{Exp:IWLcont}
\end{align}
with $M= 1,\, 2,\, \cdots $. The factor $(R_*/\chi)^{2(M-1)}$ of the first contribution appears, since $I_{i_1 \cdots i_{2N}}$ and $I^{\rm int}_{i_1 \cdots i_{2(N+ M-1)}}$ carry different powers of $R_*/\chi$. In computing the terms in the second line, which is given by $2N$ product of $\delta A^{(2)}$, we used $\delta A^{(2)}_{i_p j} {\delta A^{(2)}}_{i_q j'} {\cal P}^{j j'} =  \delta A^{(2)}_{i_p j} {\delta A^{(2)j}}_{i_q}$. The ellipses stand for the trace part and other non-linear terms in perturbations. As discussed above, these include post-Born corrections which are of similar order of magnitude. 
While we did not write it explicitly, for $M < N$, the lacking indices should be supplied by the projection tensor ${\cal P}_{ij}$. Therefore, such terms do not contribute to the traceless component. Since the order of $\delta A^{(2M)}_{i_1 \cdots i_{2M}}$ is given by 
\begin{align}
 & \left| \delta A^{(2M)}_{i_1 \cdots
 i_{2M}}  \right|
 = {\cal O}\left( ( k \chi')^{2(M-1)} \right) \times |\delta A^{(2)}|\,,  \label{Exp:orderA2M}
\end{align}
the first contribution in Eq.~(\ref{Exp:IWLcont}) amounts to 
\begin{align}
  \left| \left( \frac{R_*}{\chi} \right)^{2(M-1)} \times \delta A^{(2M)}_{i_1 \cdots i_{2M}} \right| &= {\cal O}\left( ( k R_*)^{2(M-1)} \right) \times |\delta A^{(2)}| \,. \label{eq:orderA2Mlinear}
\end{align}
In Eq.~(\ref{Exp:orderA2M}), we have replaced $\partial/\partial \bar{\theta}^i$ with $k \chi'$, where $k$ is the Fourier mode of the 3D coordinates of the centroid on the image plane. Equation (\ref{eq:orderA2Mlinear}) indicates that at the linear order of the perturbation, the traceless component of the first brackets of Eq.~(\ref{Exp:IWL}) with $n= 2N \geq 4$ is suppressed at least by $( k R_*)^{n-2}$, which is much smaller than 1 since we only consider larger scales than the typical size of galaxies. To be precise, the integrand of $\delta A^{(2)}$ in Eq.~(\ref{eq:orderA2Mlinear}) is further suppressed by $(\chi'/\chi)^{2(M-1)}$. For example, for the 4th moment, the tree-level diagram which appears by contracting the first contribution in Eq.~(\ref{Exp:IWLcont}) roughly scales as
\begin{align}
 & (k R_*)^4 \times P_{\rm L}(k) \,.
\end{align}
Meanwhile, for the 2nd moment with $N=1$, the linear contribution simply gives the deformation matrix $\delta A^{(2)}_{i_1 i_2}$.

Similarly, all the non-linear contributions which include $\delta A^{(n)}_{i_1 \cdots i_n}$ with $n \geq 3$ are suppressed by positive powers of $k R_*$. Therefore, the only contributions which are not suppressed by $k R_* \ll 1$ are products of $\delta A^{(2)}$, i.e. the second contribution in Eq.~(\ref{Exp:IWLcont}). In the end of this subsection we will discuss the loop contributions from the products of $\delta A^{(2)}$.

So far, we have not considered the contribution from the shift of the centroid.
According to Ref.~\cite{Okura:2006fi}, the centroid shift can be estimated by (see also Ref.~\cite{Viola:2011ue})
\begin{align}
     |\theta_{{\rm s} i}(\bar{\bm{\theta}}) - \bar{\theta}_{{\rm s} i}| \sim {\cal O}(\Delta \theta^j \partial \delta A^{(2)}/\partial \theta^j ) \times \Delta \theta_i\,. 
\end{align}
Repeating a similar argument, we find that the leading contribution of the centroid shift has more $(kR_*)$ than the leading contribution and is suppressed more on the large scale.
For this reason, we do not consider the contribution of the centroid shift.

Next, let us discuss the first brackets of Eq.~(\ref{Exp:IWL}) for an odd number $n$. 
Repeating the same argument, we find that the odd $n$th moment is always suppressed by $k R_*$, since the intrinsic $n$th moment of the galaxy shape function $I^{\rm int}_{i_1i_2\cdots i_n} (\bar{\bm{\theta}}_{\rm s})$ vanishes for an odd $n$ under the assumption of Eq.~(\ref{odd}). In particular, the first brackets of Eq.~(\ref{Exp:IWL}) is suppressed at least by $(k R_*)^{n-2}$ at the linear perturbation.

In summary, we showed that the only contributions that are not suppressed by $k R_* (\ll 1)$ are the products of $\delta A^{(2)}$. 
A contraction of $\delta A^{(2)}$ yields weak lensing loops, computed in the projected 2D plane. As emphasized in Sec.~\ref{SSec:nla}, these 2D loops are qualitatively different from 3D loop contributions to the galaxy shape moments, which are projected into 2D after computing loop contributions in 3D (see, e.g. Eq.~(\ref{eq:NLA})). The 3D loop contributions are included as the non-linear alignment effect. For a comparison of 2D and 3D loops, here we estimate their contributions to the angular power spectrum of the 4th shape moment, $C^{(4,4)}_l$. 
Taking the flat-sky limit, the 2D weak lensing loops in $C^{(4,4)}_l$ is roughly estimated as,
\begin{align}
    C^{(4,4){\rm 2D}{\textrm -}{\rm 1L}}_l \sim \langle (\delta A^{(2)}\delta A^{(2)})(\bm{l}) (\delta A^{(2)}\delta A^{(2)})(\bm{l}') \rangle \sim 2\int \frac{d^2 l_1}{(2\pi)^2} C_\gamma(l_1) C_\gamma(|\bm{l}-\bm{l}_1|)\,,\label{eq:aps_4th_wl}
\end{align}
where $C_\gamma(l)$ is the lensing shear for 2nd moment and $\bm l$ is a 2D vector on the Cartesian coordinate. 
We find that the contribution of 2D lensing 1-loop, given by \refeq{aps_4th_wl}, is much smaller than the one of the 3D 1-loop(NLA). 
Therefore, evaluating the forecast in this paper, we neglect the weak lensing contribution, whose linear contribution is suppressed by $kR_*$ and whose loop contributions are much smaller than the 3D loops, included in non-linear alignment effects.

In this Appendix, we have estimated the $n$th moment deformation of the galaxy shape due to the weak lensing, assuming the absence of the intrinsic deformation, i.e. assuming a circular intrinsic image. As was argued in Sec.~\ref{sec:PNG}, the angular dependent PNG can generate the intrinsic alignment of the galaxy shape. To estimate the weak lensing contribution more accurately by taking into account the intrinsic deformation, the assumption, (\ref{odd}), should be abandoned.

\section{PNG from spin-4 particle}\label{sec:appB}
In this Appendix, we compute the contribution of the PNG generated by a spin-4 particle, given in Eq.~(\ref{eq:NGIC}), to the 4th moment galaxy shape function.

\subsection{Contributions to galaxy shape function}\label{ssec:appB}
From Eq.~(\ref{eq:4thbiasex}), the cross-correlation of the matter density field with $\tilde{g}_{ijkl}$ is
\begin{align}
\langle \delta({\bm x}) \tilde{g}_{ijkl}({\bm y})\rangle
=b_{K^2}^{(4)}\left\langle \delta(\bm{x})  \left[K_{ij} (\bm{y})  K_{kl} (\bm{y}) \right]^{{\rm TL}_3,\,{\rm sym} } \right\rangle
+{\cal O}((kR_*)^2).
\end{align}
Let us compute $\left\langle \delta(\bm{x})  \left[K_{ij} (\bm{y})  K_{kl} (\bm{y}) \right]^{{\rm TL}_3,\,{\rm sym} } \right\rangle$ term in the presence of the non-Gaussian initial condition such as Eq.~(\ref{eq:NGIC}).
\begin{align}
&\left\langle \delta(\bm{x})   \left[K_{ij} (\bm{y})  K_{kl} (\bm{y}) \right]^{{\rm TL}_3,\,{\rm sym} }\right\rangle \notag\\
=&\int \frac{d^3 \bm{k}}{(2\pi)^3}e^{i \sbm{k} \cdot \sbm{r}}\mathcal{M}(k)
\int \frac{d^3 \bm{p}}{(2\pi)^3}~ [\hat{p}_i\hat{p}_j\hat{p}'_k\hat{p}'_l ]^{\rm TL_3}
\mathcal{M}(p)\mathcal{M}(|\bm{k}+\bm{p}|)B_{\Phi}(k,p,|\bm{k}+\bm{p}|) 
\\
=&\int \frac{d^3 \bm{k}}{(2\pi)^3}e^{i \sbm{k} \cdot \sbm{r
}}\mathcal{M}(k) P_{\Phi}(k)
\int \frac{p^2dp}{(2\pi)^2}\mathcal{M}^2(p)P_{\Phi}(p)
\nonumber\\
&\hspace{1cm}\int d\mu \int \frac{d\varphi}{2\pi} ~ [\hat{p}_i\hat{p}_j\hat{p}'_k\hat{p}'_l ]^{\rm TL_3}
\sum_{\ell}{\cal A}_{\ell}\mathcal{P}_{\ell}(\mu)[2+2q\mu\{n_{\Phi}+n_{\mathcal{M}}(p)\}+\mathcal{O}(q^2)] \label{eq:appb_1}
\end{align}
with $\bm{r}=\bm{x}-\bm{y},\ \hat{\bm{p}}' =-\widehat{(\bm{k}+\bm{p})},\ q=k/p,\ \mu=\hat{\bm k}\cdot\hat{\bm p},\ n_{\Phi}=n_s-4,\ n_{\mathcal{M}}(p)\equiv \partial \ln\mathcal{{M}}(p)/\partial \ln{p}$.
In the first line, we used the symmetric tensor property and
in the second line, we used Eq.(\ref{eq:NGIC}) and expanded $B_{\phi}(k,p,|\bm{ k}+\bm{p}|)$ and $\mathcal{M}(|\bm{k}+\bm{p}|)$ in the powers of $q \ll 1$.

We also apply $\hat{\bm{p}}' \sim \hat{\bm{p}}$ in the powers of $q\ll1$.
Choosing the $z$ direction of the polar coordinates along the direction of $\hat{\bm k}$ and integrating over the azimuthal angle $\varphi$, we obtain
\begin{align} 
  \int \frac{d\varphi}{2\pi} [\hat{p}_{i_1} \cdots \hat{p}_{i_n} ]^{\rm TL_3}
  \propto  [\hat{k}_{i_1} \cdots \hat{k}_{i_n} ]^{\rm TL_3} \label{eq:phi_integral}.
\end{align}
This can be understood by noticing that since the left hand side of Eq.~(\ref{eq:phi_integral}) should be independent of $\varphi$, all the tensor indicies should be along the direction of $\hat{\bm k}$, satisfying the symmetric traceless condition. 

The amplitude of the left hand side of Eq.~(\ref{eq:phi_integral}) can be determined as follows. As shown in Ref.~\citep{Ehrentraut1998} (in this paper, the spatial dimension is $d$, so here we set $d=3$), the contraction between $[\hat{p}_{i_1} \cdots \hat{p}_{i_n}]^{\rm TL_3}$ and
$[\hat{k}_{i_1} \cdots \hat{k}_{i_n}]^{\rm TL_3}$ is given by
\begin{align}
  [\hat{p}^{i_1} \cdots \hat{p}^{i_n}]^{\rm TL_3}
  [\hat{k}_{i_1} \cdots \hat{k}_{i_n}]^{\rm TL_3}
 = \frac{n!}{(2n-1)!!}{\cal P}_n(\mu).  \label{eq:formula_inn_prod}
\end{align}
Operating $[\hat{k}^{i_1} \cdots \hat{k}^{i_n} ]^{\rm TL_3}$ on the left hand side of Eq.~(\ref{eq:phi_integral}) and using Eq.~(\ref{eq:formula_inn_prod}), we find that the amplitude of the left hand side of Eq.~(\ref{eq:phi_integral}) should be ${\cal P}_n(\mu)$, i.e.
\begin{align}\label{eq:appb_2}
  \int \frac{d\varphi}{2\pi} [\hat{p}_{i_1} \cdots \hat{p}_{i_n} ]^{\rm TL_3}
  = {\cal P}_n(\mu) [\hat{k}_{i_1} \cdots \hat{k}_{i_n} ]^{\rm TL_3}.
\end{align}
Here, we used 
\begin{align}
 [\hat{k}^{i_1} \cdots \hat{k}^{i_n}]^{\rm TL_3} 
     [\hat{k}_{i_1} \cdots \hat{k}_{i_n}]^{\rm TL_3}
  = \frac{n!}{(2n-1)!!}{\cal P}_n(\hat{\bm k}\cdot\hat{\bm k}=1)= \frac{n!}{(2n-1)!!} \,.
\end{align}
Using Eq.~(\ref{eq:appb_2}) in Eq.~(\ref{eq:appb_1}), we obtain
\begin{align}
\langle \delta(\bm{x})  \left\{K_{ij} (\bm{y})  K_{kl} (\bm{y}) \right\} \rangle
&=\frac{2}{9}{\cal A}_4\int \frac{d^3 k}{(2\pi)^3}e^{i \sbm{k} \cdot \sbm{r}}\mathcal{M}(k) P_{\Phi}(k)~[\hat{k}_i\hat{k}_j\hat{k}_k\hat{k}_l ]^{\rm TL_3}
\int \frac{p^2dp}{2\pi^2}\mathcal{M}^2(p)P_{\Phi}(p)\,,
\label{eq:bare_3p}
\end{align}
where we haved used the orthonormality of the Legendre polynomials.

\subsection{Renormalization}\label{sec:appc}
Next, we define the renormalized bias parameter $b^{(4)}_\textrm{NG}$ following the discussions in Refs.~\citep{Schmidt:2012ys,Schmidt:2015xka}.
\subsubsection{Gaussian initial conditions}
First, we consider the case of Gaussian initial conditions.
Let us introduce the coarse-grained density field $\delta_{\rm L}$ and tidal field $K_{{\rm L},ij}$ 
with a coarse-graining scale $R_{\rm L}$.
Then, using a functional $F_{{\rm L},ijkl}$, we can formally express $\tilde{g}_{ijkl}(\bm{x})$ as
\begin{align}
\tilde{g}_{ijkl}(\bm{x}) = F_{{\rm L},ijkl}(\delta_{\rm L}(\bm{x});K_{{\rm L},pq}(\bm{x});\delta_{\rm s}(\bm{x})),
\end{align}
with $\delta_{\rm s}(\bm{x}) \equiv \delta(\bm{x}) - \delta_{\rm L}(\bm{x})$ being the small-scale fluctuations
on which in principle the 4th moment depends other than $\delta_{\rm L}$ and $K_{{\rm L},ij}$.
A formal Taylor expansion of $F_{{\rm L},ijkl}$ in $\delta_{\rm L}$ and $K_{{\rm L},ij}$ leads to
\begin{align}  
\tilde{g}_{ijkl}(\bm{x}) =
c^{(4)}_{K^2}(R_{\rm L}; \delta_s(\bm{x})) \left[ K_{{\rm L},ij} K_{{\rm L},kl} \right]^{{\rm TL}_3,\,{\rm sym} }(\bm{x})
+\mathcal{O} ( \delta^3_{\rm L} , \nabla^2 \delta_{\rm L})\,.
\label{eq:bare_expand}
\end{align}
In general, the coefficient $c^{(4)}_{K^2}$ depend on the short modes $\delta_{\rm s}$ and the coarse-graining scale $R_{\rm L}$ \footnote{$R_{\rm L}$ is an arbitrary coarse-graining scale, while $R_{\rm L}$ should satisfy $R_{\rm L} > R_*$, where $R_*$ corresponds to the physical size of galaxies/halos.}.
In the Gaussian case, however, the short modes and long modes are uncorrelated, so the $c^{(4)}_{K^2}$ can be regarded as an effective constant.

The spherical symmetry requires the expectation value of $\tilde{g}_{ijkl}(\bm{x}) $ to vanish.
Let us consider the following modification of the tidal field:
\begin{align}
K_{{\rm L},ij} (\bm{x},\tau) \to K_{{\rm L},ij} (\bm{x},\tau) + D(\tau) \beta_{ij};\ \ \  
\delta_{\rm L} (\bm{x},\tau) \to \delta_{\rm L} (\bm{x},\tau),
\end{align}
or equivalently the modification of the Newtonian potential,
\begin{align}
\Phi_{\rm N}(\bm{x}) \to \Phi_{\rm N} (\bm{x}) 
+ \frac{3}{4} \Omega_{\rm m0} H_0^2 (1+z) D(z) \beta_{ij}x^ix^j,
\end{align}
where $\beta_{ij}$ is a constant symmetric traceless tensor.
One can interpret this as the leading observable effect of a potential perturbation in the $k\to 0$ limit, as constant and pure-gradient potential perturbations can be removed by coordinate transformations. Alternatively, this effect can be realized in simulations by implementing an anisotropic expansion, roughly resembling a Bianchi~I spacetime \cite{Schmidt:2018hbj,Masaki:2020drx,Akitsu:2020fpg}.

Under this modification, the expectation value of $\tilde{g}_{ijkl}$ changes to
\begin{align}
\langle \tilde{g}_{ijkl}(\bm{x}) \rangle_\beta = c^{(4)}_{K^2}\left[ \beta_{ij}\beta_{kl} \right]^{{\rm TL}_3,\,{\rm sym} }
+ \mathcal{O}(\delta_{\rm L}^3 , \nabla^2 \delta_{\rm L}).
\end{align}
Notice that $\langle \tilde{g}_{ijkl}(\bm{x}) \rangle$ does not vanish owing to the presence the preferred direction $\beta_{ij}$.
Then, the renormalized bias parameters are introduced via
\begin{align}
b^{(4)}_{K^2} \equiv \left.\left[ \frac{\partial^2}{ \partial \beta_{ij}\partial\beta_{kl} } \right]^{{\rm TL}_3,\,{\rm sym} }\langle \tilde{g}_{ijkl} \rangle_\beta\right|_{\beta=0}
=c^{(4)}_{K^2} +\mathcal{O}([\delta_{\rm L}, K_{{\rm L},ij}]^2).
\end{align}
Note that here the summation is not taken. 
This definition of the bias parameter $b^{(4)}_{K^2}$ is independent of $R_{\rm L}$ by construction.
The point is that the renormalized bias is defined as the response to the locally uniform transformation of the tidal field (second derivatives of the potential).

Finally we get
\begin{align}
    \langle \delta(\bm{x}) \tilde{g}_{ijkl}(\bm{y}) \rangle
    =b^{(4)}_{K^2}\left\langle \delta(\bm{x}) \left[ K_{ij}K_{kl} \right]^{{\rm TL}_3,\,{\rm sym} }(\bm{y}) \right\rangle +\cdots.
\label{eq:ren_bias_gauss}
\end{align}
For Gaussian initial conditions, the three-point function in Eq.~(\ref{eq:ren_bias_gauss}) only arises from non-linear evolution, therefore is relevant only on small scales.

\subsubsection{Non-Gaussian initial conditions with no scaling (local-type)}
In the presence of the primordial non-Gaussianity (\ref{eq:NGIC}), Eq.~(\ref{eq:bare_3p}) yields
\begin{align}
\left\langle \delta(\bm{x})  \left[ K_{ij}K_{kl} \right]^{{\rm TL}_3,\,{\rm sym} }(\bm{y})  \right\rangle
=\frac{2}{9}{\cal A}_4 \mathcal{D}_{ijkl} \xi_{\delta \Phi} (|\bm{x}-\bm{y}|) \langle \delta^2_{\rm L} \rangle.
\end{align}
This expression strongly depends on the coarse-graining scale through $\langle \delta_{\rm L}^2 \rangle$, so the bias expansion Eq.~(\ref{eq:ren_bias_gauss}) is not sufficient for non-Gaussian initial conditions. 
In this case, we have to take into account the dependence of $g_{ijkl}$ on the small-scale fluctuations $\delta_{\rm s}$ explicitly,
since the primordial non-Gaussianity couples the long-modes with short-modes.
To do this, let us introduce the parameter $y_{\rm s}^{ijkl}$ as the hexadecapole anisotropy of the local small-scale correlation function within a region of size $R_{\rm L}$,
\begin{align}
y_{\rm s}^{ijkl}(\bm{x}) =& \frac{1}{\sigma_y^2}
\int d^3 \bm{r} ~ W_{\rm L}(|\bm{r}|)
\left[
K_{\rm s}^{ij}\left(\bm{x}-\frac{\bm{r}}{2}\right) K_{\rm s}^{kl}\left(\bm{x}+\frac{\bm{r}}{2}\right) 
\right]^{{\rm TL}_3,\,{\rm sym} },
\label{eq:y_s}
\\
\sigma_y^2 =& \int \frac{d^3 \bm{k}}{(2\pi)^3} \widetilde{W}_{\rm L}(k)  \widetilde{W}^2_{\rm s}(k) P_{\rm m}(k),
\end{align}
where $W_{\rm L}$ is an isotropic window function of the scale $R_{\rm L}$, $K_{\rm s}^{ij}({\bm x}\pm {\bm r}/2) \equiv \mathcal{D}_r^{ij}\delta_{\rm s}({\bm x}\pm {\bm r}/2)$ with $\mathcal{D}_r^{ij}$ being the derivative operator acting on $\bm{r}$ and $\widetilde{W}_{\rm s}(k) \equiv 1-\widetilde{W}_{\rm L}(k) $.
We now introduce explicitly the dependence of $g_{ijkl}(\bm{x})$ on $y^{ijkl}_{\rm s}(\bm{x})$:
\begin{align}
\tilde{g}_{ijkl}(\bm{x})
=F_{{\rm L},ijkl} (\delta_{\rm L}(\bm{x}); K_{{\rm L},pq}(\bm{x}) ;y_{\rm s}^{pqrs}(\bm{x})).
\end{align}
Expanding $F_{{\rm L},ijkl}$ to linear order in $y_{\rm s}^{pqrs}$ leads to adding a term
\begin{align}
c^{(4)}_\textrm{NG} y_{\rm s}^{ijkl} (\bm{x})
\end{align}
to the expansion on the r.h.s. of Eq.~(\ref{eq:bare_expand}). Then, we have additional contribution to
matter-shape correlation,
\begin{align}
\langle \delta(\bm{x}) \tilde{g}_{ijkl}(\bm{y}) \rangle
=&b^{(4)}_{K^2} \left\langle \delta(\bm{x})  \left[ K_{{\rm L},ij} K_{{\rm L},kl} \right]^{{\rm TL}_3,\,{\rm sym} }(\bm{y})\right\rangle
+c^{(4)}_\textrm{NG}\langle \delta(\bm{x}) y_{\rm s}^{ijkl}(\bm{y}) \rangle
+\cdots.
\label{eq:with_counter_term}
\end{align}
The Fourier transform of Eq. (\ref{eq:y_s}) is given by
\begin{align}
    y_{\rm s}^{ijkl}(\bm{k})
    =
    \frac{1}{\sigma_y^2}
    \int\frac{d^3 \bm{p}_1}{(2\pi)^3} \left[\hat{p}_1^i\hat{p}_1^j\hat{p}_2^k\hat{p}_2^l \right]^{{\rm TL}_3,\,{\rm sym} }
    \widetilde{W}_{\rm L}\left(\left|-\bm{p}_1+\frac{1}{2}\bm{k}\right|\right)
    \delta_{\rm s}(\bm{p}_1)\delta_{\rm s}(\bm{p}_2),
\end{align}
where we introduced ${\bm p}_1$ and ${\bm p}_2$ satisfying ${\bm p}_1+{\bm p}_2={\bm k}$.
We then obtain
\begin{align}
    \langle \delta(\bm{x}) y^{ijkl}_{\rm s} (\bm{y}) \rangle
    =&\frac{1}{\sigma^2_y}
    \int \frac{d^3 \bm{k}}{(2\pi)^3} e^{-i\sbm{k}\cdot \sbm{r}}\mathcal{M}(k)
    \int \frac{d^3 \bm{p}_1}{(2\pi)^3} \widetilde{W}_{\rm L}\left(\left|-\bm{p}_1+\frac{1}{2}\bm{k}\right|\right)
    \left[\hat{p}_1^i\hat{p}_1^j\hat{p}_2^k\hat{p}_2^l \right]^{{\rm TL}_3,\,{\rm sym} }
     \nonumber\\
     &\times \mathcal{M}_{\rm s}(p_1)\mathcal{M}_{\rm s}(|\bm{k}+\bm{p}_1|)
     B_\Phi(k, p_1, |\bm{k}+\bm{p}_1|)
\end{align}
with $\bm{r} \equiv \bm{x}-\bm{y}$
and $\mathcal{M}_{\rm s}(k) \equiv \mathcal{M}(k)\widetilde{W}_{\rm s}(k)$.
Expanding this integrand in power of $q_1=k/p_1$ and performing the angle integral with respect to $\bm{p}_1$, we have
\begin{align}
    \langle \delta(\bm{x}) y^{ijkl}_{\rm s} (\bm{y}) \rangle =&
    \frac{1}{\sigma_y^2}
    \int \frac{d^3 \bm{k}}{(2\pi)^3} e^{i\sbm{k}\cdot \sbm{r}}\mathcal{M}(k) P_\Phi(k) 
    \left[\hat{k}^i\hat{k}^j\hat{k}^k\hat{k}^l \right]^{\rm TL_3}
    \int\frac{p_1^2 dp_1}{(2\pi)^2} \widetilde{W}_{\rm L}(p_1) \mathcal{M}_{\rm s}^2(p_1)P_\Phi(p_1)
    \nonumber \\
    &\ \times \int_{-1}^1 d\mu_1 \mathcal{P}_4(\mu_1) 
    \sum_\ell {\cal A}_\ell \mathcal{P}_\ell(\mu_1)
    \left[2 +\mathcal{O}(q_1)\right]\,,
\end{align}
where we have used $\widetilde{W}_{\rm L}\left( \left|-\bm{p}_1 +\frac{1}{2}\bm{k} \right|  \right) = \widetilde{W}_{\rm L}(p_1) + {\cal O}(q_1) $ and $\hat{p}_2 =-\hat{p}_1 + {\cal O}(q_1)$.
Following the same strategy of the calculation of Eq.~(\ref{eq:bare_3p}), we obtain
\begin{align}
\langle \delta(\bm{x}) y_s^{ijkl}(\bm{y}) \rangle = \frac{2}{9} {\cal A}_4 \mathcal{D}_{ijkl} \xi_{\Phi\delta}(|\bm{x}-\bm{y}|).
\end{align}
Thus, there are two non-Gaussian terms which are proportional to ${\cal A}_4\mathcal{D}_{ijkl}\xi_{\Phi\delta}$, 
one of which explicitly depends on $R_{\rm L}$.
A renormalized bias $b^{(4)}_\textrm{NG}$ should consist of $R_{\rm L}$-independent combination of these contributions.
In other words, 
physically, $b^{(4)}_\textrm{NG}$ should correspond to the response of the 4th moment to a specific ($R_{\rm L}$-independent) transformation of the density field as implied in the previous subsection.

In fact, under the initial condition given by Eq.~(\ref{eq:NGIC}) 
the local power spectrum is modulated like Eq.~(\ref{eq:alps}),
which means that the long-wavelength potential perturbation leads an anisotropic modulation of the local initial matter power spectrum described by
\begin{align}
    P^\textrm{ini}_{{\rm m},\alpha} (\bm{k}_{\rm S}; {\bm x}) = \left[ 1+  \alpha_{{\rm L}\,pqrs}({\bm x})
    \left[ \hat{k}_{\rm S}^p\hat{k}_{\rm S}^q\hat{k}_{\rm S}^r\hat{k}_{\rm S}^s \right]^{\rm TL_3}\right]
        P^\textrm{ini}_{{\rm m},\textrm{iso}}(k_{\rm S}),
\end{align}
with
\begin{align}
    \alpha_{{\rm L}\,pqrs}({\bm x}) \equiv \frac{35}{8}
    \int \frac{d^3{\bm k}_{\rm L}}{(2\pi)^3}  {\cal A}_4
    \left[\hat{k}_{{\rm L},p} \hat{k}_{{\rm L},q} \hat{k}_{{\rm L},r} \hat{k}_{{\rm L},s} \right]^{\rm TL_3}\Phi(\bm{k}_{\rm L})e^{i\sbm{k}_{\rm L} \cdot \sbm{x}}.
\end{align}
Note that we treat $\alpha_{{\rm L},pqrs}$ as a locally constant number.
Here, we assumed $f_4(k_{\rm L}/k_{\rm S}) \simeq 1$. 
This implies that in terms of the density field the presence of the anisotropic non-Gaussianity alters the local density field such that
\begin{align}
\delta_\alpha(\bm{k}_{\rm S}) = 
\left[ 1+\frac12 \alpha_{{\rm L}\,pqrs}
    \left[ \hat{k}_{\rm S}^p\hat{k}_{\rm S}^q\hat{k}_{\rm S}^r\hat{k}_{\rm S}^s \right]^{\rm TL_3} \right]
\delta(\bm{k}_{\rm S})\,.
\label{eq:AppE_local_delta_local}
\end{align}
This is an anisotropic (hexadecapole), scale-independent rescaling of the density field.
After this transformation, the expectation value of the 4th moment galaxy shape changes to
\begin{align}
\langle \tilde{g}_{ijkl} \rangle_{\alpha}
=
b^{(4)}_{K^2}  \left\langle \left[K_{{\rm L},ij} K_{{\rm L},kl} \right]^{{\rm TL}_3,\,{\rm sym} } \right\rangle_\alpha
+c^{(4)}_\textrm{NG}\left\langle y_{\rm s}^{ijkl} \right\rangle_\alpha,
\end{align}
where
\begin{align}
& \left\langle\left[K_{{\rm L},ij} K_{{\rm L},kl} \right]^{{\rm TL}_3,\,{\rm sym} } \right\rangle_\alpha 
=
\int \frac{d^3 \bm{k}}{(2\pi)^3} 
\left[ \hat{k}_i \hat{k}_j \hat{k}_k \hat{k}_l \right]^{\rm TL_3}
\left(1+ \alpha_{{\rm L}\,pqrs}\left[ \hat{k}^p\hat{k}^q\hat{k}^r\hat{k}^s \right]^{\rm TL_3} \right)
\widetilde{W}^2_{\rm L}(k)P_{\rm m}(k),
\label{eq:del_psi_average}
\\
& \left\langle y_{\rm s}^{ijkl} \right\rangle_\alpha = 
\frac{1}{\sigma_y^2} \int \frac{d^3 \bm{k}}{(2\pi)^3} 
\left[ \hat{k}_i \hat{k}_j \hat{k}_k \hat{k}_l \right]^{\rm TL_3}
\left(1+ \alpha_{{\rm L}\,pqrs}\left[\hat{k}^p\hat{k}^q\hat{k}^r\hat{k}^s\right]^{\rm TL_3} \right)
\widetilde{W}_{\rm L}(k)\widetilde{W}^2_{\rm s}(k)P_{\rm m}(k).
\label{eq:y_s_average}
\end{align}
To proceed the computation of Eq.~(\ref{eq:y_s_average}), we use the following identity:
\begin{align}
\int \frac{d^2\hat{k}}{4\pi}
\hat{k}_{i_1}\cdots\hat{k}_{i_n}
=\frac{1}{(n+1)!!} [\delta_{i_1 i_2}\cdots\delta_{i_{n-1} i_n}]^{\rm sym}\,,
\end{align}
where $ [\delta_{i_1 i_2}\cdots\delta_{i_{n-1} i_n}]^{\rm sym}$ means to symmetrize the expression $\delta_{i_1 i_2}\cdots\delta_{i_{n-1} i_n}$ in the indices $i_1\cdots i_n$.
For instance, the explicit expression for $n=4$ case is given by
\begin{align}
 [\delta_{i_1 i_2}\delta_{i_3 i_4}]^{\rm sym}
= \delta_{i_1 i_2}\delta_{i_3i_4} +\delta_{i_1 i_3}\delta_{i_2i_4}  + \delta_{i_1i_4}\delta_{i_2i_3}.
\end{align}
After some algebra, we get
\begin{align}
\left\langle  \left[ K_{{\rm L},ij} K_{{\rm L},kl} \right]^{{\rm TL}_3,\,{\rm sym} } \right\rangle_\alpha  
\Big/ \left\langle \delta^2_{\rm L} \right\rangle =
\left\langle y_{\rm s}^{ijkl} \right\rangle_\alpha = \frac{4!}{9!!} \alpha_{{\rm L}\,ijkl}.
\end{align}
Then, the expectation of the 4th moment shape function is modified as
\begin{align}
\langle \tilde{g}_{ijkl} \rangle_\alpha
=\left[ \frac{8}{315}b^{(4)}_{K^2}\langle \delta_{\rm L}^2 \rangle +\frac{8}{315}c_\textrm{NG}
\right] \alpha_{{\rm L}\,ijkl}.
\end{align}
This tells us that the linear response of the mean 4th moment of galaxies, through which we define the renormalized bias $b^{(4)}_\textrm{NG}$, is given by
\begin{align}
b^{(4)}_\textrm{NG}
\equiv\left.\frac{\partial \langle \tilde{g}_{ijkl} \rangle_\alpha}{\alpha_{{\rm L}\,ijkl}}\right|_{\alpha=0}
=\frac{8}{315}b^{(4)}_{K^2}\langle \delta_{\rm L}^2 \rangle +\frac{8}{315}c^{(4)}_\textrm{NG}.
\end{align}
This means that the counter term should be given by
\begin{align}
c^{(4)}_\textrm{NG}
=\frac{315}{8}b^{(4)}_\textrm{NG}-
b^{(4)}_{K^2}\langle \delta_{\rm L}^2 \rangle.
\end{align}
Plugging this into Eq. (\ref{eq:with_counter_term}), we obtain
\begin{align}
\langle \delta(\bm{x}) \tilde{g}_{ijkl}(\bm{y}) \rangle
=&
\frac{35}{4}b^{(4)}_\textrm{NG}{\cal A}_4\mathcal{D}_{ijkl}\xi_{\delta \Phi}(|\bm{x}-\bm{y}|).
\end{align}

\subsubsection{Non-Gaussian initial conditions with scaling}
In this case, the modulation of the local initial matter power spectrum becomes
\begin{align}
    P^\textrm{ini}_{{\rm m},\alpha} (\bm{k}_{\rm S};{\bm x}) = \left[ 1+  h_{4}\left( \frac{k_{\rm p}}{k_{\rm S}} \right)\alpha_{{\rm L}\,pqrs}({\bm x})
    \left[ \hat{k}_{\rm S}^p\hat{k}_{\rm S}^q\hat{k}_{\rm S}^r\hat{k}_{\rm S}^s \right]^{\rm TL_3}\right]
        P^\textrm{ini}_{{\rm m},\textrm{iso}}(k_{\rm S}),
\end{align}
with
\begin{align}
    \alpha_{{\rm L}\,pqrs}({\bm x}) \equiv \frac{35}{8}
    \int \frac{d^3{\bm k}_{\rm L}}{(2\pi)^3} {\cal A}_4 g_{4}\left( \frac{k_{\rm L}}{k_{\rm p}} \right) 
    \left[\hat{k}_{{\rm L},p} \hat{k}_{{\rm L},q} \hat{k}_{{\rm L},r} \hat{k}_{{\rm L},s} \right]^{\rm TL_3}\Phi(\bm{k}_{\rm L})e^{i\sbm{k}_{\rm L} \cdot \sbm{x}}.
\end{align}
In terms of the density field, the presence of the anisotropic non-Gaussianity alters the local density field such that
\begin{align}
\delta_\alpha(\bm{k}_{\rm S}) = 
\left[ 1+\frac{1}{2}  h_{4}\left( \frac{k_{\rm p}}{k_{\rm S}} \right)\alpha_{{\rm L}\,pqrs}
    \left[ \hat{k}_{\rm S}^p\hat{k}_{\rm S}^q\hat{k}_{\rm S}^r\hat{k}_{\rm S}^s \right]^{\rm TL_3} \right]
\delta(\bm{k}_{\rm S}).
\end{align}
Obviously, in this case the local density field is modulated with the scale-dependence, 
unlike \eqref{eq:AppE_local_delta_local} where the local density field is rescaled uniformly.
In this case, defining the renormalized bias with respect to the locally uniform modulation leads to the final expression in the Fourier space,
\begin{align}
    \langle \delta(\bm{k}) \tilde{g}_{ijkl}(\bm{k}') \rangle
    =&
    \left[ \hat{k}_i \hat{k}_j \hat{k}_k \hat{k}_l \right]^{\rm TL_3}\frac{35}{4}b^{(4)}_\textrm{NG}{\cal A}_4
    g_{4}\left( \frac{k}{k_{\rm p}} \right)\mathcal{M}^{-1}(k)P_{\rm m}(\bm{k})(2\pi)^3\delta(\bm{k}+\bm{k}').
\end{align}

\acknowledgments
K.~K. is supported by JSPS KAKENHI Grant  No. JP19J22018. K.~A. is supported by JSPS KAKENHI Grant No. JP19J12254. F.~S. acknowledges support from the Starting Grant (ERC-2015-STG 678652) ``GrInflaGal'' of the European Research Council.
Y.~U. is supported by Grant-in-Aid for Scientific Research (B) under Contract No. 19H01894, Grant-in-Aid for Scientific Research on Innovative Areas under Contract Nos.~16H01095 and 18H04349, and the Deutsche Forschungsgemeinschaft (DFG, German Research Foundation) - Project number 315477589 - TRR 211.

\bibliography{main}
\end{document}